\documentclass[manuscript]{aastex}

\newcommand{\um}{$\mu$m}

\newcommand{\Spitzer}{{\it {Spitzer}}}
\newcommand{\kms}{km~s$^{-1}$}
\newcommand{\Msun}{M$_{\odot}$}

\newcommand{\hii}{\mbox{$\mathrm{H\,{\scriptstyle {II}}}$}}

\newcommand{\meth}{CH$_3$OH}
\newcommand{\degreesym}{$^{\circ}$}

\shorttitle{\meth\, Masers toward Green Sources}
\shortauthors{Chambers et al.}

\begin{document}

\title{Methanol Maser Emission from Galactic Center Sources with Excess 4.5~\um\, Emission}

\author{E. T. Chambers\altaffilmark{1}, F. Yusef-Zadeh\altaffilmark{1}, and D. Roberts\altaffilmark{1,2}}
\altaffiltext{1}{Department of Physics and Astronomy, Northwestern University, 
  Evanston, IL  60208}
\altaffiltext{2}{Adler Planetarium \& Astronomy Museum, 1300 S. Lake Shore Drive, Chicago, IL  60605}
\email{e-chambers@northwestern.edu, zadeh@northwestern.edu, doug-roberts@northwestern.edu}

\begin{abstract}

We present a study of signatures of on-going star formation in a sample of
protostellar objects with enhanced 4.5~\um\, emission (`green' sources) near
the Galactic center.  To understand how star formation in the Galactic center
region compares to that of the Galactic disk, we used the Expanded Very Large
Array to observe radiatively excited Class~II 6.7~GHz \meth\, masers and
collisionally excited Class~I 44~GHz \meth\, masers, both tracers of high-mass
star formation, toward a sample of 34 Galactic center and foreground `green'
sources.  We find that 33$\pm$15\% of Galactic center sources are coincident
with 6.7~GHz masers, and that 44$\pm$17\%\, of foreground sources are
coincident with 6.7~GHz masers.  For 44~GHz masers, we find correlation rates
of 27$\pm$13\% and 25$\pm$13\% for Galactic center green sources and
foreground green sources, respectively. Based on these \meth\, maser detection
rates, as well as correlations of green sources with other tracers of star
formation, such as 24~\um\, emission and infrared dark clouds (IRDCs), we find
no significant difference between the green sources in the Galactic center and
those foreground to it.  This suggests that once the star formation process
has begun, the environmental differences between the Galactic center region
and the Galactic disk have little effect on its observational signatures.  We
do find, however, some evidence that may support a recent episode of star
formation in the Galactic center region.

\end{abstract}


\section{Introduction}

Studying star formation in the Central Molecular Zone (CMZ) of the Galaxy
(within a few hundred pc of the Galactic center) can help provide insight into
the structure and evolution of our Milky Way Galaxy, as well as provide a
template for studying star formation in the nuclei of other galaxies.  The
CMZ, which contains $\sim$~5~$\times$~10$^{7}$~\Msun\, of molecular gas
\citep{pier00}, is home to several of the prominent star-forming regions
toward the Galactic center, such as the Sgr~A, Sgr~B, and Sgr~C \hii\,
complexes.  These regions are evolved enough to contain UV-emitting stars, but
the CMZ also shows evidence of earlier stages of star formation
\citep{lisz09,ball10}.  Among the best tracers of the early stages of star
formation are \meth\, masers and regions of extended, enhanced 4.5~\um\,
emission.

A new method of identifying star formation activity in its early stages is by
selecting sources with enhanced 4.5~\um\, emission.  Such sources, commonly
called green fuzzies \citep{cham09} or extended green objects
\citep[EGOs;][]{cyga08}, are named for how they appear in {\it Spitzer}/IRAC
\citep{fazi04} 3-color images (8.0~\um\, in red, 4.5~\um\, in green, and
3.6~\um\, in blue).  The enhancement at 4.5~\um\, likely arises from a
shock-excited H$_2$ or CO spectral feature in the 4.5~\um\, band (Marston et
al. 2004, Noriega-Crespo et al. 2004).  More recent work by \citet{debu10} and
\citet{fost11} has shown that the 4.5~\um\, emission from some green sources
is caused by shock-excited H$_2$ emission, while others show no prominent
spectral features.  Despite its uncertain origin, enhanced 4.5~\um\, emission,
frequently referred to as `green' emission in this paper, is a reliable tracer
of early protostellar activity \citep{cham09,cyga08,cyga09}.

Another tracer of star formation activity is one of the brightest known maser
lines, the \meth\, transition at 6.7~GHz.  This Class~II \meth\, maser
transition, which is thought to be radiatively excited by a central high-mass
protostellar object \citep{crag92}, exclusively traces high-mass
($\geq$~8~\Msun) star formation \citep{wals01, mini03}.  Class~II masers are
often found very close to protostars, supporting this idea \citep{casw97,
  elli05}.  Collisionally excited Class~I \meth\, masers are also thought to
be reliable tracers of star formation.  Unlike Class~II masers, however, they
are not restricted to high-mass star-forming regions, as evidenced by the
detection of Class~I \meth\, masers toward low- and intermediate-mass star
forming regions \citep{kale06,kale10}.  Class I masers tend to be found at
larger angular distances from protostars than Class~II masers
\citep{kurt04,elli05}, and are well-correlated with the outflows associated
with star formation \citep{plam90, kurt04, voro06}.  Recent work has shown
that Class~I \meth\, masers are also associated with regions of shocked gas
where expanding \hii\, regions collide with neighboring molecular clouds
\citep{voro10}.  Thus, Class~I masers are associated with two evolutionary
phases of star formation.  Because green sources are known to be in early
evolutionary phases, Class~I masers associated with them are likely to be
associated with protostellar outflows rather than \hii\, regions.

The first to propose a correlation between sources with enhanced 4.5~\um\,
emission and \meth\, masers was \citet{yuse07}, who found that at least 1/3 of
the green sources in the Galactic center are associated with \meth\, maser
emission.  Subsequent studies \citep{cham09, chen09, cyga09} have shown that
\meth\, masers are strongly associated with 4.5~\um\, excess sources in the
Galactic disk.  Moreover, these green sources are also highly associated with:
(1) 24~\um\, emission, indicative of heated dust around a protostar, and (2)
infrared dark clouds (IRDCs), which are known to harbor the early stages of
high-mass star formation (see Section~\ref{irdcs}).  Because both Class~I and
Class~II \meth\, maser emission are generated during high-mass star formation,
their association with 4.5~\um\, excess sources helps to confirm that the
green sources are indeed tracing the early stages of high-mass star formation.

\begin{figure}
\includegraphics[angle=-90,width=\textwidth,clip=true]{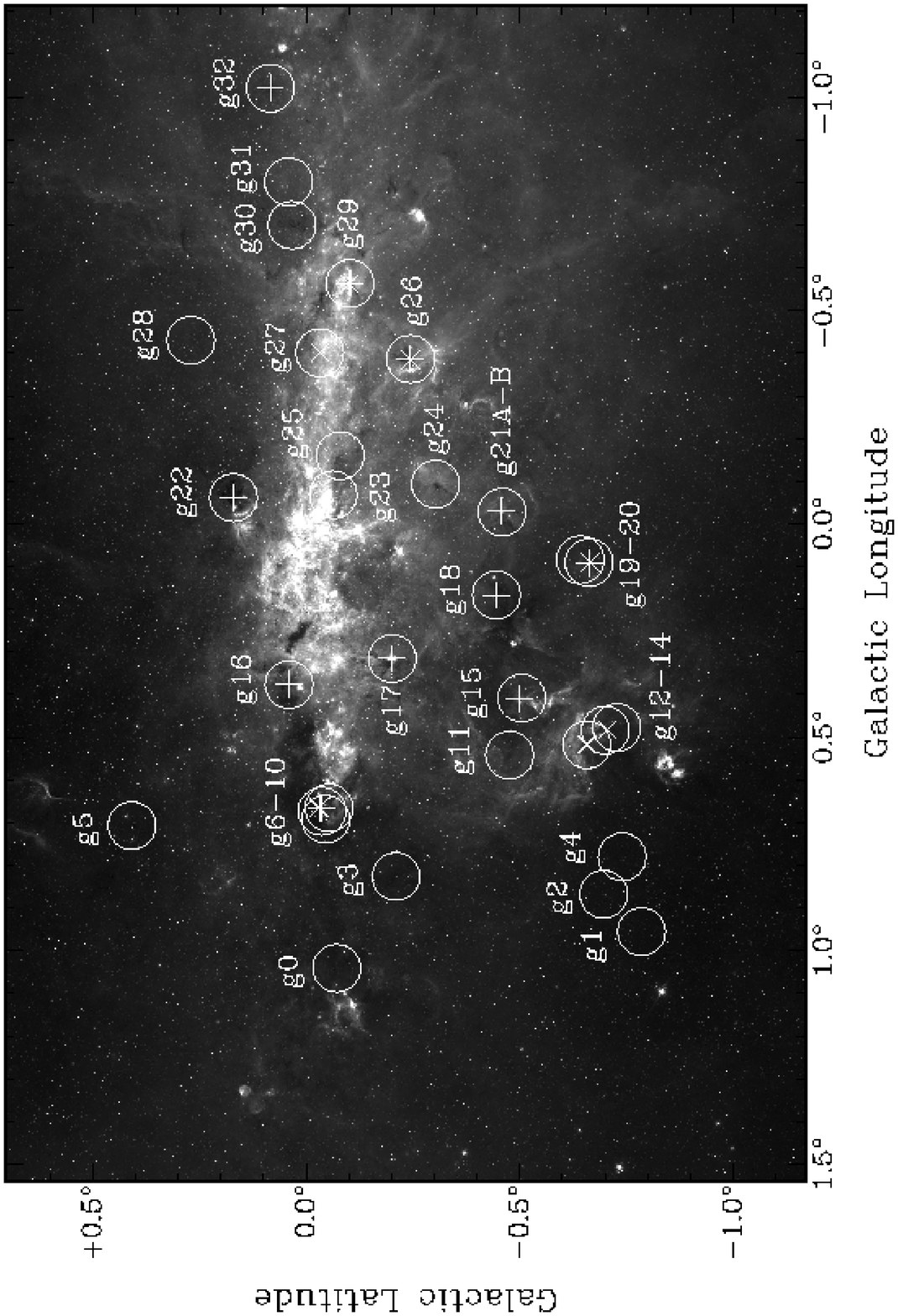}

\caption{{\it Left}: \Spitzer/IRAC 8.0~\um\, image of the Galactic center
  region.  Circles mark the positions of the sources with
  enhanced 4.5~\um\, emission that comprise our source list.  Plus signs ($+$)
  designate the positions of 6.7~GHz masers that are associated with the green
  sources, and cross signs ($\times$) designate the positions of 44~GHz masers
  that are associated with the green sources.\label{galcen_sources}}
\end{figure}

In a detailed examination of the star formation in the Galactic center,
\citet{yuse09}\, (Y-Z09 hereafter) identified and examined 34 green sources
toward the central region of the Galaxy ($|\ell|<1.05^{\circ}$ and
$|b|<0.8^{\circ}$; see Fig.~\ref{galcen_sources}) using \Spitzer/IRAC data.
Y-Z09 identified these sources, by eye, using the empirical `green' ratio
$I(4.5)/[I(3.6)^{1.2} \times I(5.8)]^{0.5}$, which helps select sources that
are enhanced at 4.5~\um\, relative to 3.6~\um\, and 5.8~\um.  Using available
published \meth\, maser data, primarily from targeted observations of known
star-forming regions and maser sites, Y-Z09 found that $\sim$~40\%\, of these
green sources are associated with Class~I and/or Class~II \meth\, maser
emission.  Due to the relatively high detection limit ($\sim$~0.3--5~Jy) of
these observations, however, it is possible that some masers have gone
undetected.  Here, we present our more sensitive ($<$~0.1~Jy), targeted, 6.7
and 44~GHz \meth\, maser observations toward these green sources using the
National Radio Astronomy Observatory's Expanded Very Large Array
(EVLA)\footnote[1]{The National Radio Astronomy Observatory is a facility of
  the National Science Foundation operated under cooperative agreement by
  Associated Universities, Inc.}.  In addition to \meth\, masers, we also
search for other star formation indicators that correlate with our sample of
4.5~\um\, emission sources, such as 24~\um\, emission and IRDCs.  We
investigate whether star formation in the CMZ differs from that in the
Galactic disk.

\section{Observations}

We used the new 4.0~-~8.0~GHz (C-band) capability of the EVLA to obtain
spectral line observations of the 6.6669~GHz 5(1,5)-6(0,6)~A$^{++}$ \meth\,
maser transition \citep{ment91}\, toward a sample of 34 green sources in the
central few degrees of the Galaxy (Fig.~\ref{galcen_sources}). The sample
selection and individual sources are described in Section~\ref{notes}.  To
observe all 34 sources, we made 31 snapshot observations (one field contains 3
green sources, and another contains 2 green sources) of $\sim$~80 seconds each
in 2007 December as part of Program AY184.  In total, 12 EVLA antennas were
used in the B configuration.  The spectra have 140~\kms\, (3.125~MHz)
bandwidth, 256 channels, and a channel width of 0.54~\kms\, (12.2~kHz).  The
primary beam of the EVLA at this frequency is $\sim$~7\arcmin, the synthesized
beams in the processed images are $\sim$~4\arcsec~x~1\arcsec, and the position
angles of the synthesized beams range from $-$14\degreesym\, to 2\degreesym,
with a mean of $-$7\degreesym.

To calibrate the 6.7~GHz data, we use standard procedures in AIPS along with a
few modifications to the standard recipe.  The calibrators are 1331+305 for
flux calibration and 1730-130 for phase calibration.  Doppler tracking was
used for these observations.  After the data were obtained, however, it was
determined that the path lengths of the fibers connecting the LO reference
signals to the antennas were not taken into account.  After the standard flux
and phase calibration, an additional phase modification is applied (using
CLCOR in AIPS) to each antenna to correct for this error.  Because the sources
are within a few degrees on the sky, only one constant offset is necessary,
rather than a more time/source specific offset.  This offset modifies the
phase only, and thus should not affect the positional or velocity accuracy of
the data.  For those sources with bright emission, self-calibration is
performed on the channel of peak emission (as identified in the AIPS task
POSSM), and subsequently applied to all other channels.  PBCOR is used to
correct the flux for primary beam attenuation.  The final sensitivity of these
6.7~GHz data is $\sim$~50~mJy~beam$^{-1}$ in each spectral channel.

In order to identify \meth\, masers at wide range of velocities, we made two
observations of each field, centered at what were intended to be
$\pm$~50~\kms.  Because Doppler tracking was used, however, an error in the
online system caused the observed frequencies to be calculated using the
submission date of the observe file rather than the actual date of
observation.  Using the online DOPSET tool and the date of the observations,
we calculated what the observing frequency should have been, and found that an
offset of 16.3-16.8~\kms\, was introduced by this error, depending on the
source.  Thus, the central velocities of the two observations for each field
are $\sim$~$-$33~\kms\, and $\sim$~$+$67~\kms, and our observations cover the
entire velocity range from $-$103~\kms\, to $+$136~\kms.

To supplement our observations, we also use a portion of the Methanol
Multibeam (MMB) survey catalog of 6.7~GHz masers \citep[][C10
  hereafter]{casw10}.  In their survey of the Galactic plane
(345\degreesym~$<~\ell~<$~6\degreesym, $|b|<$2\degreesym), C10 used the Parkes
64-m telescope to search for 6.7~GHz maser emission with a sensitivity of
$\sim$~170~mJy (1$\sigma$).  In order to pinpoint the location of the maser
emission to within 0.4\arcsec, follow-up observations of the Parkes detections
were carried out using the Australia Telescope Compact Array (ATCA).

We used the EVLA in the C configuration to observe the 44.069~GHz
7(0,7)-6(1,6)~A$^{++}$ \meth\, maser emission \citep{mori85}\, toward our
sample of 34 green sources in 2008 May as part of Program AY184.  As with the
6.7~GHz observations, these 34 sources are observed in 31 fields.  A total of
26 antennas were used, including 15 EVLA antennas.  At this frequency, the
primary beam is $\sim$~1\arcmin, the synthesized beams in our data are
$\sim$~1\arcsec~x~0.5\arcsec, and the position angles of the synthesized beams
range from $-$8\degreesym\, to 12\degreesym, with a mean of $-$5\degreesym.
The spectra have 84~\kms\, (12.5~MHz) bandwidth and 64 channels, resulting in
a channel width of 1.3~\kms\, (195~kHz).  Each source was observed for
$\sim$~2~minutes in each of two velocity ranges, from $\sim$~$-$89 to
$-$7~\kms\, and $\sim$~$-$2 to $+$80~\kms.  Because the velocity range from
$-$7~\kms\, to $-$2~\kms\, was not covered by these observations, it is
possible that masers exist in this range and were undetected by our
observations.  C10 found that only one of the 29 Class~II masers in the region
containing our green sources is in this velocity range.  As a result, it seems
unlikely that a significant number of masers have gone undetected in our
search.  Because the the possibility remains, however, our 44~GHz maser
detection rates should be considered lower limits.

Standard flux (1331+305) and phase (17443-31165) calibration are applied to
the data in AIPS.  Because these are high-frequency observations,
fast-switching was used for frequent phase monitoring.  Self-calibration is
performed on the channel of peak emission (as identified in the AIPS task
POSSM), and subsequently applied to all other channels.  PBCOR is used to
correct the flux for primary beam attenuation.  Because these observations
combined data from VLA and EVLA antennas, Doppler tracking was not used (as
recommended by NRAO).  During the data reduction process, Doppler offsets are
calculated using the online DOPSET tool. The final sensitivity of these
observations is $\sim$~70~mJy~beam$^{-1}$ in each spectral channel.

\section{Results}

To better understand excess 4.5~\um\, emission sources, and the star
formation occurring toward the Galactic center, we compiled several of their
characteristics.  Three of the 34 green sources (g1, g2, and g4) were found to
be positionally coincident with known planetary nebulae
\citep[][Y-Z09]{vand01,jaco04}. Because we are interested in studying regions
of star formation, we exclude these three sources from subsequent analysis.
To examine the remaining 31 sources, we compare their (1) 6.7~GHz Class~II
\meth\, maser emission, (2) 44~GHz Class~I \meth\, maser emission, (3)
24~\um\, emission using \Spitzer/MIPS \citep{riek04}\, data from Y-Z09, (4)
Galactic location, (5) mass, as estimated by Y-Z09, and (6) association with
IRDCs, using \Spitzer/IRAC data \citep{stol06, aren08, rami08}.  A summary of
these results can be found in Table~\ref{green-summary}.

\subsection{Association of 6.7~GHz \meth\, Masers with Green Sources}

Class~II 6.7~GHz \meth\, masers are radiatively excited and are known to trace
the early stages of high-mass star formation \citep[e.g.,][]{ment91, crag92,
  wals01, mini03}. Because of their strong association with high-mass star
formation, we searched for 6.7~GHz \meth\, masers toward our sample of green
sources.  We detect a total of 18 maser sites in the 31 observed fields.  The
masers are identified, by eye, in EVLA data cubes before self-calibration.
The typical 1$\sigma$\, sensitivity in these cubes is $\sim$~50~mJy.  Spectra
of the 18 maser sites are shown in Figures~\ref{g0}-\ref{g32}, and their
positions, velocities, and peak intensities can be found in
Table~\ref{six-summary}.  If a maser site displays more than one velocity
feature, the position and velocity of the peak intensity are listed.

All of our maser detections, except for one, are present in the C10 catalog.
Comparing the positions of the masers, they have a mean offset of 0.9\arcsec.
The maser positions from our data are identified by the position of the peak
pixel rather than an elliptical Gaussian fit, possibly resulting in an offset
of one or two pixels (our data have a pixel size of 0.3\arcsec).  C10 cite a
positional uncertainty of 0.4\arcsec\, for their catalog.  Together, the
positional uncertainties can account for the mean offset of 0.9\arcsec, and we
consider the masers that we detect to be positional matches to the C10 masers.
Moreover, we find an average offset in peak velocity of only 0.2~\kms\,
(compared to our channel width of 0.54~\kms) between the two sets of masers.
As a result, we are confident that we are detecting the same maser sources as
C10.

We detect one maser toward green source g31 (G359.199$+$0.041) that is not in
the C10 catalog, located at $\alpha, \delta$\, (J2000) $=
17^{h}43^{m}37.4^{s}, -29^{\circ}36'10.3''$\, with a velocity of $-$4.1~\kms.
This maser has a peak flux of $\sim$~1.4~Jy, not much greater than the 1~Jy
value at which the C10 survey is close to 100\% complete \citep{gree09}.
Because 6.7~GHz masers can be variable by factors of a few over timescales of
months \citep{goed04}, it is not surprising that this maser escaped detection
by C10.

There are four \meth\, masers, detected by C10, that are in our fields of view
but went undetected in our observations (0.167$-$0.446, 0.376+0.040,
358.980+0.084, and 359.970$-$0.457 in the naming scheme of C10).  Each of these
masers displayed variability of at least a factor of 2 in the different C10
observations, and their lowest observed intensities range from $<$~0.2~Jy to
1.3~Jy.  Their variability makes it plausible that these masers were in a low
or dormant state during our observations, resulting in their non-detection in
our observations.

Because of its excitation mechanism, 6.7~GHz maser emission is likely found
close to protostars.  Recent work \citep[e.g.,][]{cyga09} has shown that
6.7~GHz masers are indeed close (typically within a few arcseconds) to
central, star-forming objects within enhanced 4.5~\um\, sources.  To determine
which green sources in our sample are associated with 6.7~GHz masers, we
select a search radius of 10\arcsec.  We find that 12 of 31 green sources
are positionally coincident with at least one 6.7~GHz maser
(Fig.~\ref{galcen_sources}).

\subsection{Association of 44~GHz \meth\, Masers with Green Sources}
Class~I \meth\, masers are collisionally excited, and are thought to form both
in the outflows associated with star formation \citep{plam90,kurt04, voro06}
and at the intersections of expanding \hii\, regions and their neighboring
molecular clouds \citep{voro10}.  Because they are reliable tracers of star
formation, we searched for 44~GHz \meth\, maser emission toward our enhanced
4.5~\um\, emission sources.  We detect 8 masers in 31 fields.  As with the
6.7~GHz maser observations, the 44~GHz masers are identified by eye in EVLA
data cubes prior to self-calibration.  These data have a typical 1$\sigma$
sensitivity of $\sim$~70~mJy.  Spectra of the 8 maser sites are shown in
Figures~\ref{g0}-\ref{g32}, and their positions, velocities, and peak
intensities can be found in Table~\ref{forty-summary}.

Because we are searching for 44~GHz masers toward green sources, which are
known to be in the early stages of star formation, we will likely identify
masers that are associated with outflows rather than \hii\, regions.  The
collisionally excited masers formed in outflows are typically found at larger
angular separations \citep[up to tens of arcseconds;][]{cyga09} from central
protostellar objects than radiatively excited masers.  Indeed, \citet{cyga09}
find some 44~GHz masers beyond the extent of the 4.5~\um\, emission used to
identify their green sources.  We select a radius of 30\arcsec\, as the
maximum separation to associate a 44~GHz maser with a green source.  We find
that 8 of 31 green sources are positionally coincident with at least one
44~GHz maser (Fig.~\ref{galcen_sources}). It is possible that these sources
are not physically associated, especially if they are located at the Galactic
center distance of $\sim$~8.5~kpc (where a separation of 30\arcsec\,
corresponds to a linear size scale of $\sim$~1.2~pc).  Because the chance of a
random alignment of these two signs of star formation activity is low,
however, we use the angular separation of 30\arcsec\, even for the Galactic
center sources.

We find that some green sources are associated with both 44~GHz and 6.7~GHz
masers. Of the 12 green sources with 6.7~GHz \meth\, masers, 5 have 44~GHz
masers.  Of the 8 sources with associated 44~GHz masers, 5 have 6.7~GHz maser
counterparts.

\subsection{24~\um\, Emission toward Green Sources}
Bright 24~\um\, emission is often associated with star formation.  This
emission may be from the heated dust in a protostar/disk system \citep{muze04,
  whit04, beut07}, or it could arise from the heated dust in an \hii\, region.
\citet{cham09}\, find a high correlation of green sources with 24~\um\,
emission, further supporting the idea that it is a reliable tracer of star
formation.  To determine if our sample of green sources is coincident with
this additional star formation indicator, we visually inspected {\it
  Spitzer}/MIPS 24~\um\, data of the Galactic center (Y-Z09).  In regions
where the MIPS data are saturated (e.g., near Sgr~A$^*$), lower resolution
{\it MSX} data at 21.34~\um\, are used to replace the missing MIPS data (see
Y-Z09 for details).  We find that 24 of the 31 green sources are coincident
with 24~\um\, emission.  Thus, we can be reasonably sure that these sources
are indeed protostellar in nature.  Of the seven green sources that do not
display 24~\um\, emission, two green sources are close to \hii\, regions that
are very bright at 24~\um: Sgr~C and Sh2-20 \citep{shar59,dutr03}.  The bright
emission from these sources may be overwhelming any 24~\um\, emission toward
the green sources.  Thus, the 24~\um\, detection rate toward green sources of
77\% (24 of 31) should be considered a lower limit.

\subsection{The Galactic Location of Green Sources}
The green sources in our sample are divided into two categories -- one
consisting of sources likely to be in the Galactic center region at a distance
of $\sim$~8.5~kpc, and the other likely to be foreground to the Galactic
center.  As described in Y-Z09, the green sources were divided into these two
categories based on their Galactic latitude.  Sources with $|b|>10$\arcmin\,
are assumed to be in the foreground, and sources with $|b|<10$\arcmin\, are
assumed to be at the Galactic center distance.  Based on this separation
method, we find that 16 and 15 green sources in our sample are foreground
sources and Galactic center sources, respectively.  While this distance
estimate may not be exact, it is a reasonable first approximation.  Based on
the scale height of 24~\um\, sources and young stellar objects in the Galactic
center ($\sim$~8\arcmin; Y-Z09), it is plausible that the green sources at low
Galactic latitudes are at the distance of the Galactic center.  With kinematic
velocities, one could attempt to derive kinematic distances to these sources.
Because the sources are close in projection to the Galactic center, however,
the derived kinematic distances would have very large errors.  As a result, we
have not made distance estimates based on kinematics.

\subsection{Masses of Green Sources \label{masses}}
The masses of the green sources were determined by Y-Z09, who performed SED
fits of the sources using YSO models \citep{robi06, whit03a, whit03b} and a
linear regression fitter \citep{robi07}.  The fits to individual sources range
from well-constrained to poorly constrained, and have typical errors of
$\sim$~25\%.  The current masses derived from these fits range from 2.1 to
29.9~\Msun, with a median mass of 10.3~\Msun.  Because these green sources are
likely still accreting, their final masses may be larger than their current
mass, and the masses we list are a lower limit to the final masses of
individual stars.  Moreover, the green sources may consist of multiple sources
at higher resolution, so the derived masses may represent the mass of a
cluster of protostars rather than individual protostars.  Nevertheless, the
derived masses of the green sources, along with their positional coincidence
with Class~II \meth\, maser emission, make it likely that the green sources
harbor high-mass protostars.

\subsection{Association of Green Sources with IRDCs \label{irdcs}}
IRDCs, which are identified as absorption features against the Galactic IR
background, are dense (n~$>$~10$^5$~cm$^{-3}$, N~$\sim$~10$^{24}$~cm$^{-2}$),
cold ($<$ 25 K; Egan et al.  1998; Carey et al. 1998, 2000), and have
characteristic sizes and masses of $\sim$~5 pc and $\sim$~few 10$^3$~\Msun\,
(Simon et al. 2006).  These large reservoirs of molecular gas harbor the
earliest stages of star and cluster formation.  Within IRDCs are compact cores
with characteristic sizes of $\sim$~0.5~pc and masses of $\sim$~120~\Msun,
comparable to compact cores associated with high-mass star formation
\citep{rath06}.  Some IRDC cores contain embedded young stars or
protostars, and a few of these embedded young stellar objects will evolve into
high-mass stars \citep[e.g.,][]{beut05,rath05}.

\citet{cham09}\, found that the cores within IRDCs span a range of
evolutionary stages, and can be separated into three broad categories: (1)
`quiescent' cores, which display no bright IRAC (3-8~\um) or 24~\um\, emission
and are in a pre-protostellar state, (2) `active' cores, which contain
extended, enhanced 4.5~\um\, emission (a `green fuzzy') coincident with
24~\um\, emission, and (3) `red' cores, which display bright 8~\um\, emission,
indicative of polycyclic aromatic hydrocarbon (PAH) emission and an \hii\,
region.  We hypothesize that our sample of enhanced 4.5~\um\, sources are
similar in their protostellar nature to the green fuzzies found within IRDCs.
To test this possibility, we examined 8~\um\, IRAC images of the Galactic
center region to determine if our green sources are associated with IRDCs.  We
find that 30 of our 31 green sources are associated with IRDCs, as identified
by eye.  Images of these sources are shown in Figures~\ref{g0}-\ref{g32}.

\subsection{Notes on Individual Sources}\label{notes}

Our source list is comprised of the 34 green sources identified by Y-Z09
(named g0 through g32; g21 is parsed into g21A and g21B).  These sources were
selected for their proximity, in projection, to the Galactic center (all
sources have $|\ell|<1.05^{\circ}$ and $|b|<0.8^{\circ}$).  In addition, all
34 sources were selected, by eye, for their enhanced 4.5~\um\, emission, as
identified by the ratio: $I(4.5)/[I(3.6)^{1.2} \times I(5.8)]^{0.5}$, referred
to as the `green ratio' hereafter.  Y-Z09 found that this empirical green
ratio of 4.5~\um\, intensity to that determined by a power-law interpolation
between 3.6~\um\, and 5.8~\um\, intensities is successful at identifying
sources with enhanced 4.5~\um\, emission.

\begin{figure*}[p]
  \centering
  G1.041$-$0.072\\
  \includegraphics[width=0.4\textwidth,clip=true]{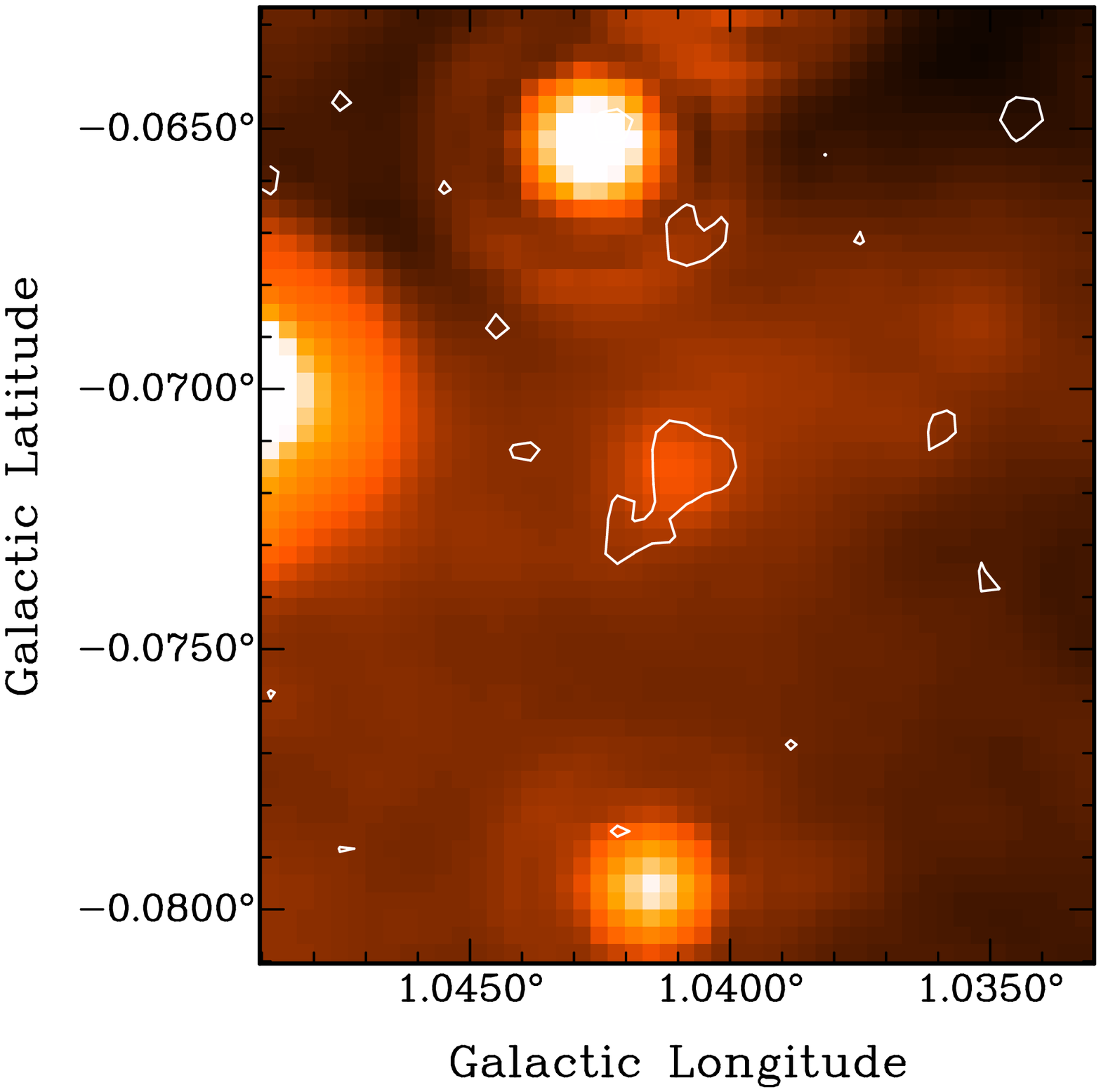}
  \hspace{-11.9cm}
  \includegraphics[width=0.4\textwidth,clip=true]{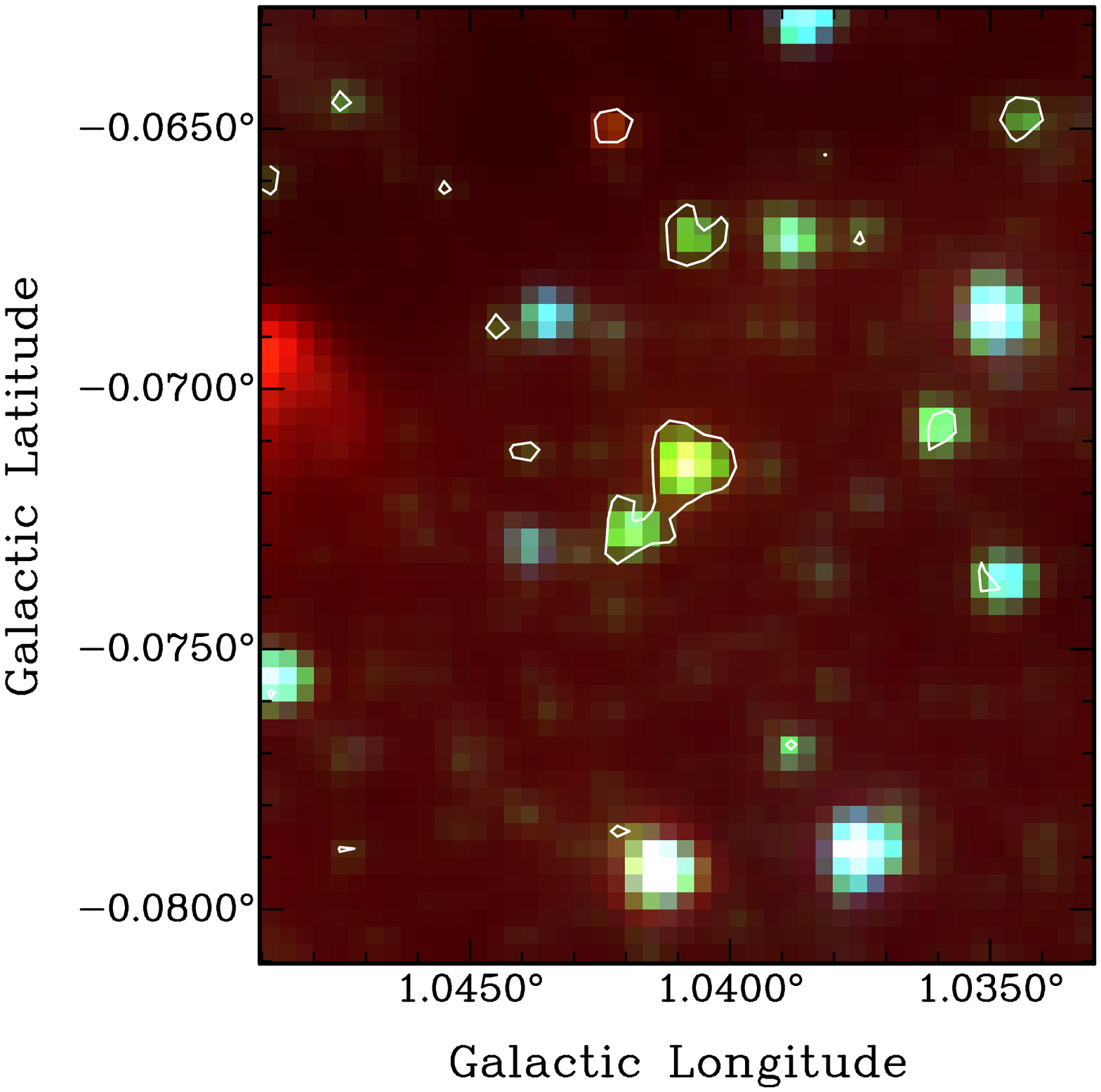}\\
  \caption{{\it Left}:  \Spitzer/IRAC 3-color image (with 8.0~\um\, in red,
  4.5~\um\, in green, and 3.6~\um\, in blue) of source g0.  {\it Right}:  
  \Spitzer/MIPS 24~\um\, image of source g0.  The contours in both images
  are at a green ratio value (see Section~\ref{notes}) of 0.50.\label{g0}}
\end{figure*}

\begin{figure*}[p]
  \centering
  G0.955$-$0.786\\
  \includegraphics[width=0.4\textwidth,clip=true]{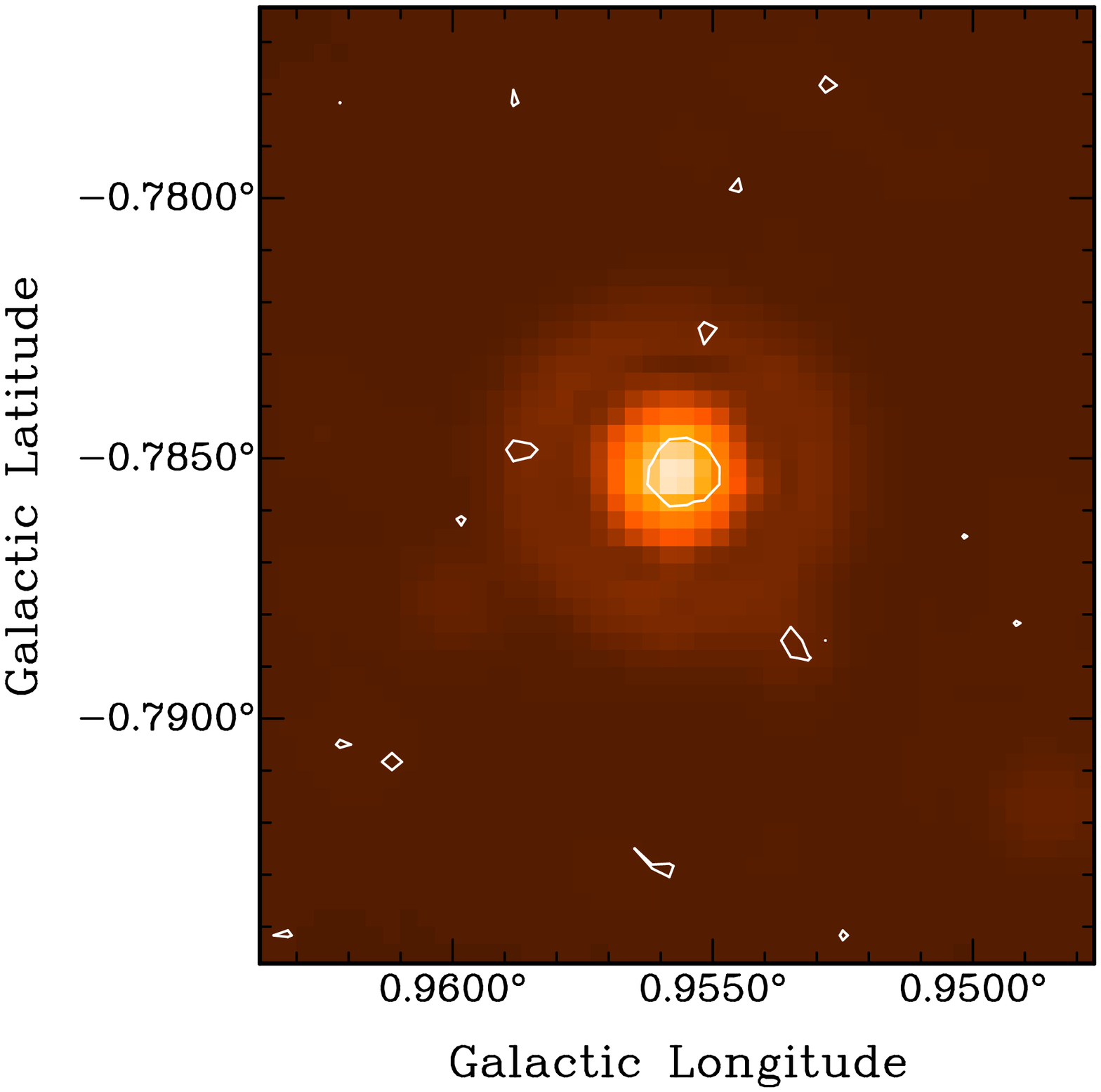}
  \hspace{-11.9cm}
  \includegraphics[width=0.4\textwidth,clip=true]{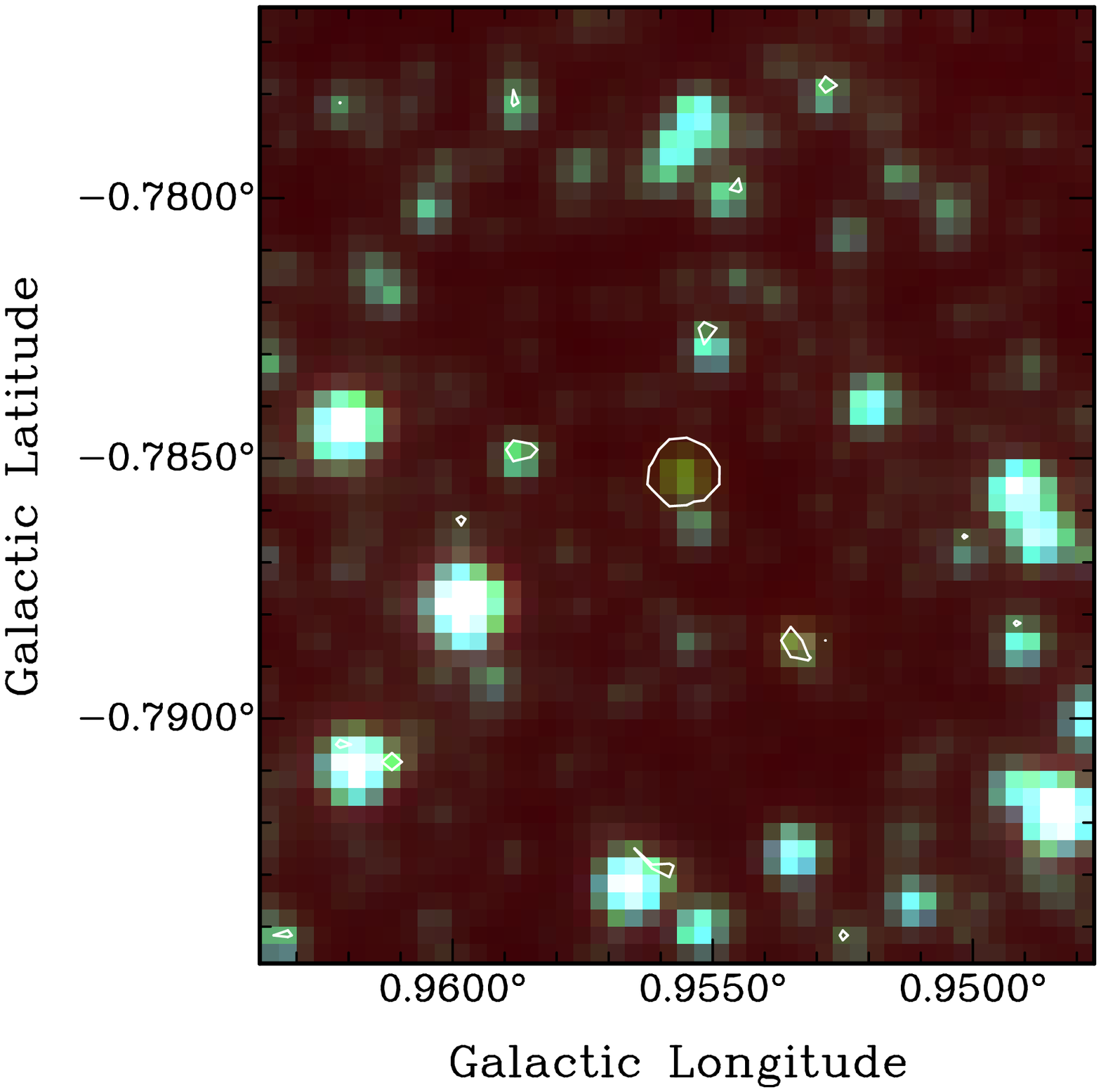}\\
  \caption{IRAC 3-color ({\it left}) and 24~\um\, ({\it right}) images of source g1.
  The contours in both images are at a green ratio value of 0.40.\label{g1}}
\end{figure*}

\begin{figure*}[p]
  \centering
  G0.868$-$0.697\\
  \includegraphics[width=0.4\textwidth,clip=true]{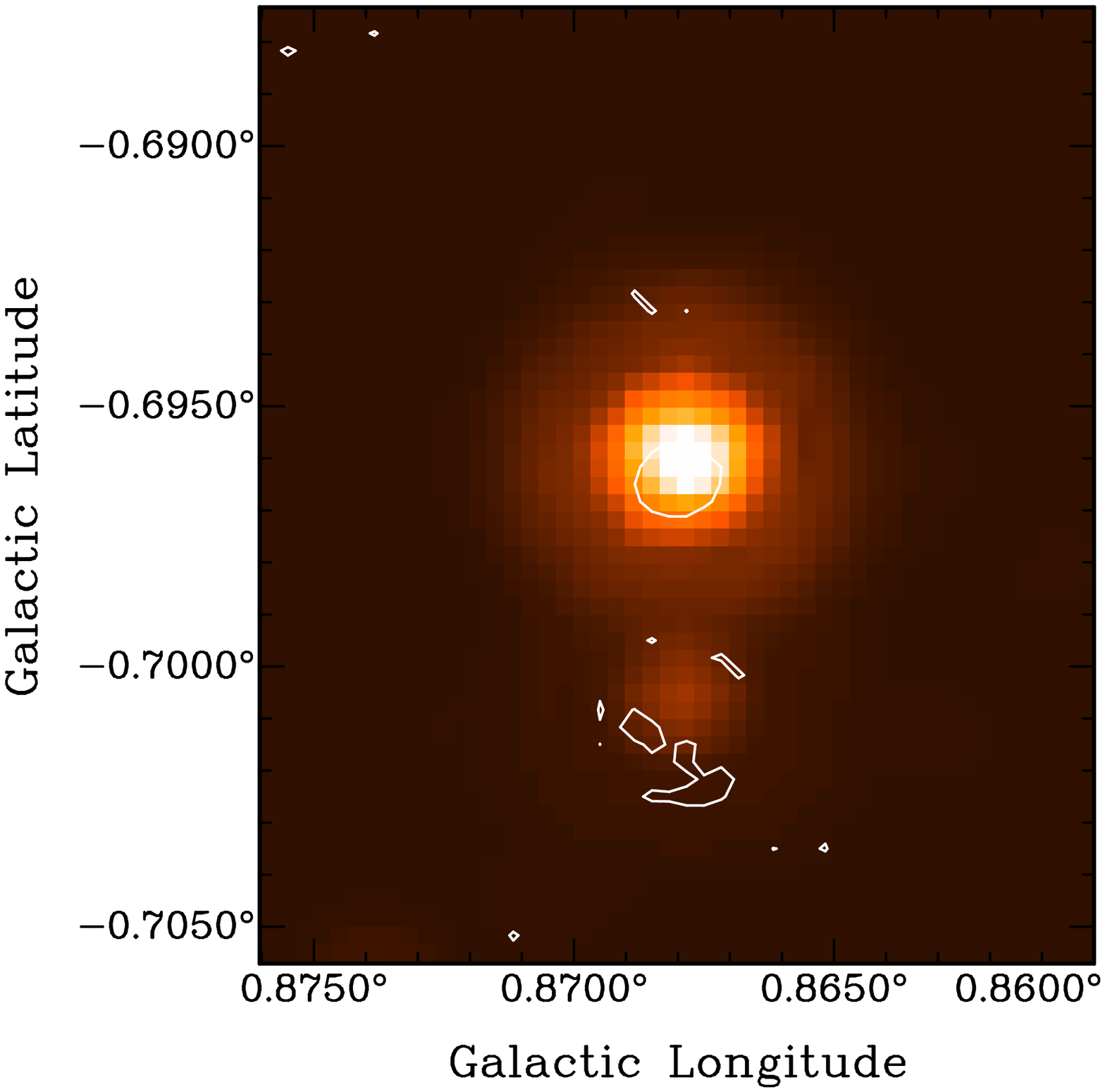}
  \hspace{-11.9cm}
  \includegraphics[width=0.4\textwidth,clip=true]{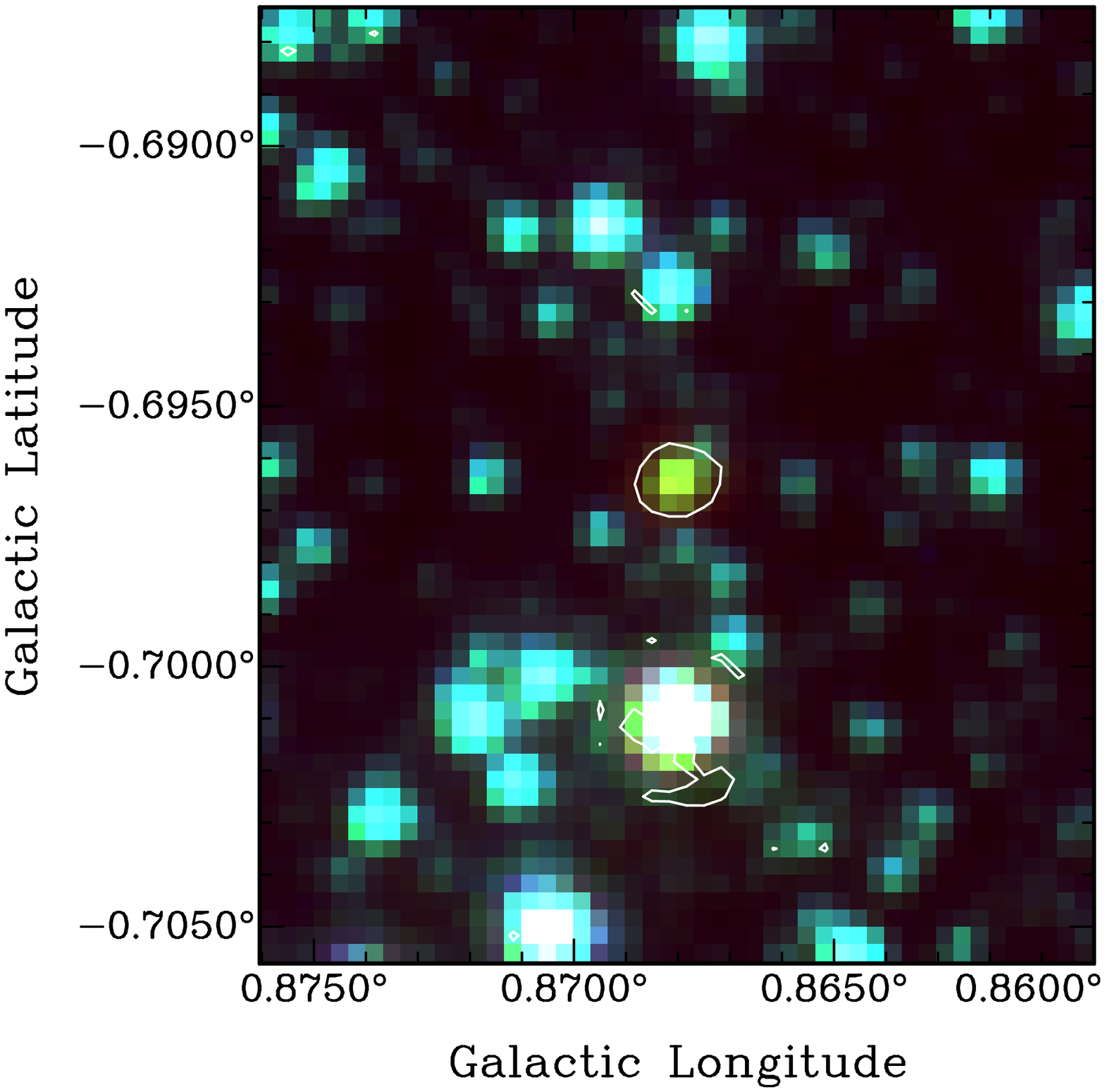}\\
  \caption{IRAC 3-color ({\it left}) and 24~\um\, ({\it right}) images of source g2.
  The contours in both images are at a green ratio value of 0.45.\label{g2}}
\end{figure*}

\begin{figure*}[p]
  \centering
  G0.780$-$0.740\\
  \includegraphics[width=0.4\textwidth,clip=true]{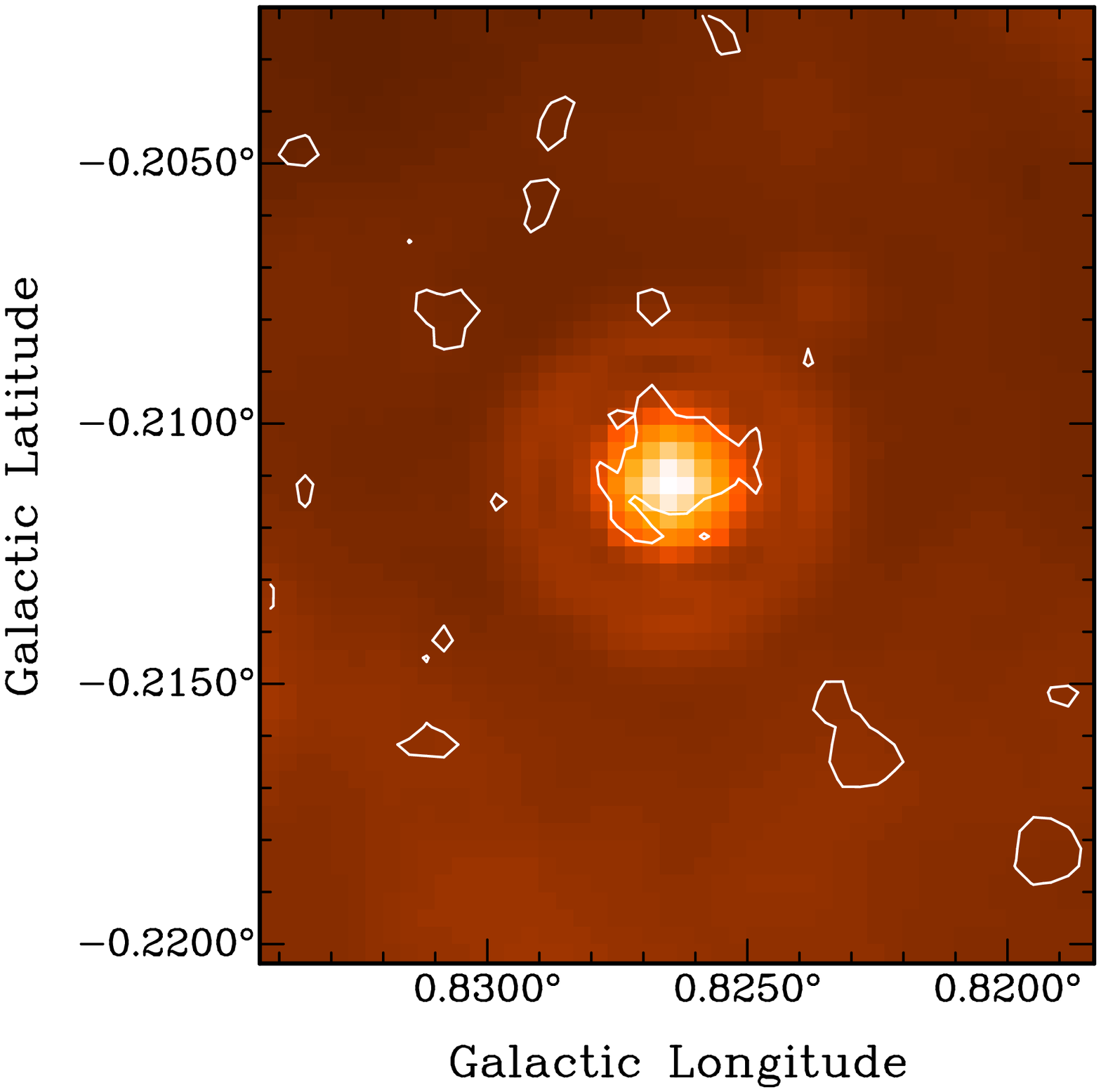}
  \hspace{-11.9cm}
  \includegraphics[width=0.4\textwidth,clip=true]{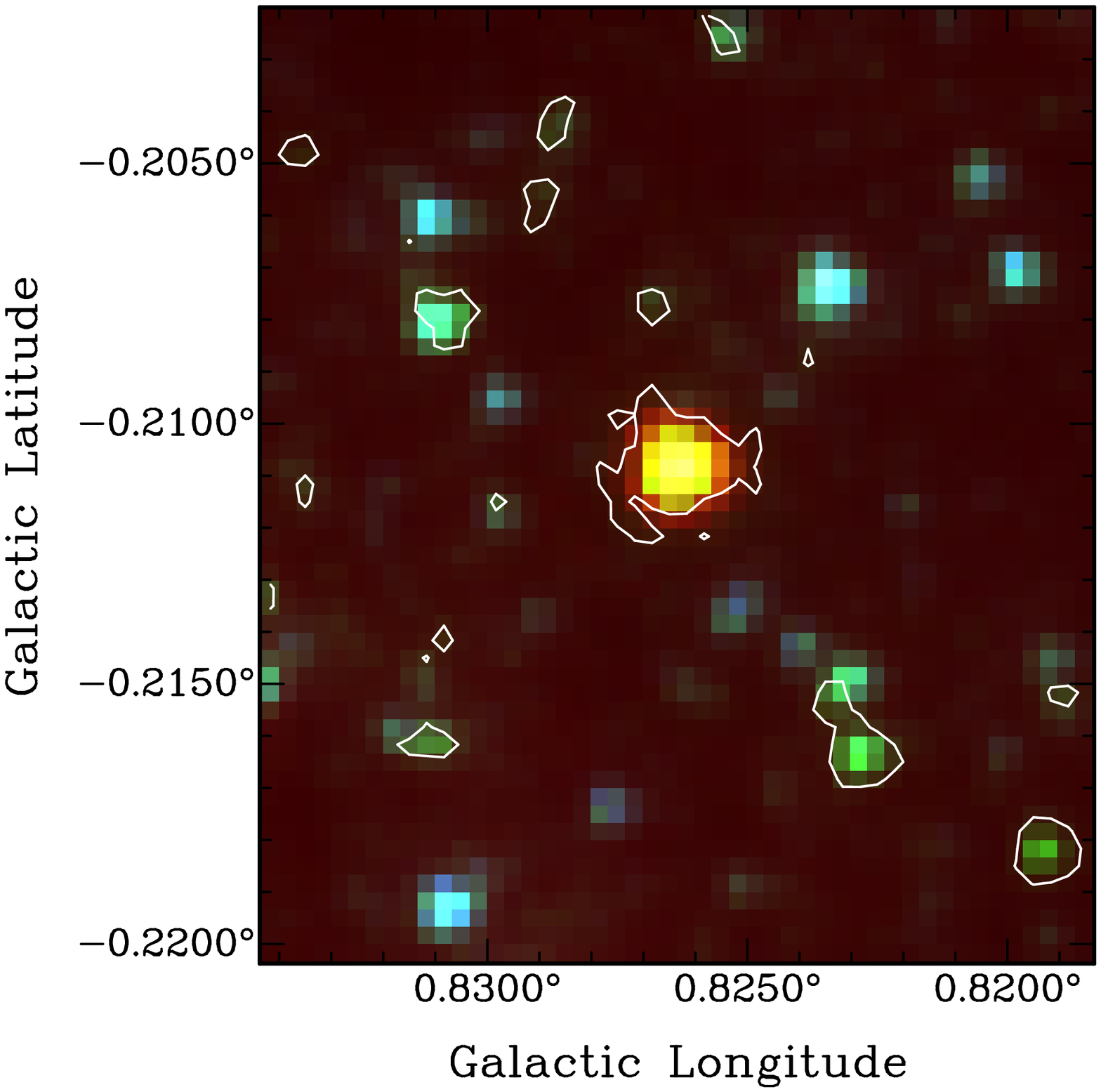}\\
  \caption{IRAC 3-color ({\it left}) and 24~\um\, ({\it right}) images of source g4.
  The contours in both images are at a green ratio value of 0.45.\label{g4}}
\end{figure*}

\begin{figure*}[p]
  \centering
  G0.826$-$0.211\\
  \includegraphics[width=0.4\textwidth,clip=true]{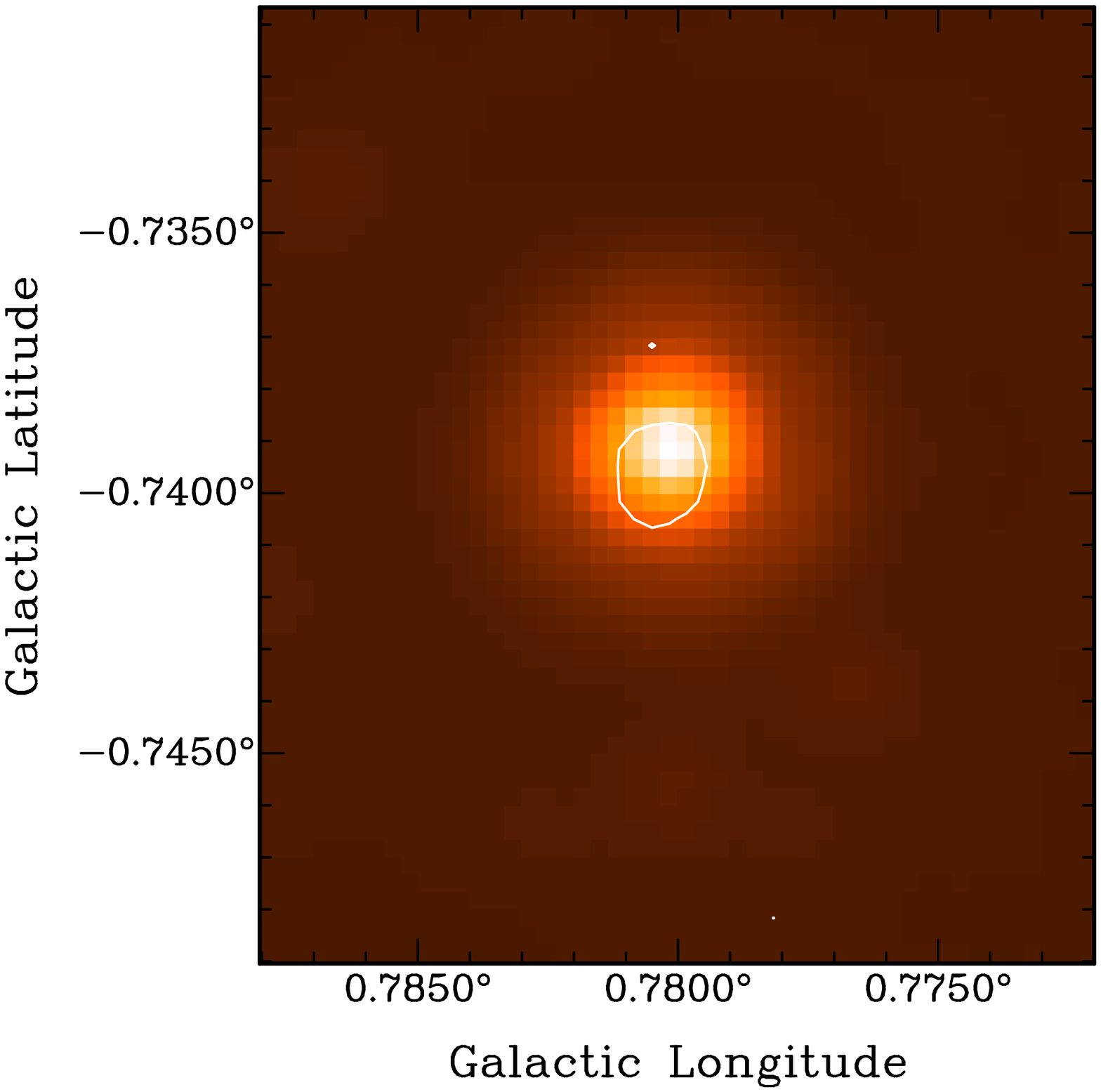}
  \hspace{-11.9cm}
  \includegraphics[width=0.4\textwidth,clip=true]{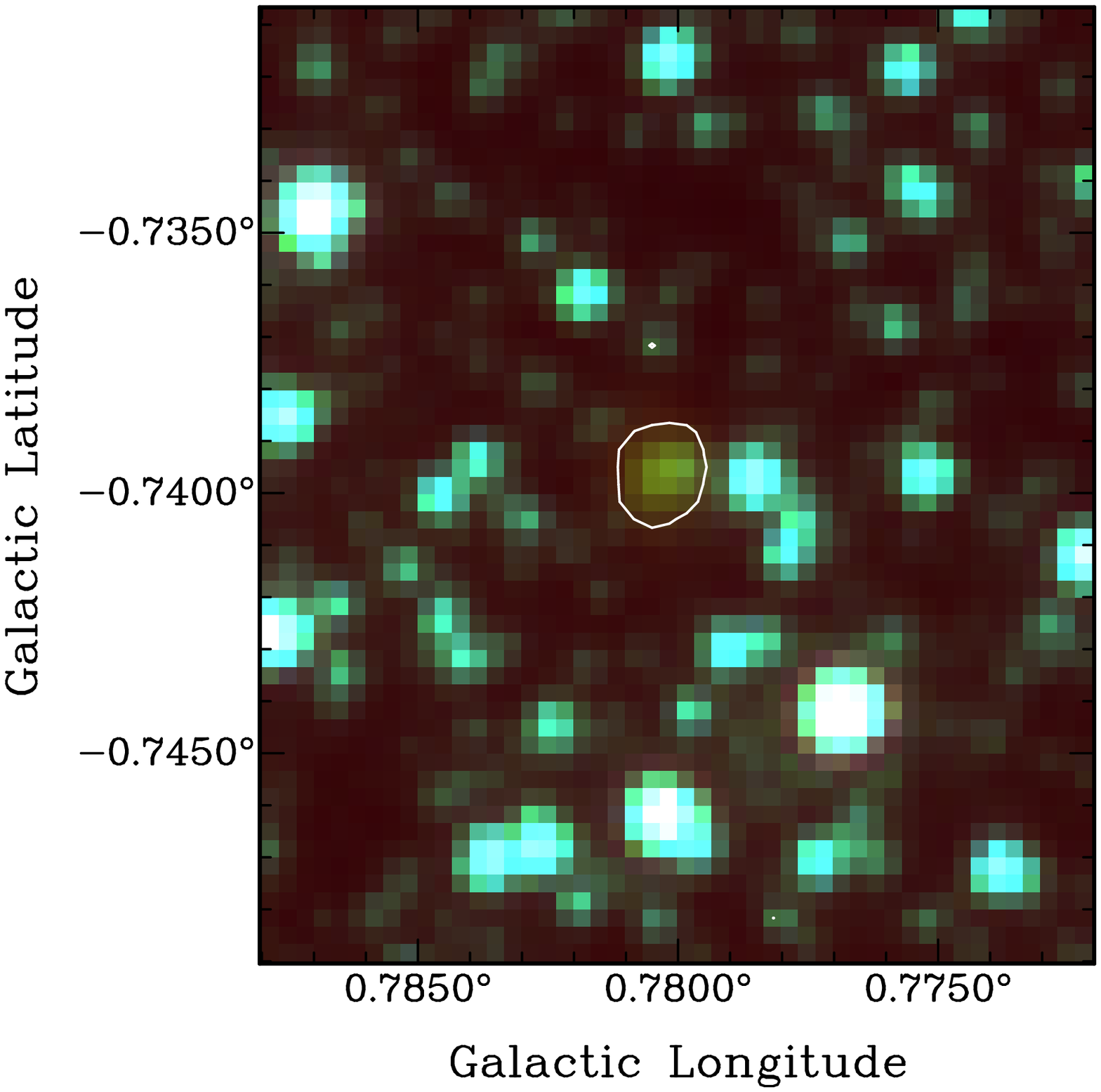}\\
  \caption{IRAC 3-color ({\it left}) and 24~\um\, ({\it right}) images of source g3.
  The contours in both images are at a green ratio value of 0.50.\label{g3}}
\end{figure*}

\begin{figure*}[p]
  \centering
  G0.708$+$0.408\\
  \includegraphics[width=0.4\textwidth,clip=true]{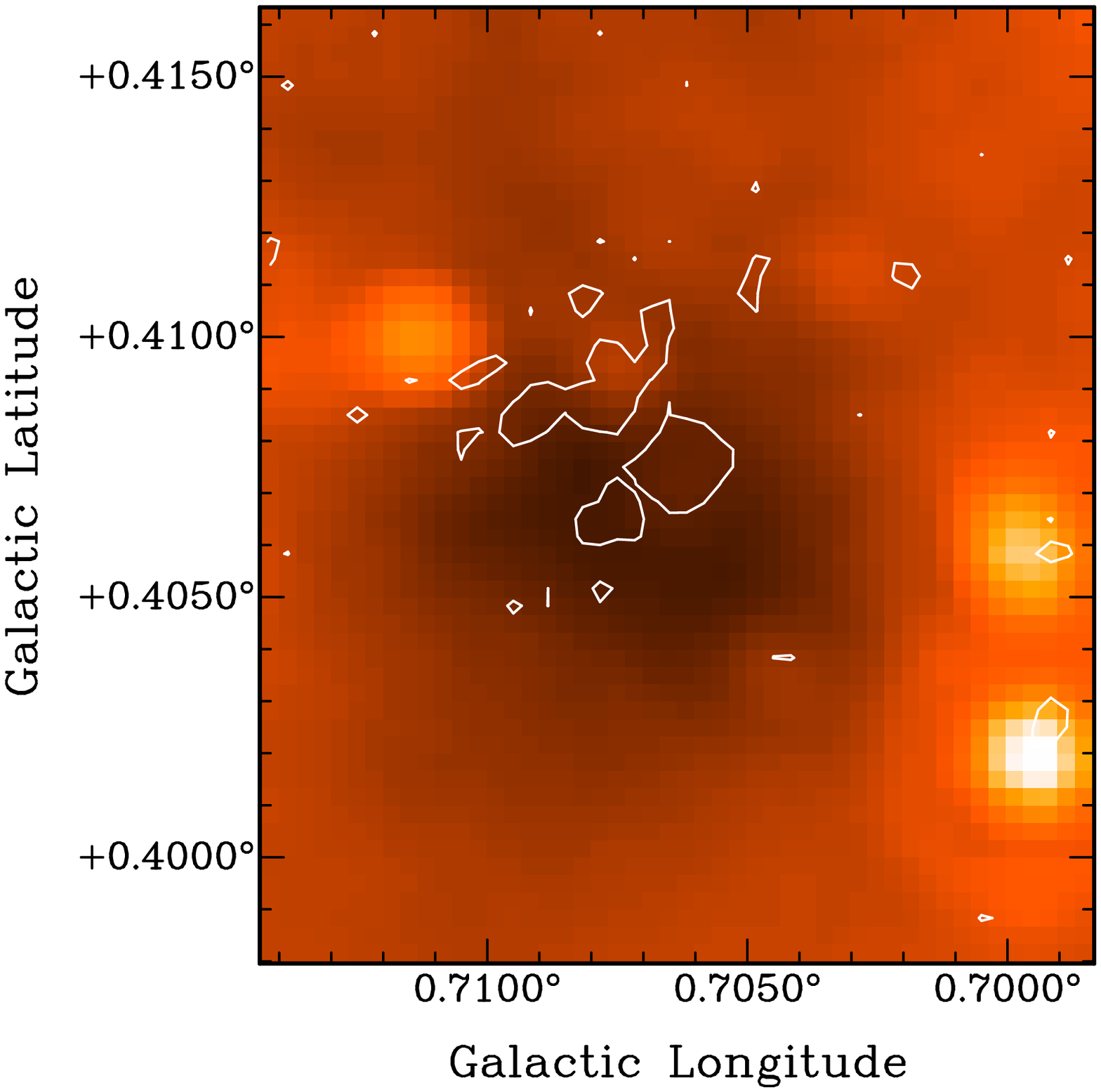}
  \hspace{-11.9cm}
  \includegraphics[width=0.4\textwidth,clip=true]{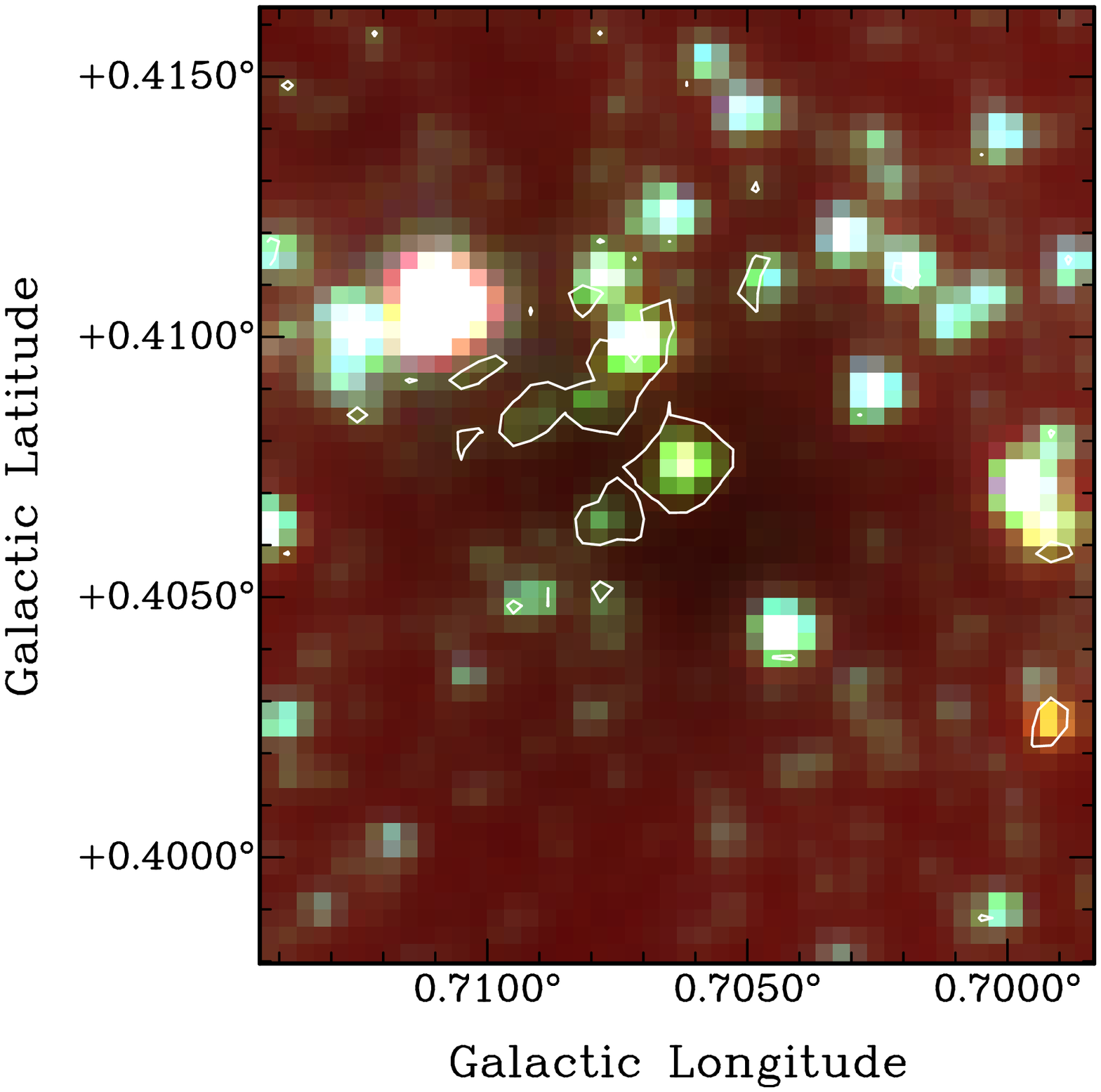}\\
  \caption{IRAC 3-color ({\it left}) and 24~\um\, ({\it right}) images of source g5.
  The contours in both images are at a green ratio value of 0.40.\label{g5}}
\end{figure*}

\begin{figure*}[p]
  \centering
  G0.693$-$0.045\\
  \includegraphics[width=0.4\textwidth,clip=true]{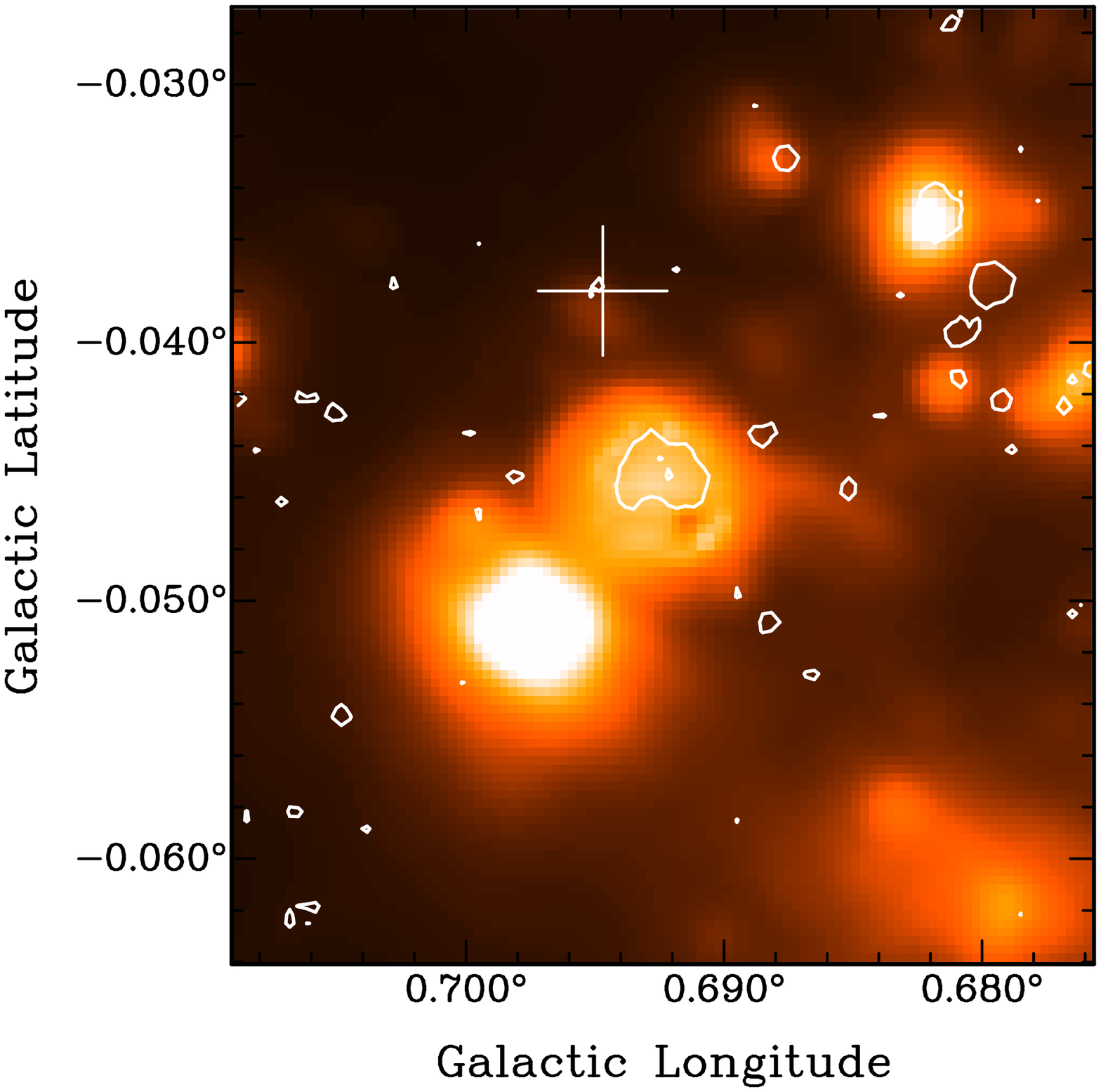}
  \hspace{-12.12cm}
  \includegraphics[width=0.4\textwidth,clip=true]{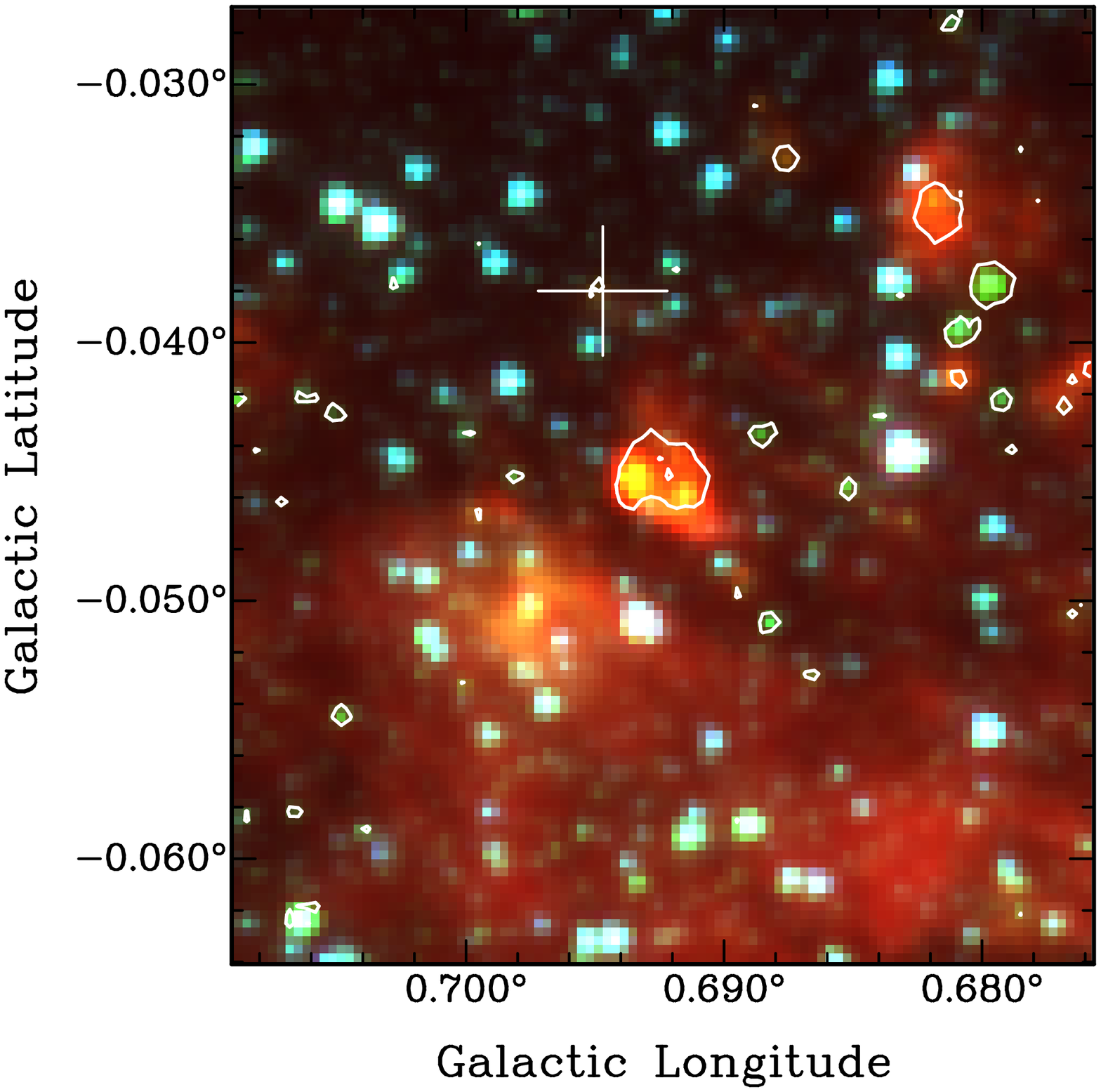}\\
  \includegraphics[angle=-90,width=0.4\textwidth,clip=true]{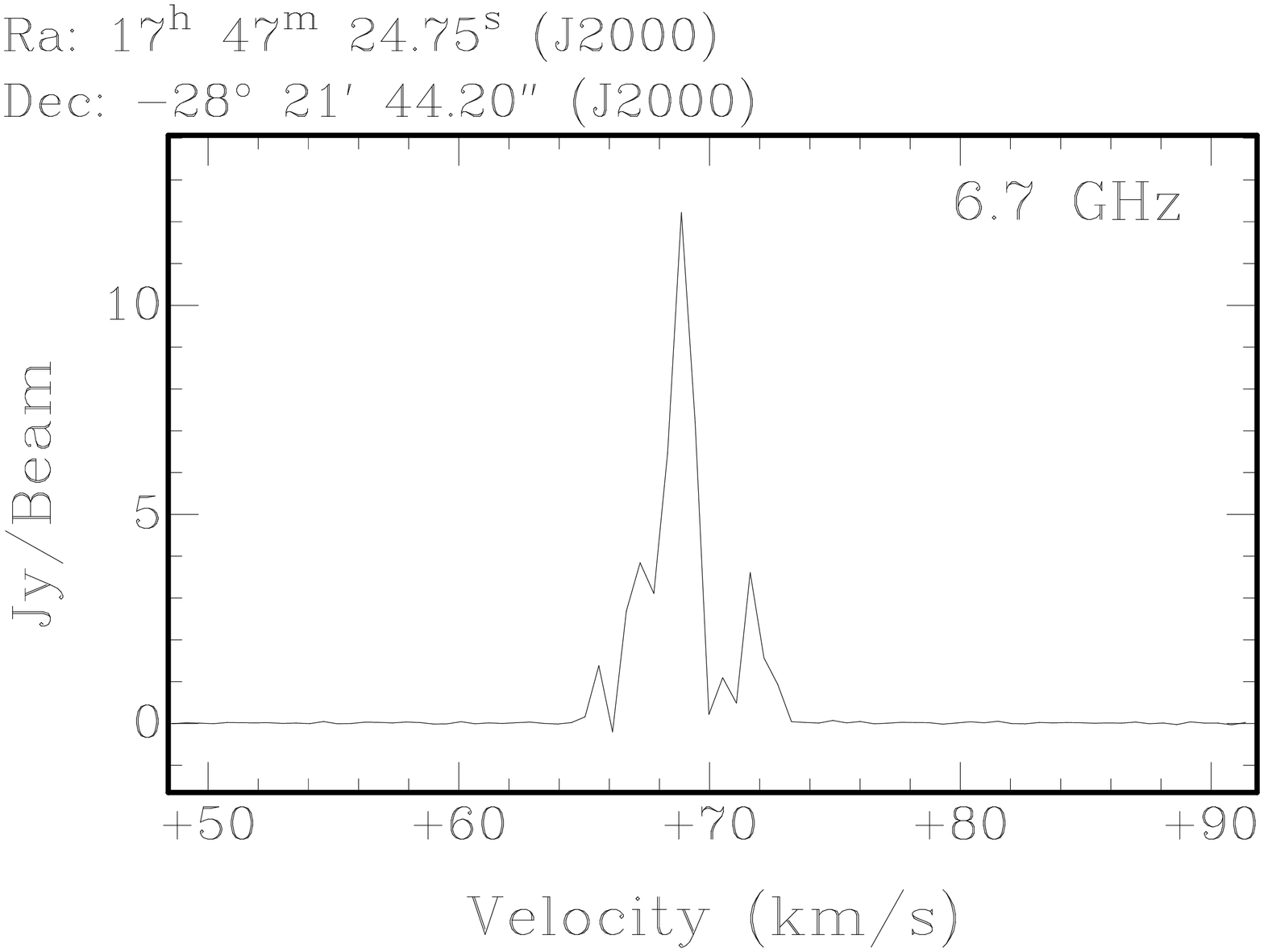}\\
  \caption{{\it Top}: IRAC 3-color ({\it left}) and 24~\um\, ({\it right})
    images of source g6.  The contours in both images are at a green ratio value
    of 0.50, and the plus sign ($+$) designates the position of a 6.7~GHz \meth\,
    maser detected with the EVLA.  Because the location of the maser
    emission is $>$10\arcsec\, from the green source, we do not classify it as
    an association.  {\it Bottom}: Spectrum of 6.7~GHz maser
    emission at the position of the cross. \label{g6}}
\end{figure*}

\begin{figure*}[p]
  \centering
  G0.679$-$0.037, G0.667$-$0.037, G0.667$-$0.035\\
  \includegraphics[width=0.4\textwidth,clip=true]{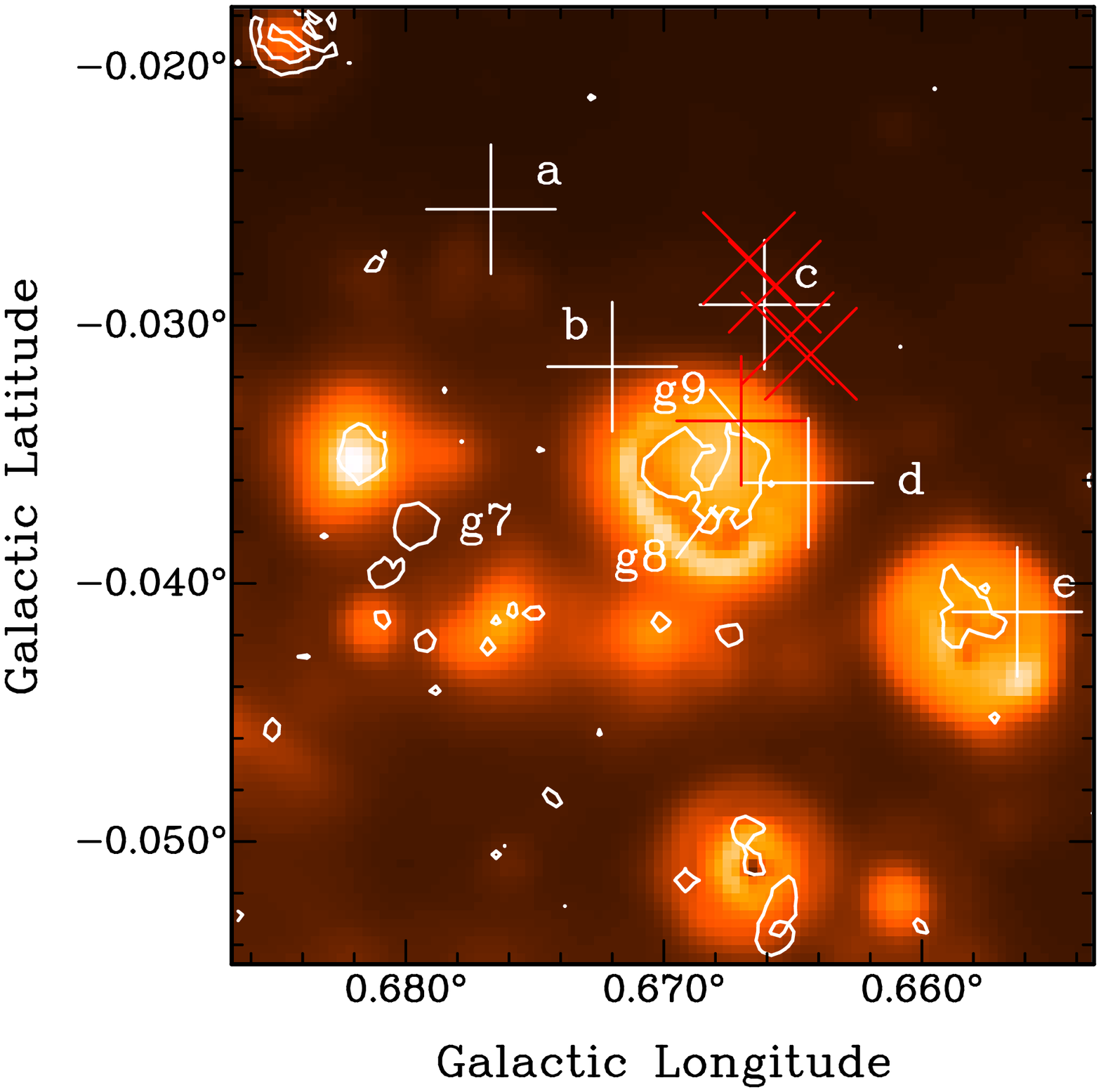}
  \hspace{-12.12cm}
  \includegraphics[width=0.4\textwidth,clip=true]{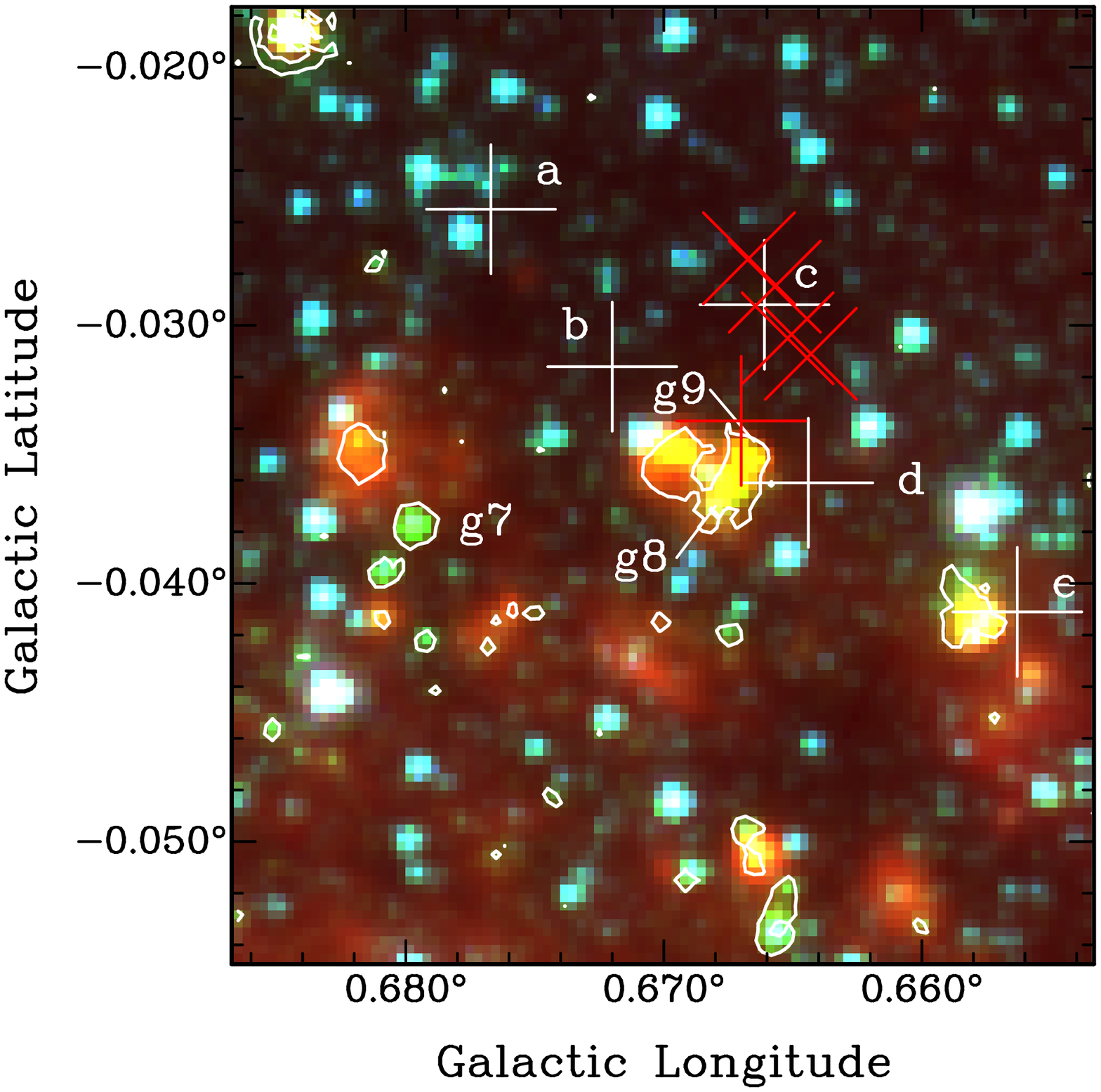}\\
  \includegraphics[angle=-90,width=0.3\textwidth,clip=true]{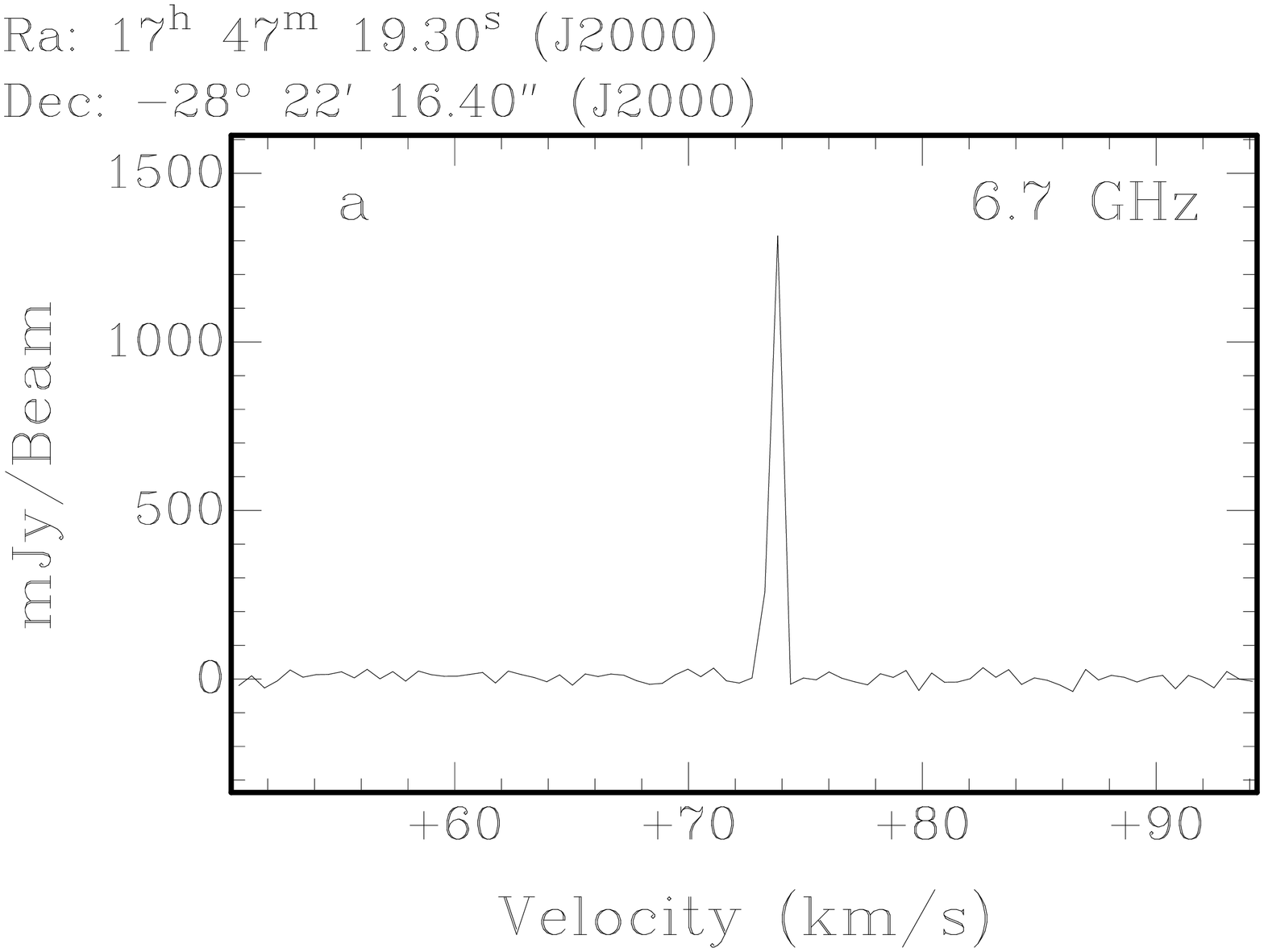}
  \includegraphics[angle=-90,width=0.3\textwidth,clip=true]{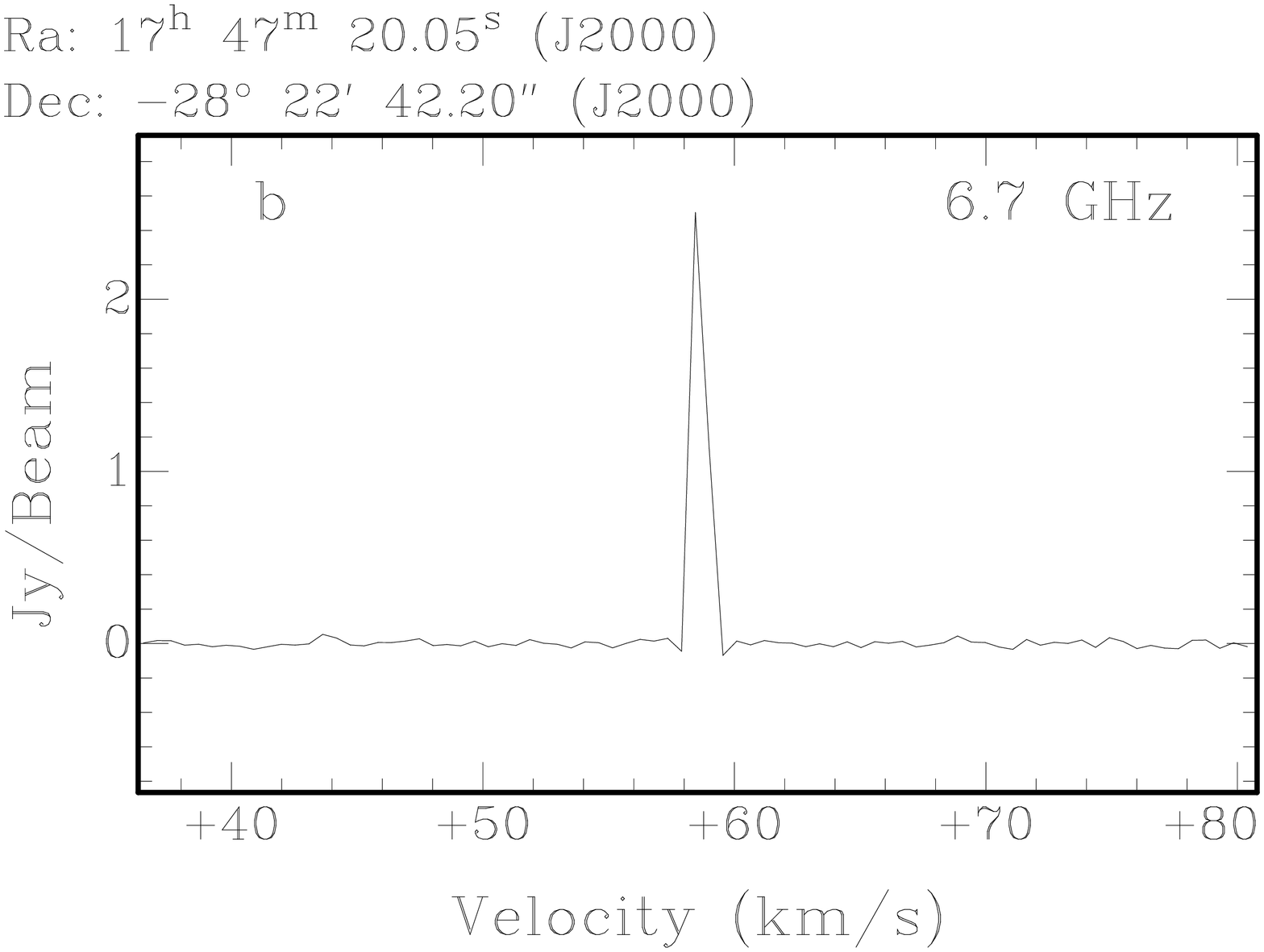}\\
  \includegraphics[angle=-90,width=0.3\textwidth,clip=true]{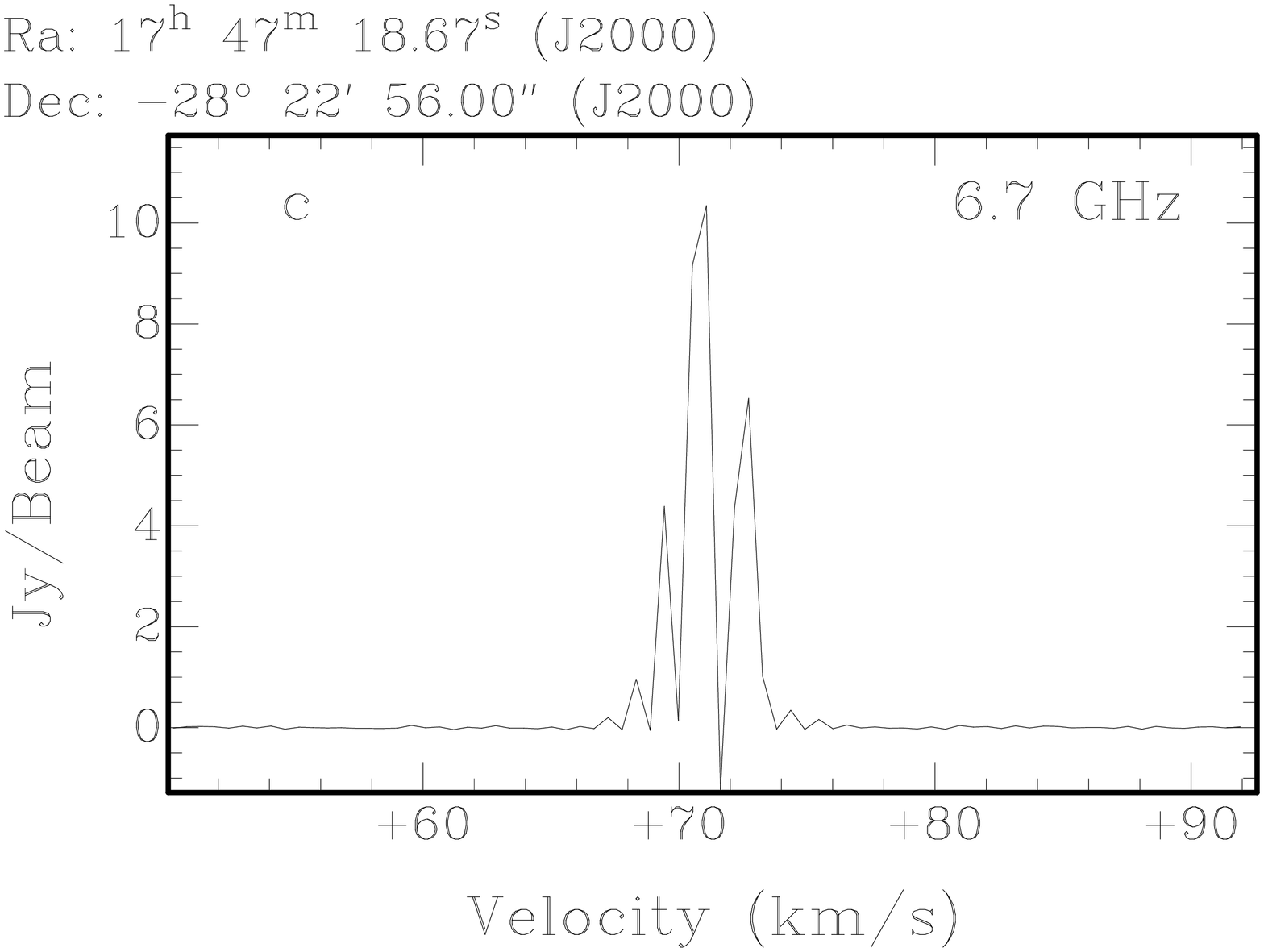}
  \includegraphics[angle=-90,width=0.3\textwidth,clip=true]{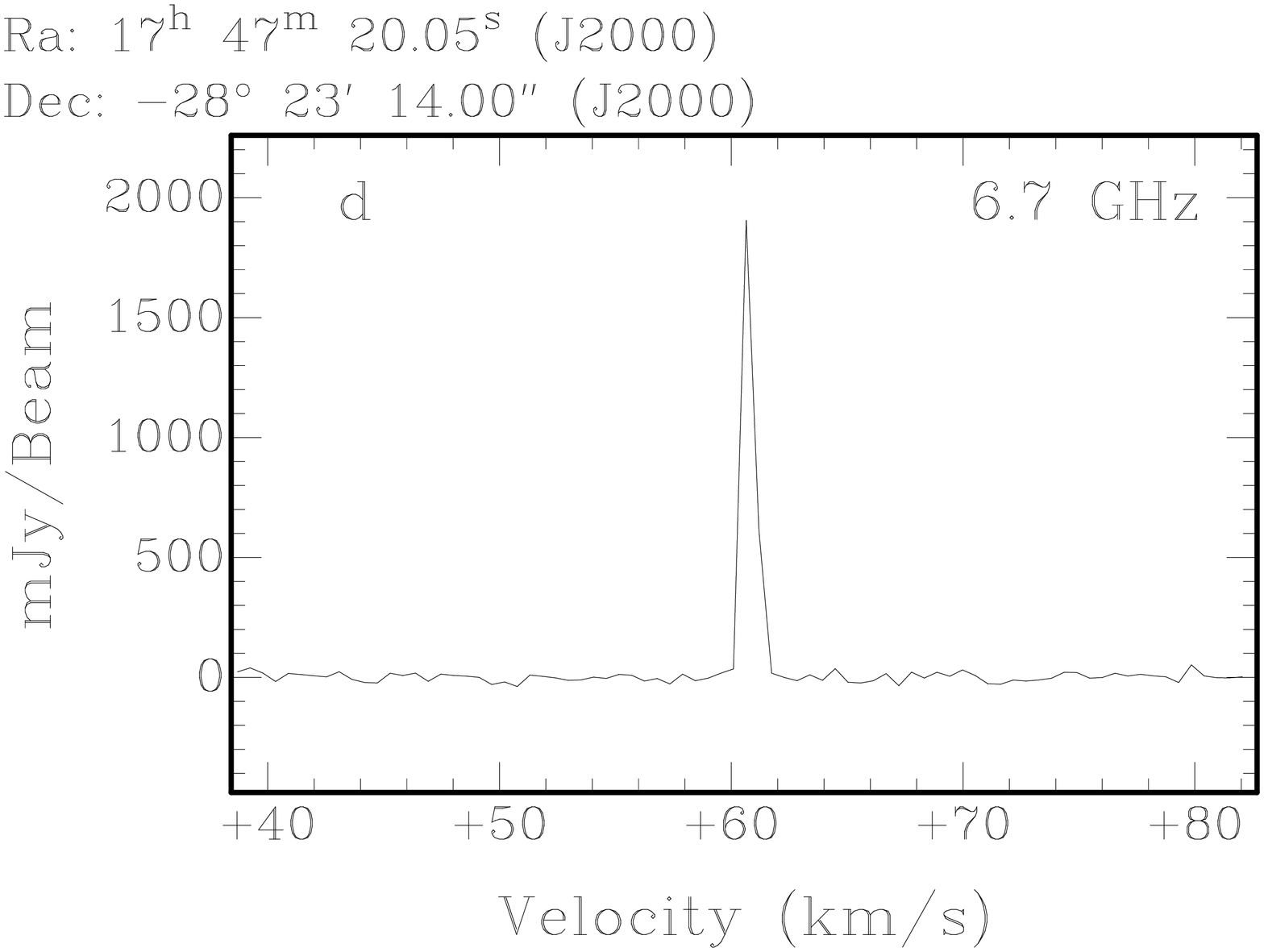}\\
  \includegraphics[angle=-90,width=0.3\textwidth,clip=true]{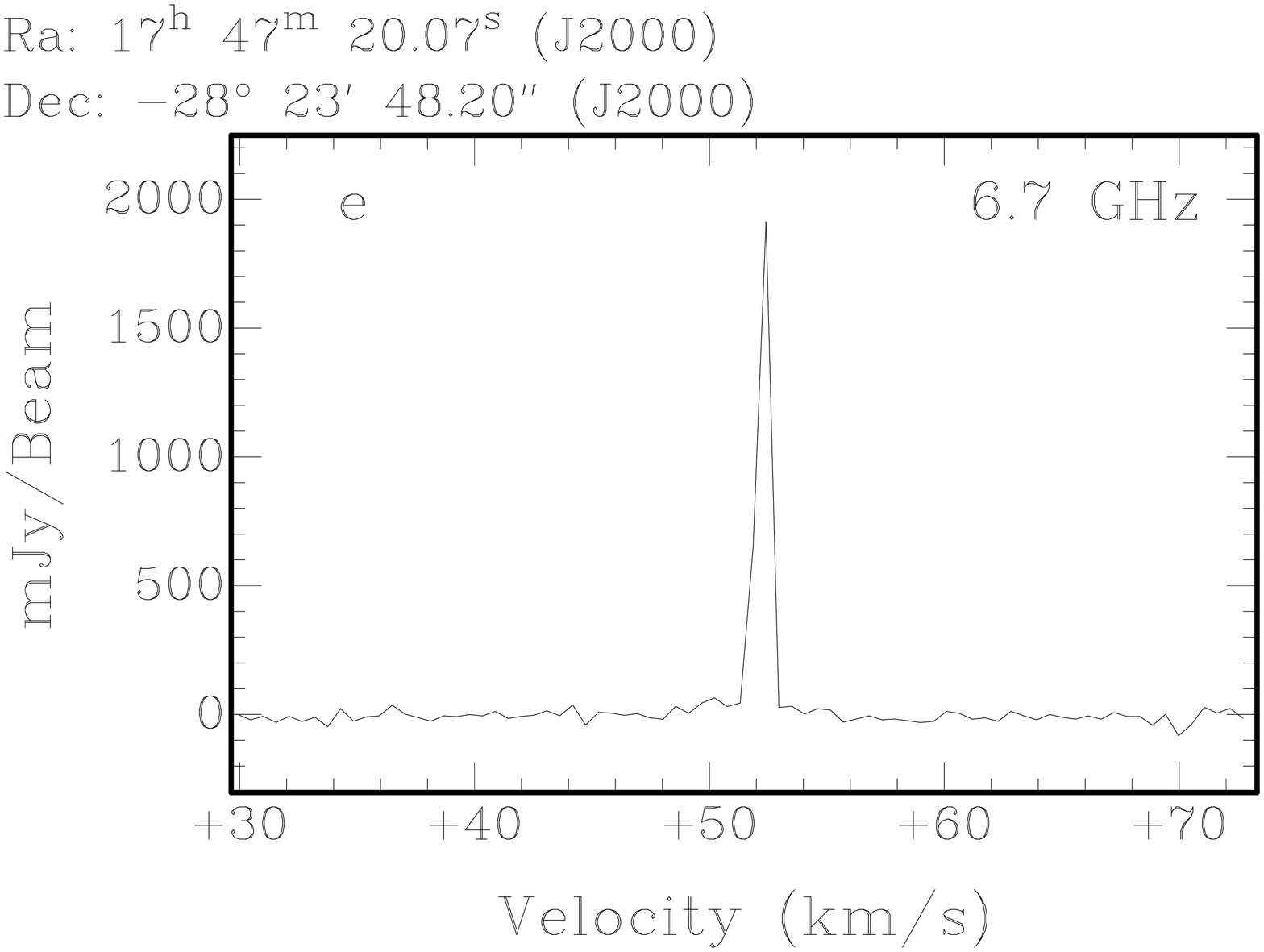}\\
  \caption{{\it Top}: IRAC 3-color ({\it left}) and 24~\um\, ({\it right})
    images of sources g7, g8, and g9.  The contours in both images are at a
    green ratio value of 0.50. White plus signs ($+$) designate the positions
    of 6.7~GHz masers detected with the EVLA, red plus signs designate the
    positions of C10 6.7~GHz masers, and red cross signs ($\times$) designate
    the positions of 44~GHz masers from Y-Z09.  Because the locations of the
    6.7~GHz EVLA masers are $>$10\arcsec\, from the green sources, we do not
    classify them as associations.  {\it Bottom}: Spectra of 6.7~GHz maser
    emission in the g7, g8, and g9 regions obtained with the
    EVLA. \label{g789}}
\end{figure*}

\begin{figure*}[p]
  \centering
  G0.665$-$0.053\\
  \includegraphics[width=0.4\textwidth,clip=true]{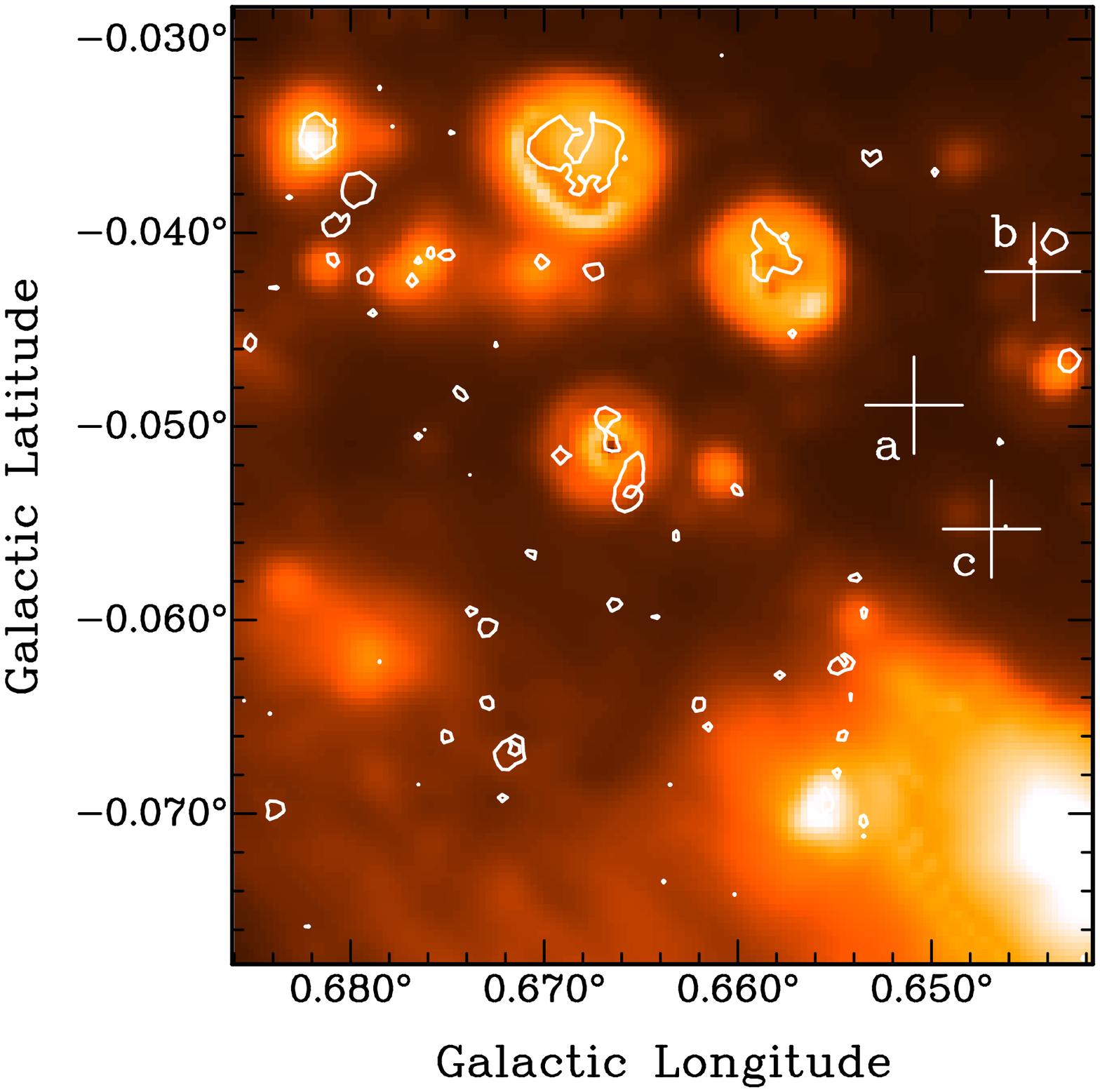}
  \hspace{-11.9cm}
  \includegraphics[width=0.4\textwidth,clip=true]{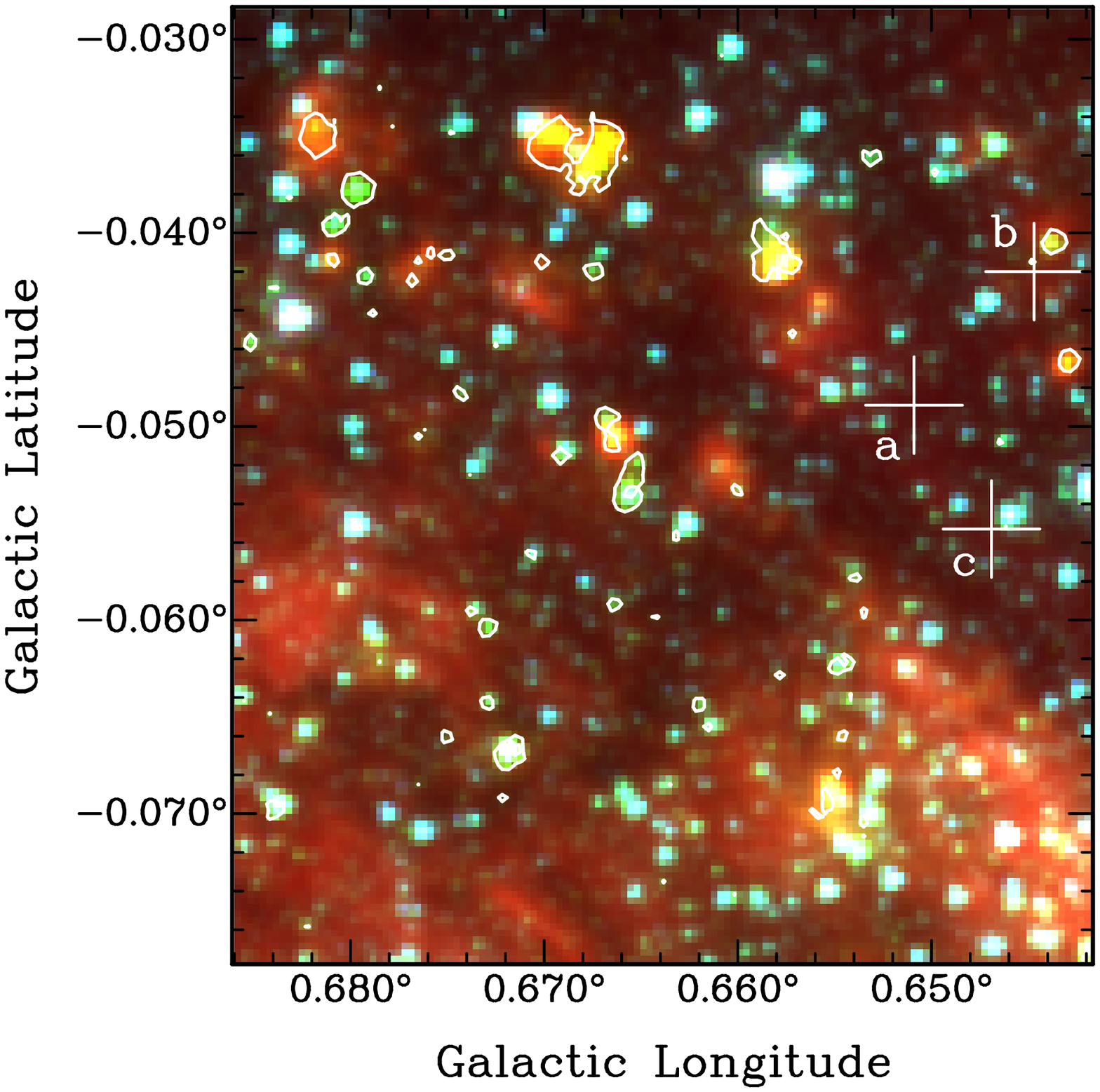}\\
  \includegraphics[angle=-90,width=0.4\textwidth,clip=true]{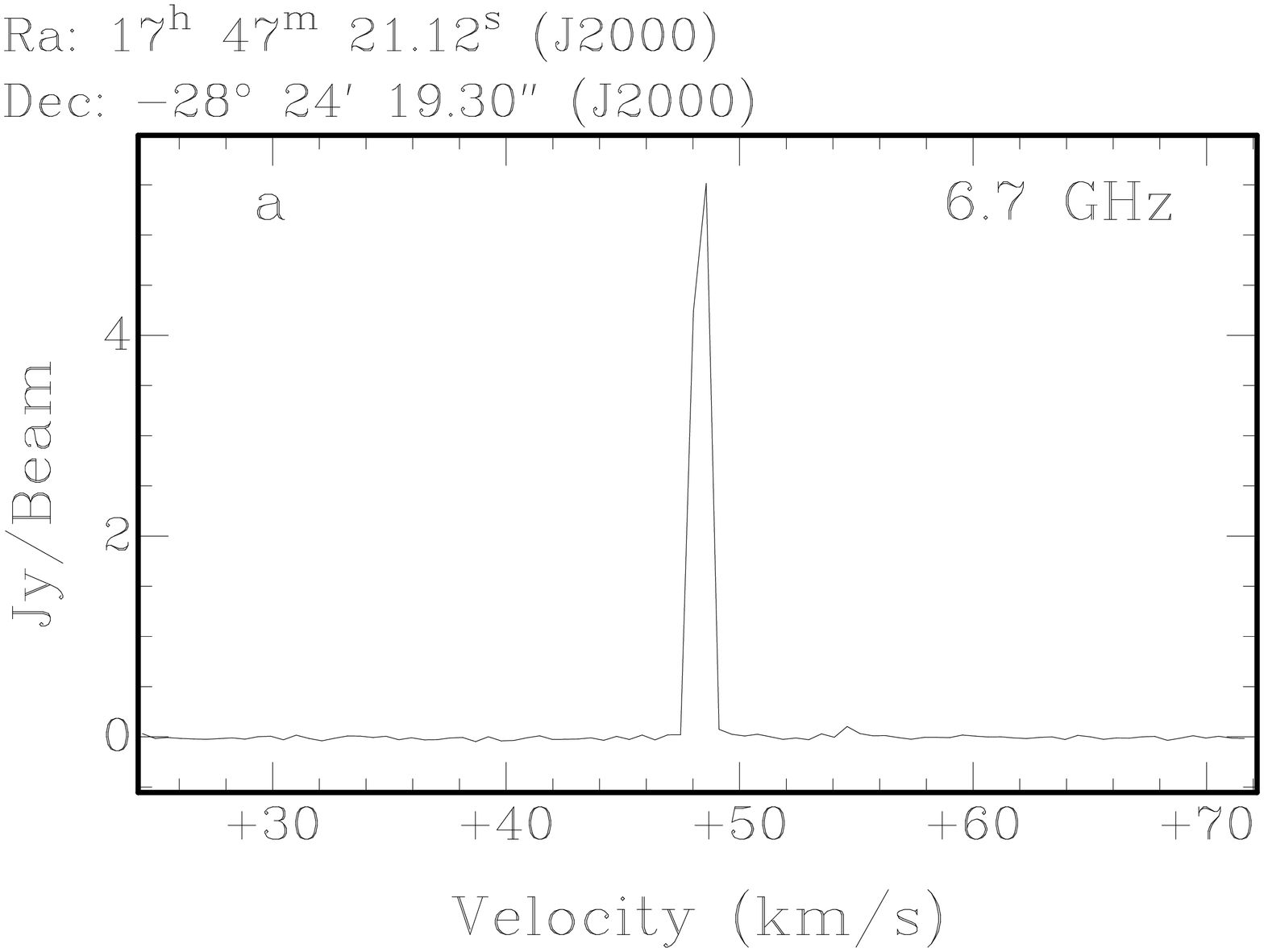}
  \includegraphics[angle=-90,width=0.4\textwidth,clip=true]{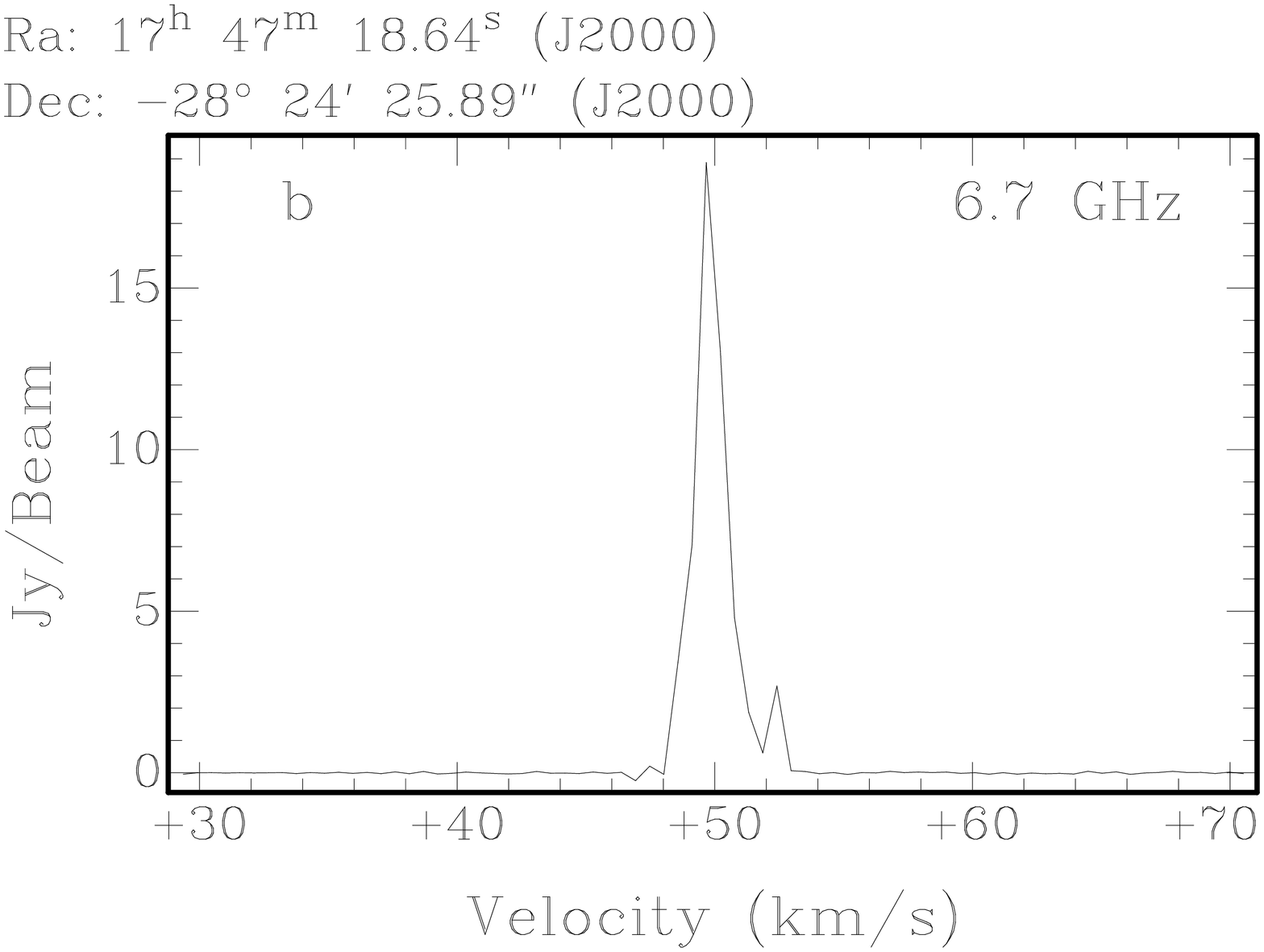}\\
  \includegraphics[angle=-90,width=0.4\textwidth,clip=true]{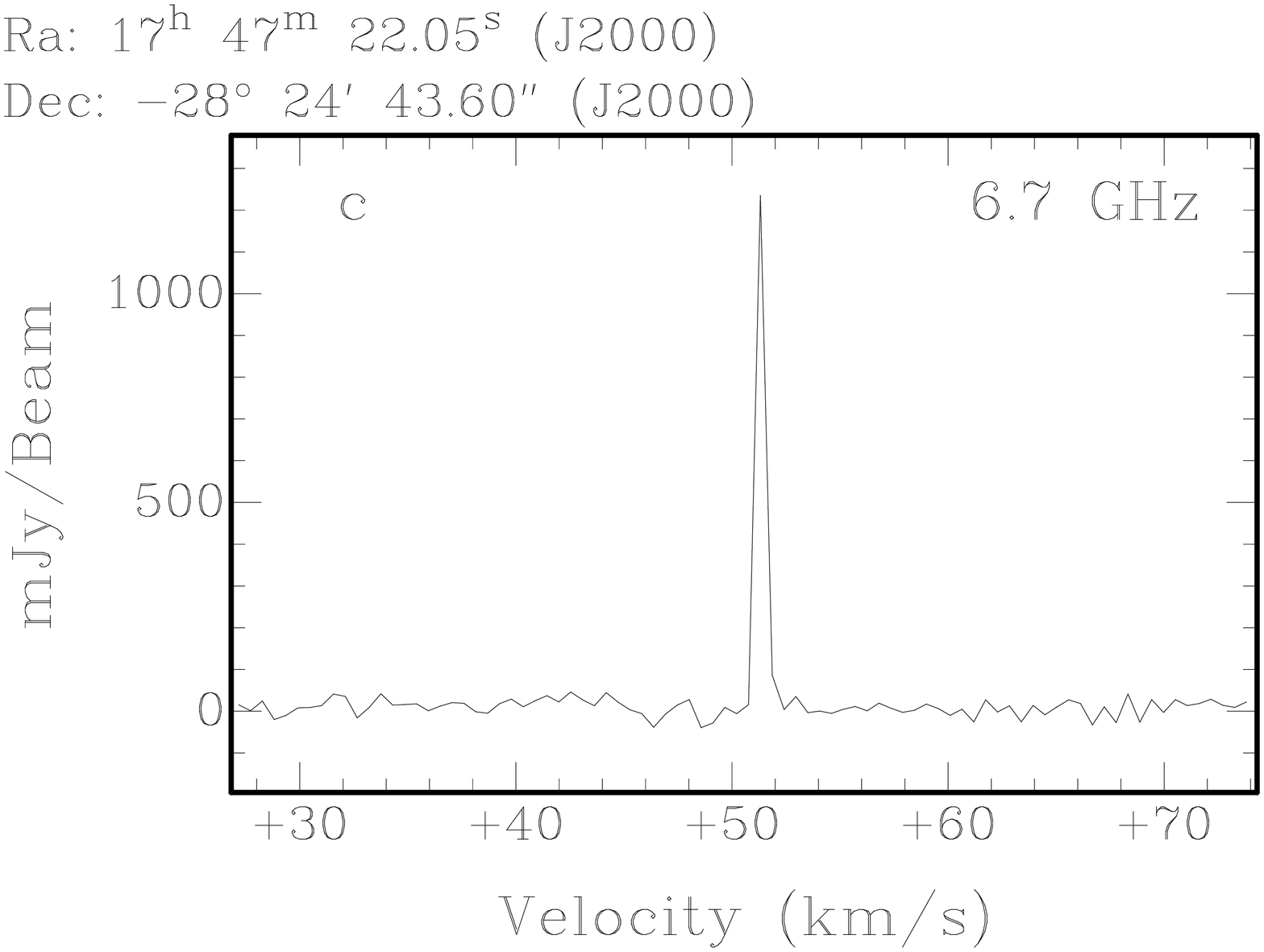}\\
  \caption{{\it Top}: IRAC 3-color ({\it left}) and 24~\um\, ({\it right})
    images of source g10.  The contours in both images are at a
    green ratio value of 0.50. White plus signs ($+$) designate the positions
    of 6.7~GHz masers detected with the EVLA.   Because the locations of the
    6.7~GHz EVLA masers are $>$10\arcsec\, from the green source, we do not
    classify them as associations.  {\it Bottom}:  Spectra of 6.7~GHz
    maser emission in the g10 field obtained with the EVLA. \label{g10}}
\end{figure*}

\begin{figure*}[p]
  \centering
  G0.542$-$0.476\\
  \includegraphics[width=0.4\textwidth,clip=true]{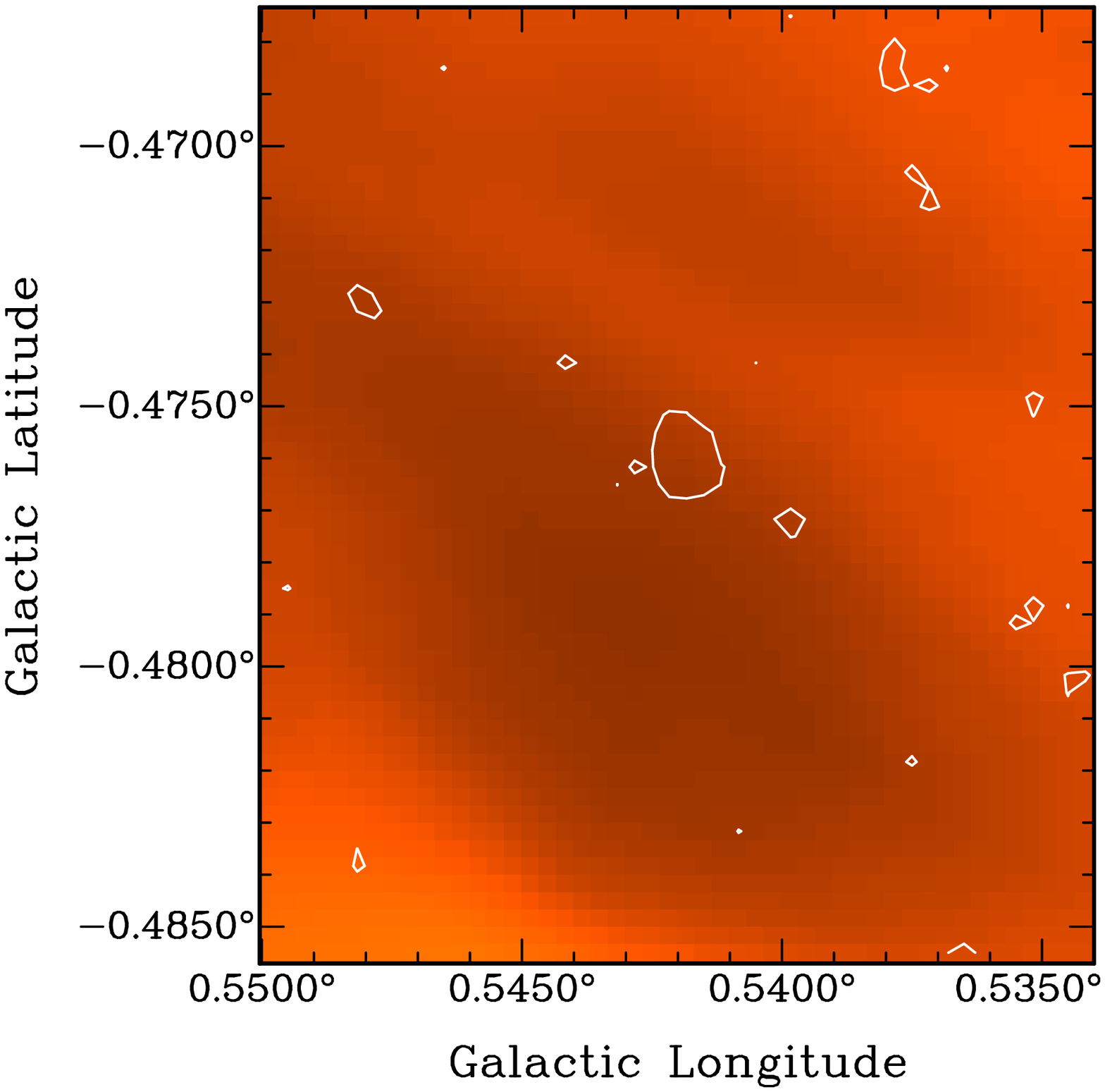}
  \hspace{-11.9cm}
  \includegraphics[width=0.4\textwidth,clip=true]{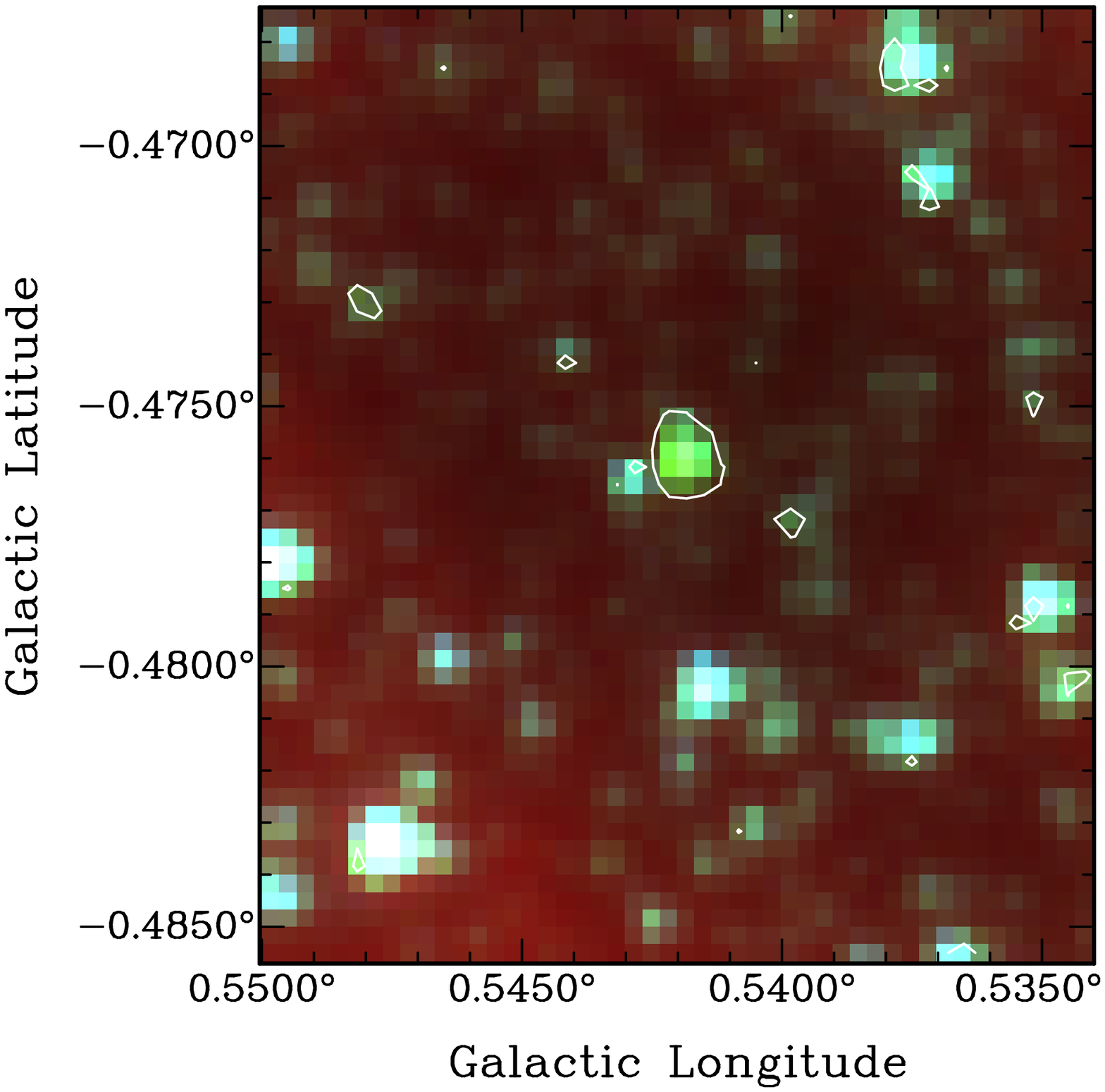}\\
  \caption{IRAC 3-color ({\it left}) and 24~\um\, ({\it right})
    images of source g11.  The contours in both images are at a
    green ratio value of 0.35.\label{g11}}
\end{figure*}

\begin{figure*}[p]
  \centering
  G0.517$-$0.657\\
  \includegraphics[width=0.4\textwidth,clip=true]{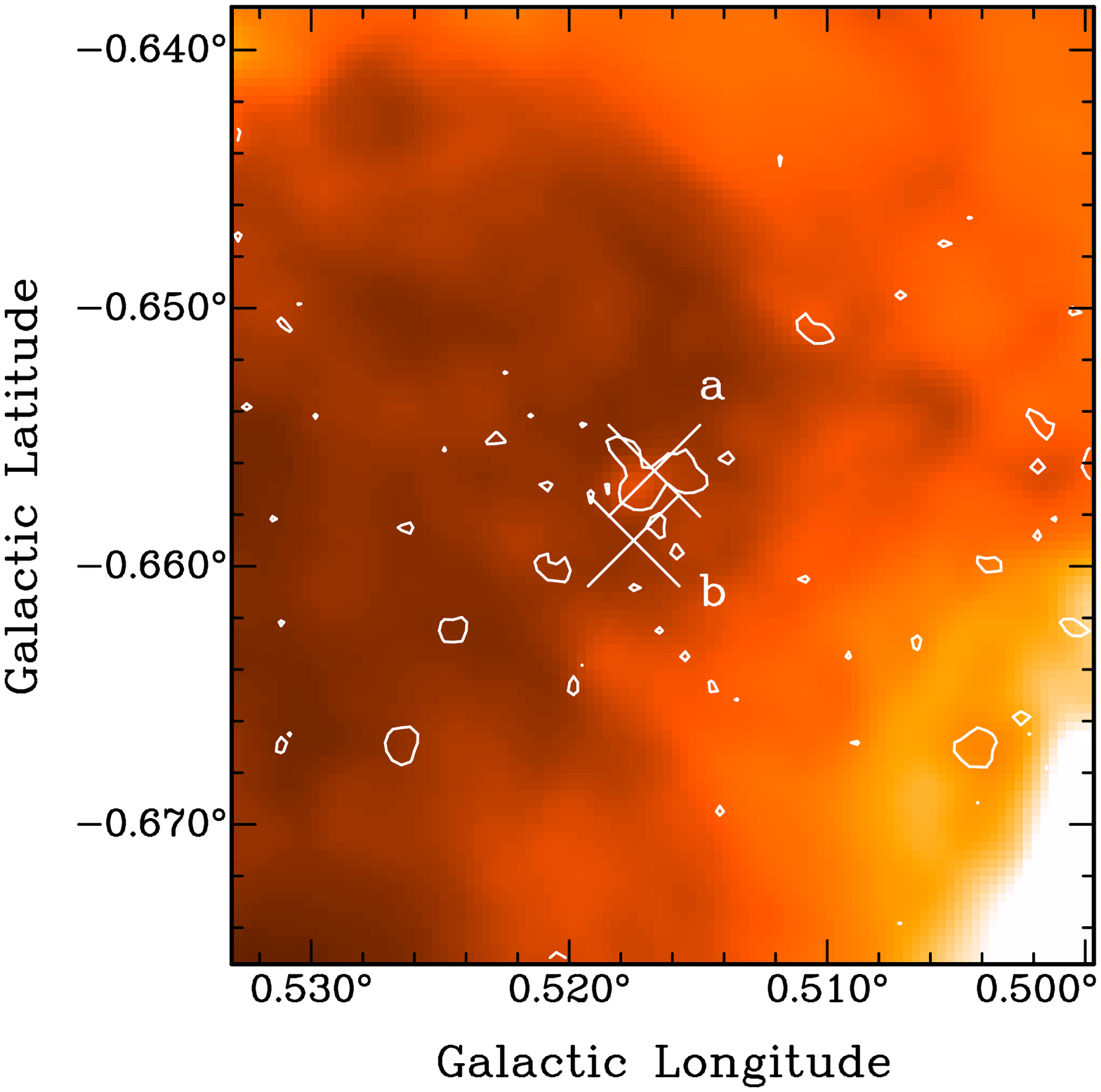}
  \hspace{-11.9cm}
  \includegraphics[width=0.4\textwidth,clip=true]{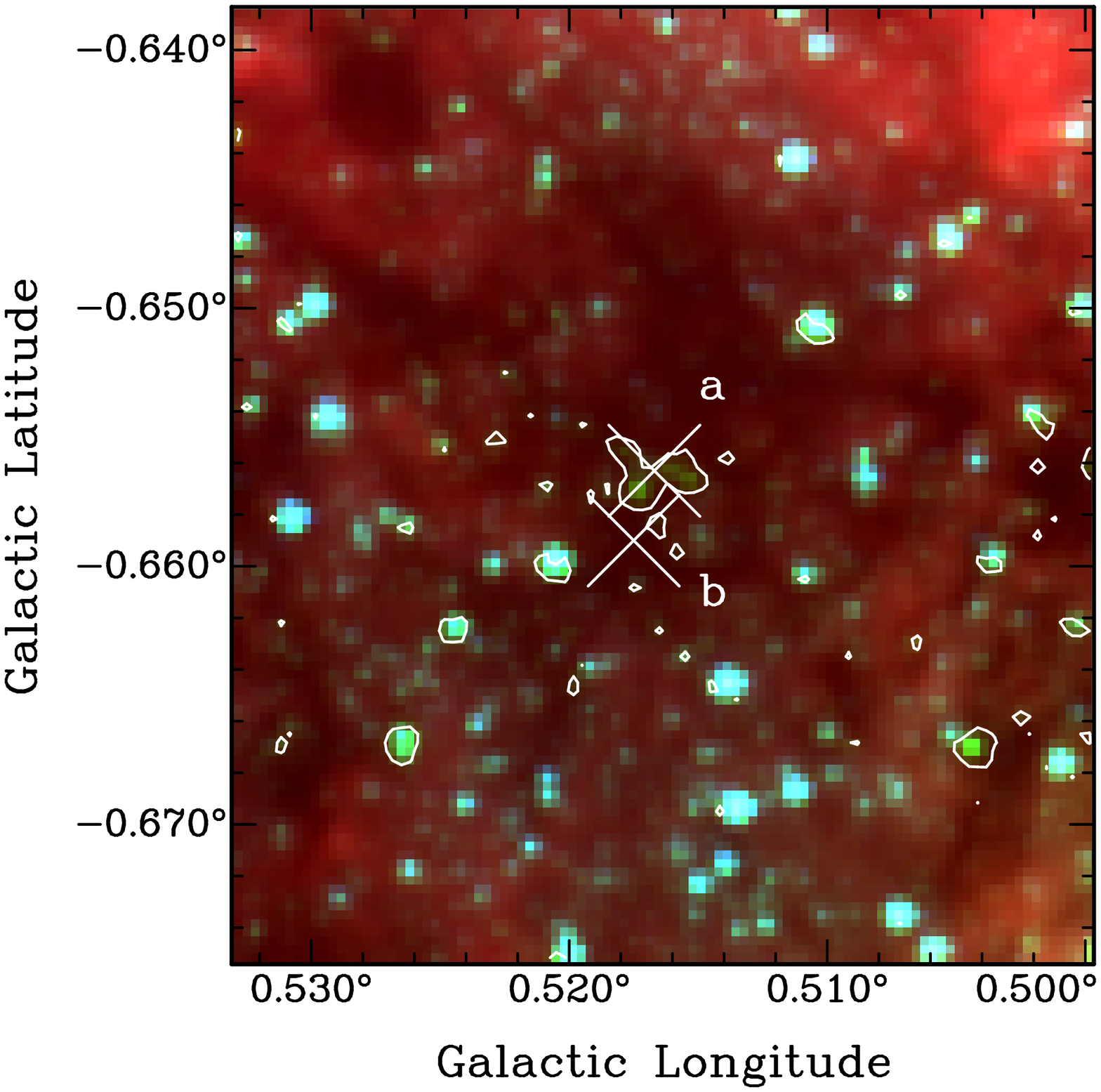}\\
  \includegraphics[angle=-90,width=0.4\textwidth,clip=true]{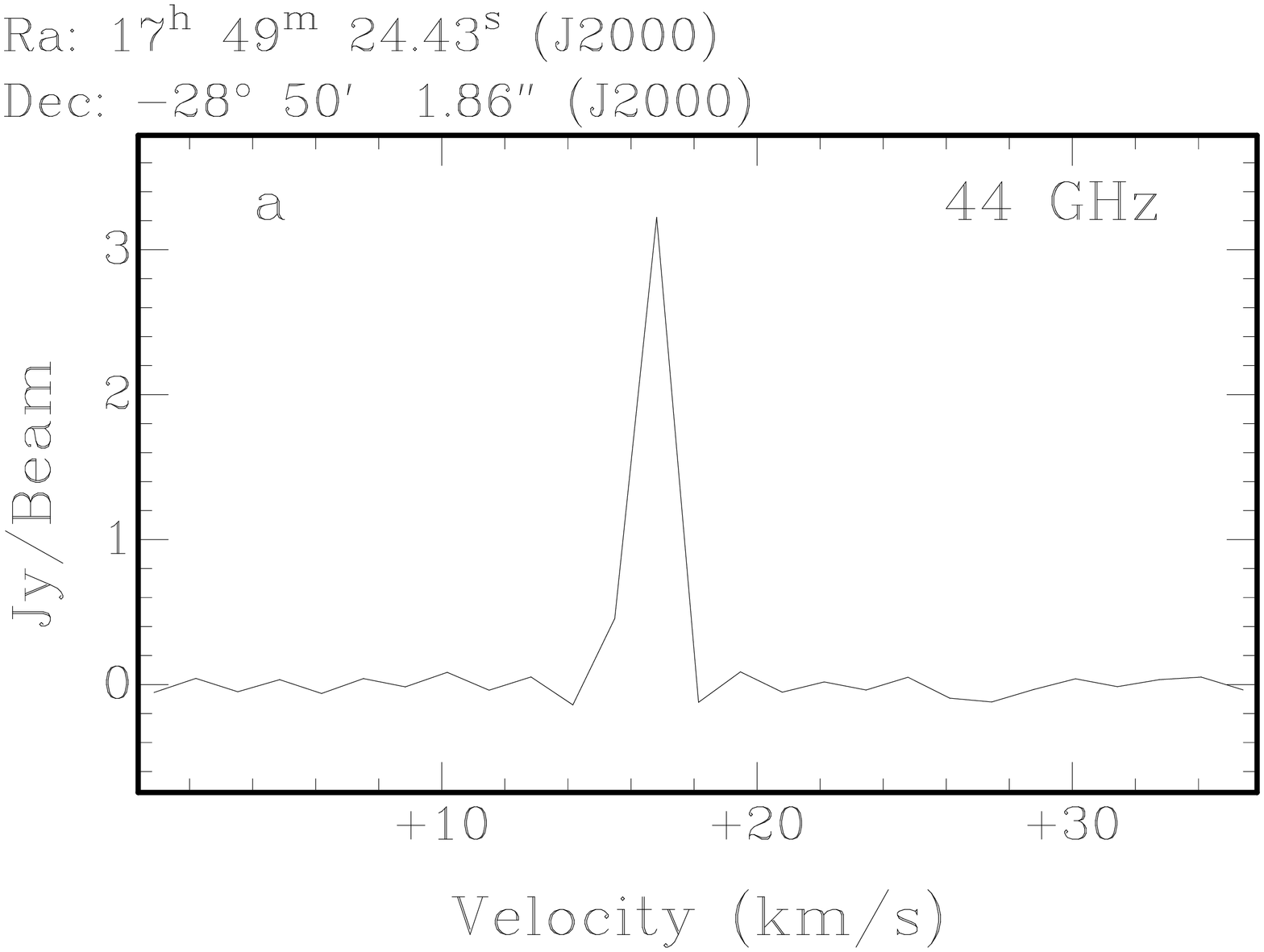}
  \includegraphics[angle=-90,width=0.4\textwidth,clip=true]{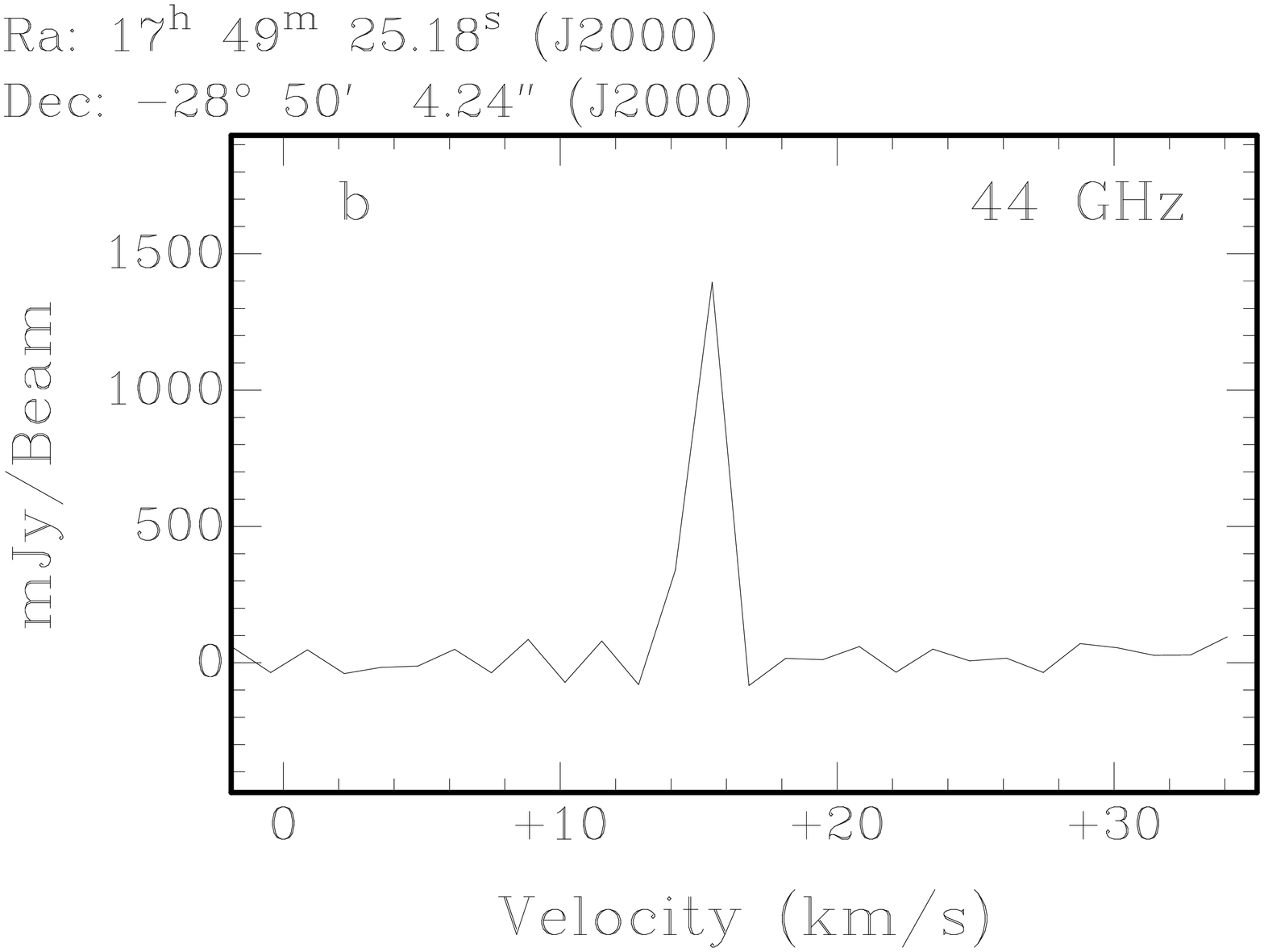}\\
  \caption{{\it Top}: IRAC 3-color ({\it left}) and 24~\um\, ({\it right})
    images of source g12.  The contours in both images are at a green
    ratio value of 0.45. White cross signs ($\times$) designate the positions
    of 44~GHz masers detected with the EVLA. {\it Bottom}: Spectra of 44~GHz
    maser emission in the g12 field obtained with the EVLA.\label{g12}}
\end{figure*}

\begin{figure*}[p]
  \centering
  G0.483$-$0.701\\
  \includegraphics[width=0.4\textwidth,clip=true]{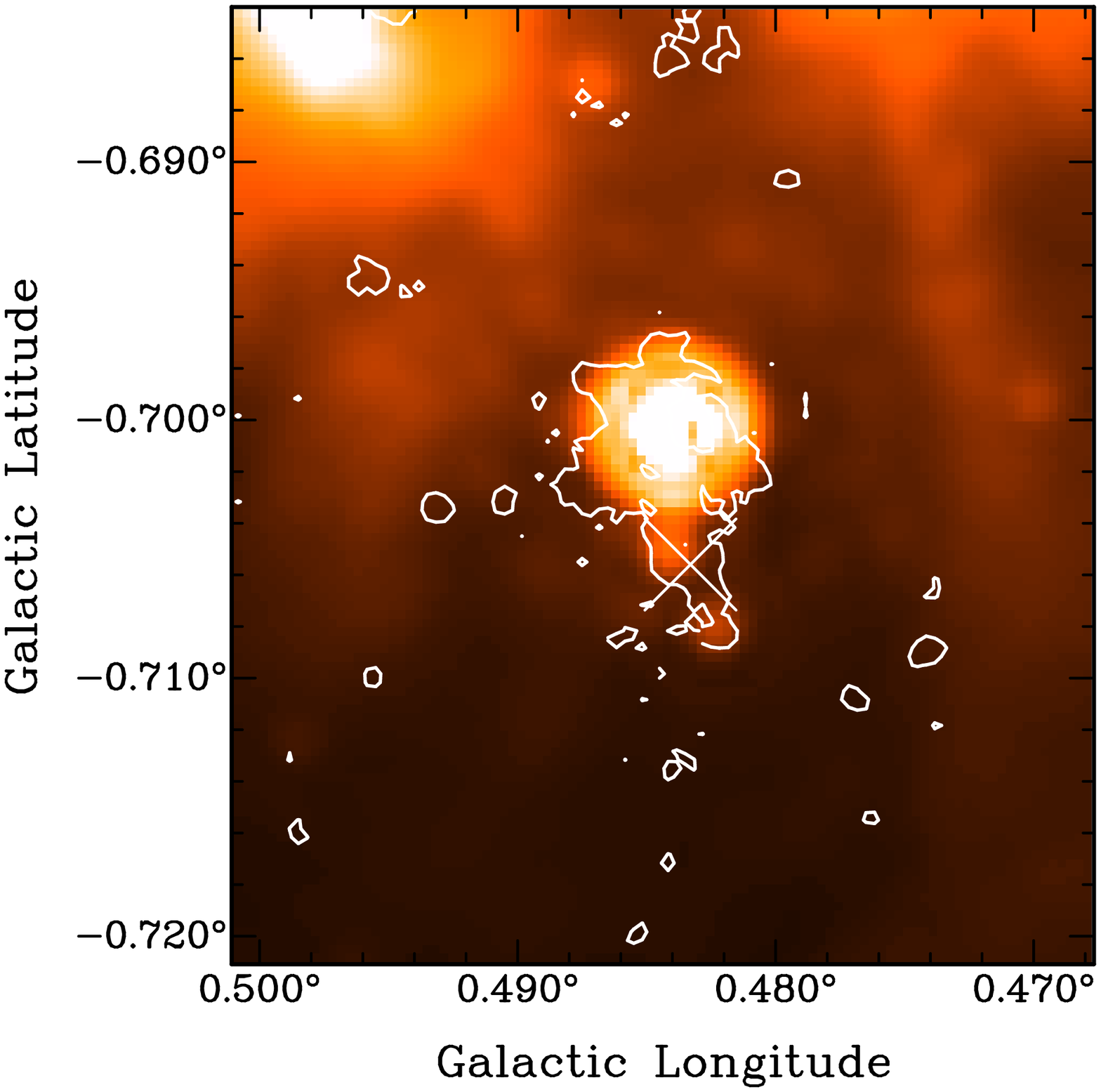}
  \hspace{-11.9cm}
  \includegraphics[width=0.4\textwidth,clip=true]{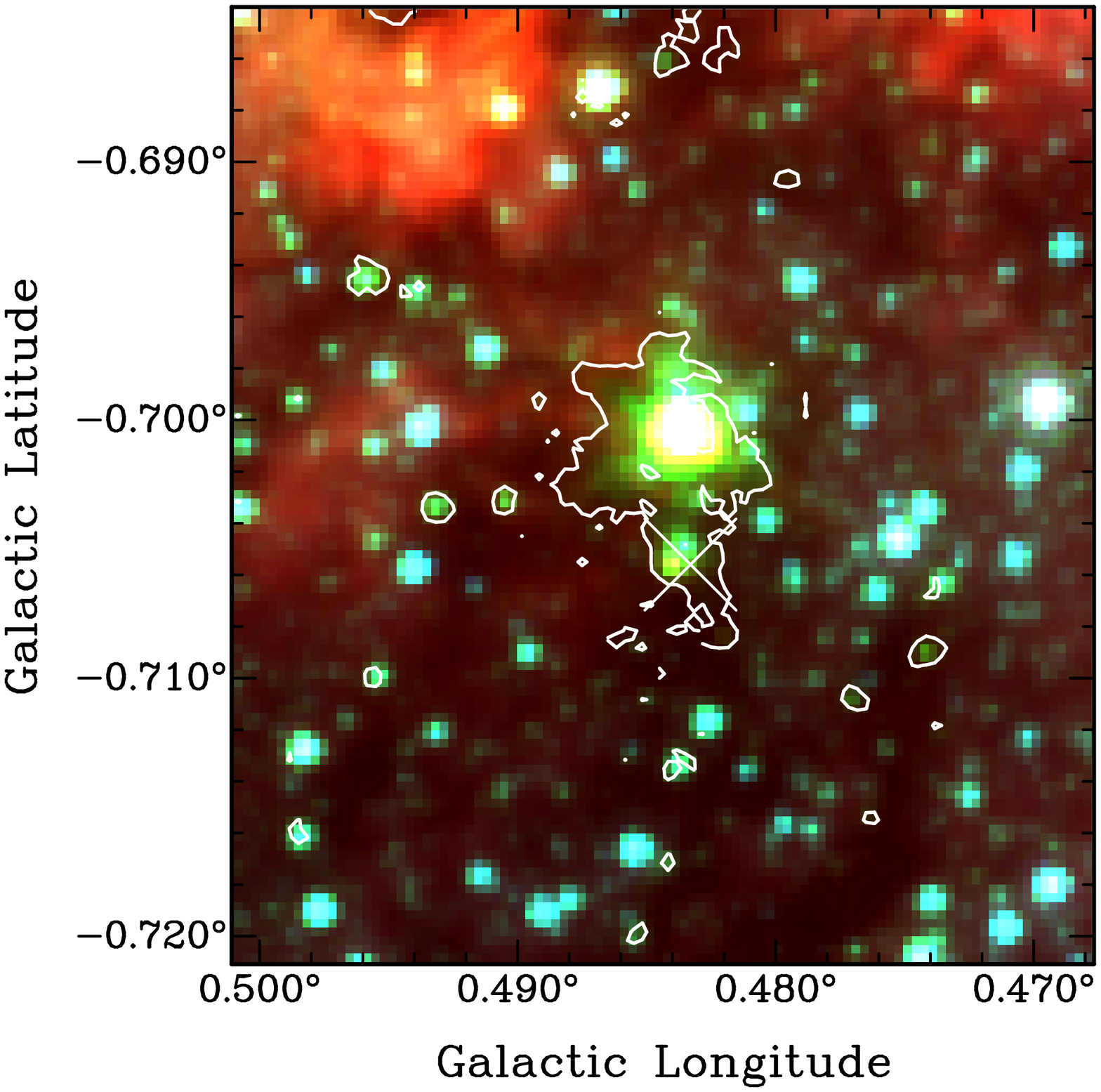}\\
  \includegraphics[angle=-90,width=0.4\textwidth,clip=true]{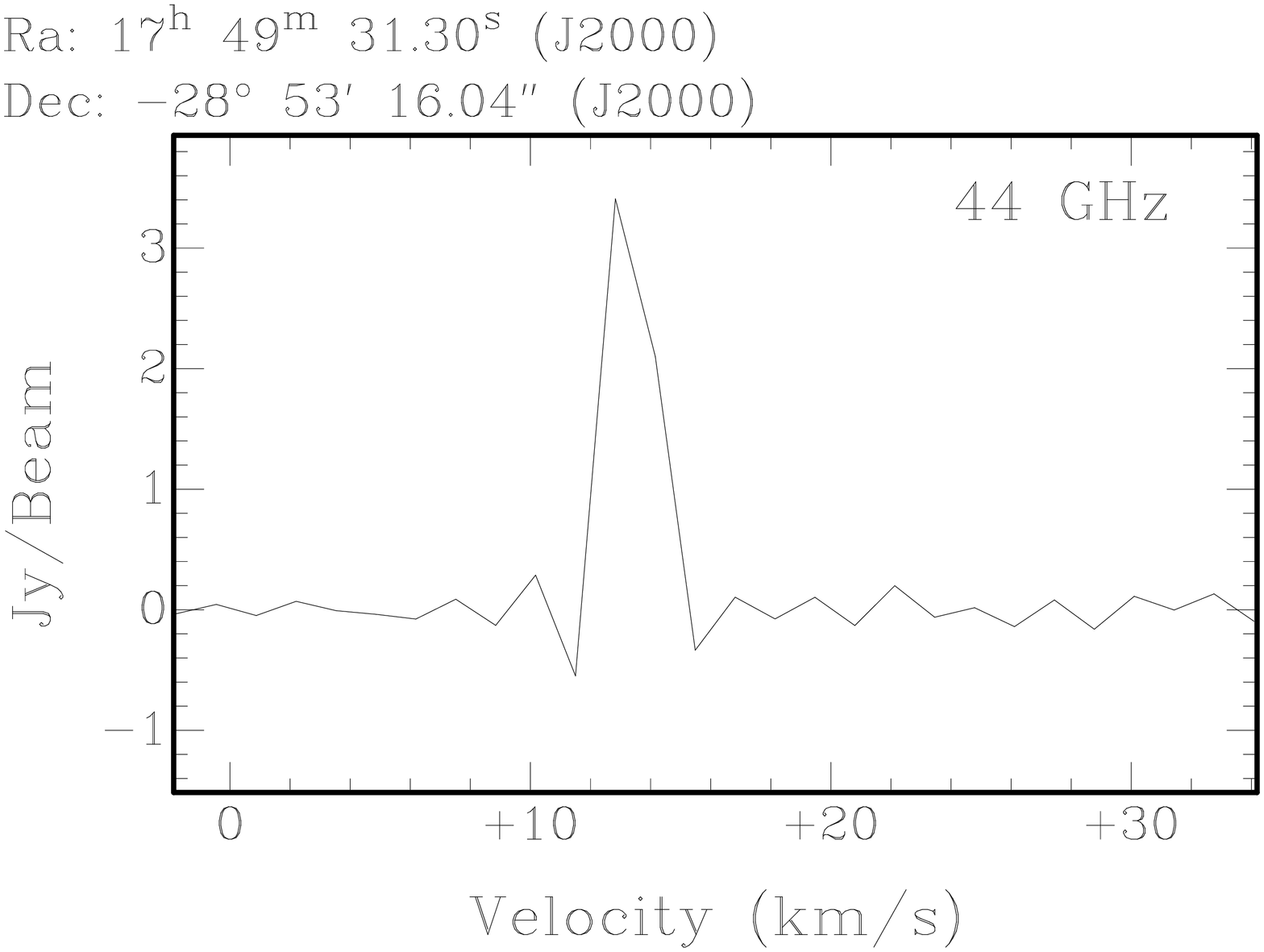}\\
  \caption{{\it Top}: IRAC 3-color ({\it left}) and 24~\um\, ({\it right})
    images of source g13.  The contours in both images are at a green
    ratio value of 0.50. The white cross sign ($\times$) designates the position
    of 44~GHz maser emission detected with the EVLA. {\it Bottom}: Spectrum of 44~GHz
    maser emission in the g13 field obtained with the EVLA.\label{g13}}
\end{figure*}

\begin{figure*}[p]
  \centering
  G0.477$-$0.727\\
  \includegraphics[width=0.4\textwidth,clip=true]{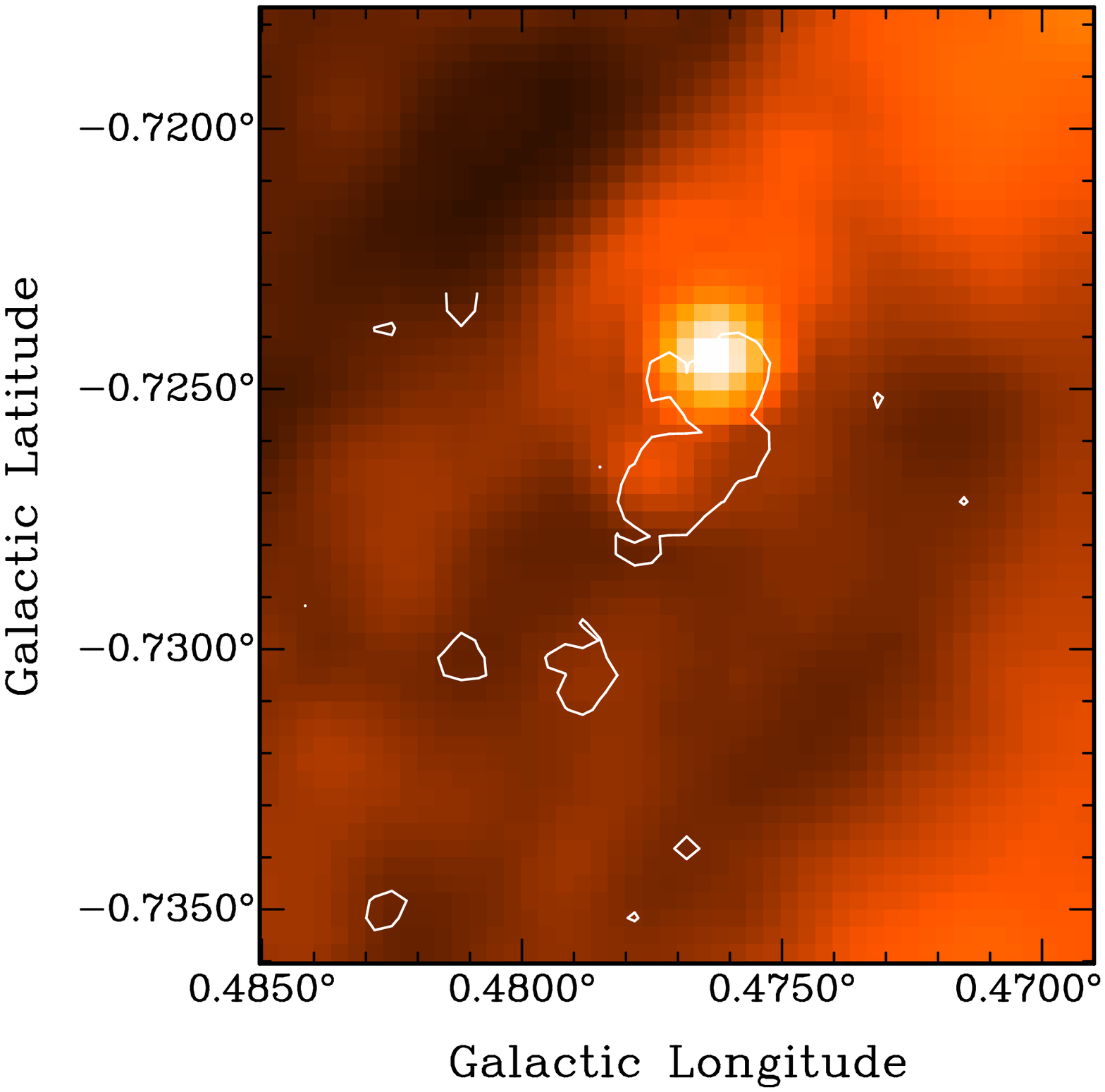}
  \hspace{-11.9cm}
  \includegraphics[width=0.4\textwidth,clip=true]{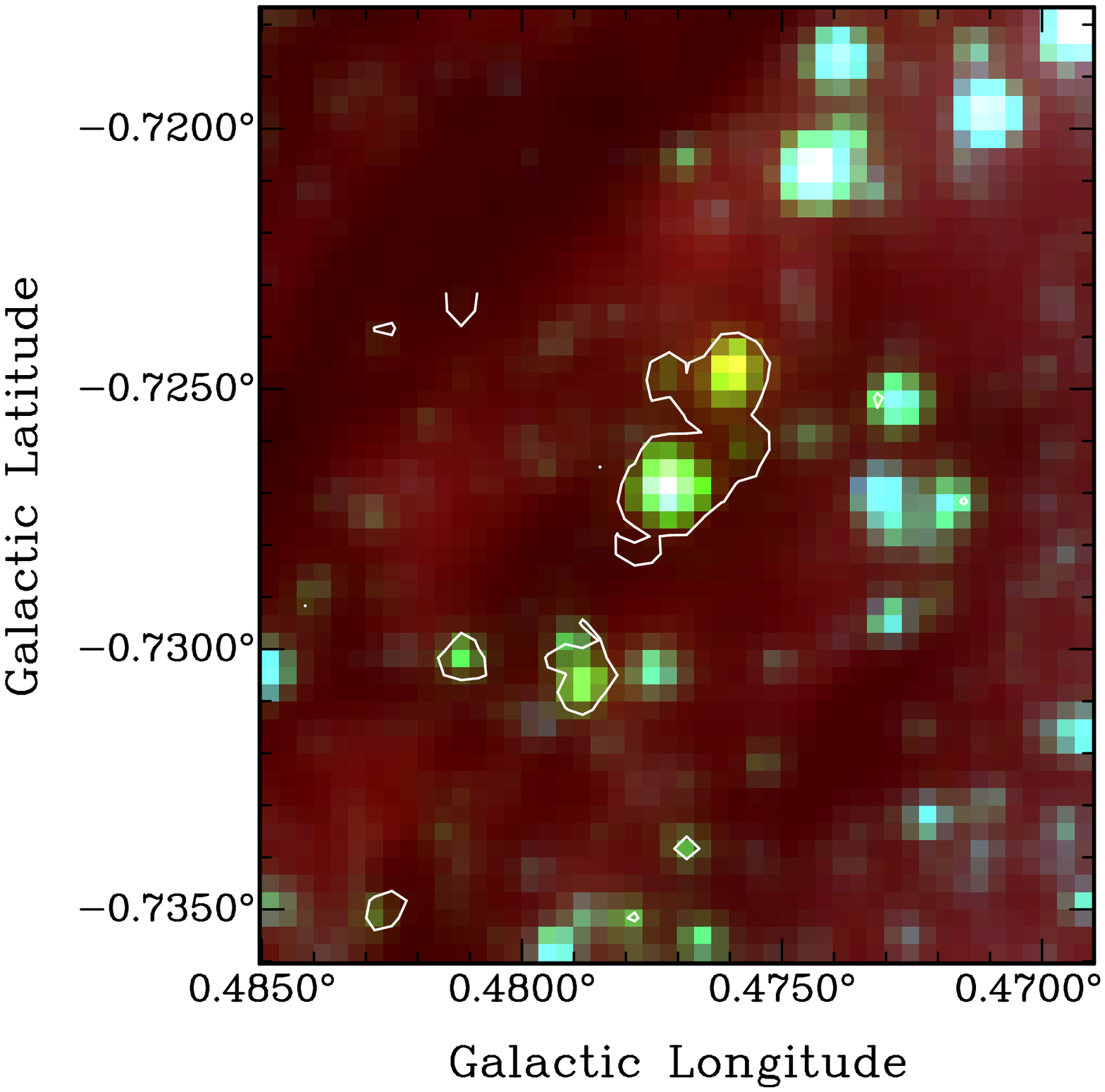}\\
  \caption{IRAC 3-color ({\it left}) and 24~\um\, ({\it right})
    images of source g14.  The contours in both images are at a 
    green ratio value of 0.55.\label{g14}}
\end{figure*}

\begin{figure*}[p]
  \centering
  G0.408$-$0.504\\
  \includegraphics[width=0.4\textwidth,clip=true]{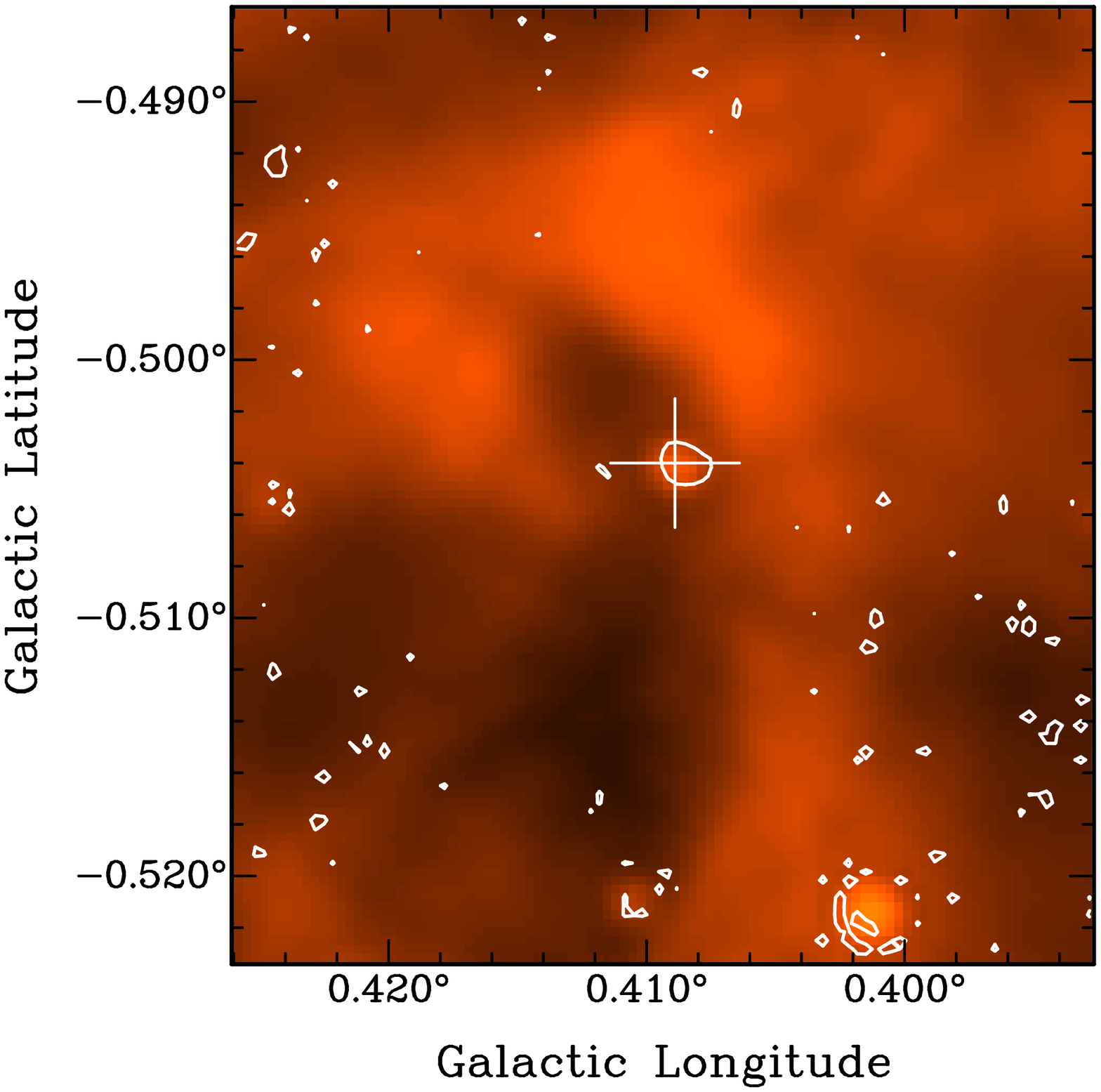}
  \hspace{-11.9cm}
  \includegraphics[width=0.4\textwidth,clip=true]{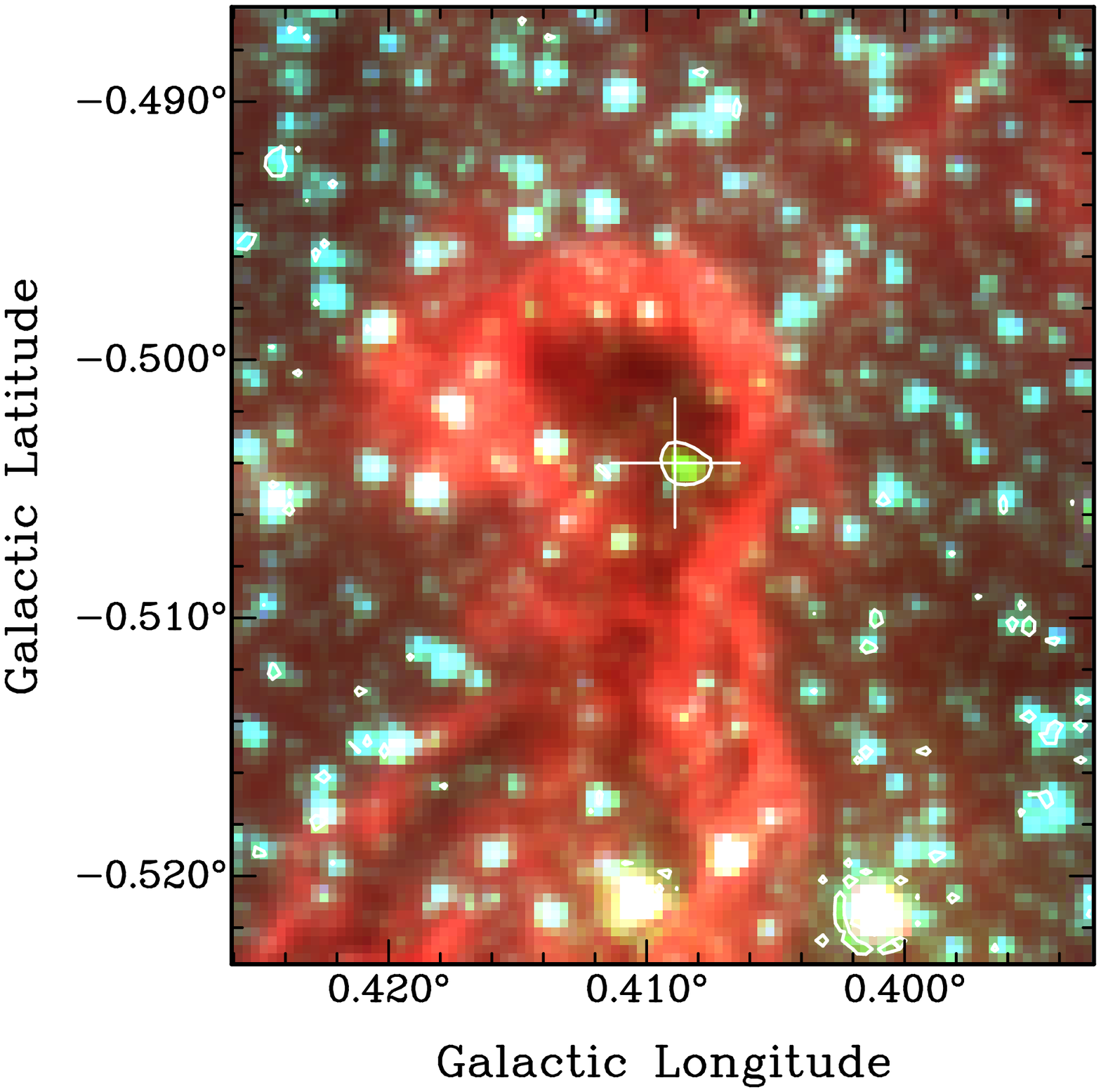}\\
  \includegraphics[angle=-90,width=0.4\textwidth,clip=true]{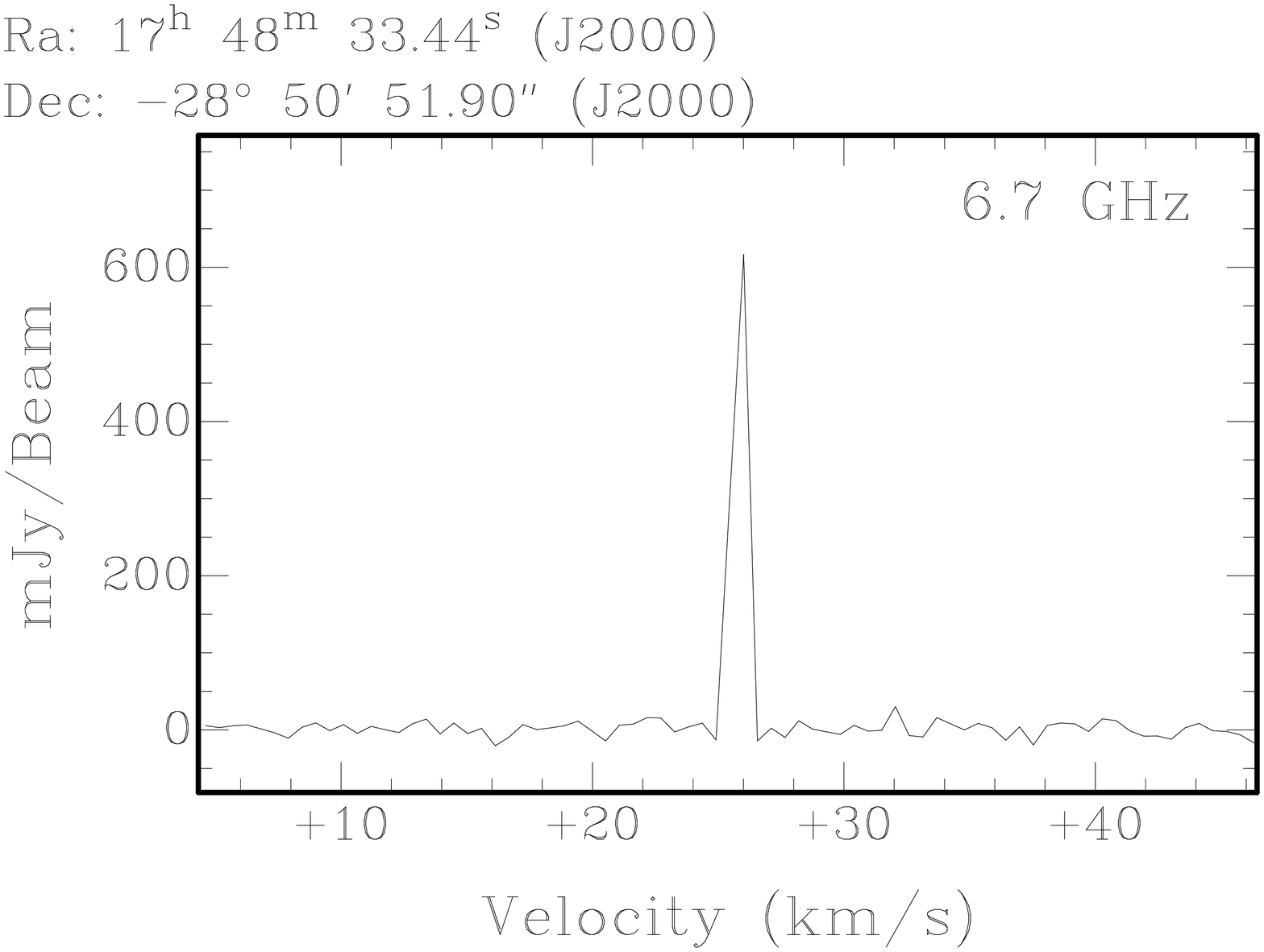}\\
  \caption{{\it Top}: IRAC 3-color ({\it left}) and 24~\um\, ({\it right})
    images of source g15.  The contours in both images are at a green 
    ratio value of 0.35. The white plus sign ($+$) designates the position of
    6.7~GHz maser emission detected with the EVLA. {\it Bottom}: Spectrum of
    6.7~GHz maser emission in the g15 field obtained with the
    EVLA.\label{g15}}
\end{figure*}

\begin{figure*}[p]
  \centering
  G0.376$+$0.040\\
  \includegraphics[width=0.4\textwidth,clip=true]{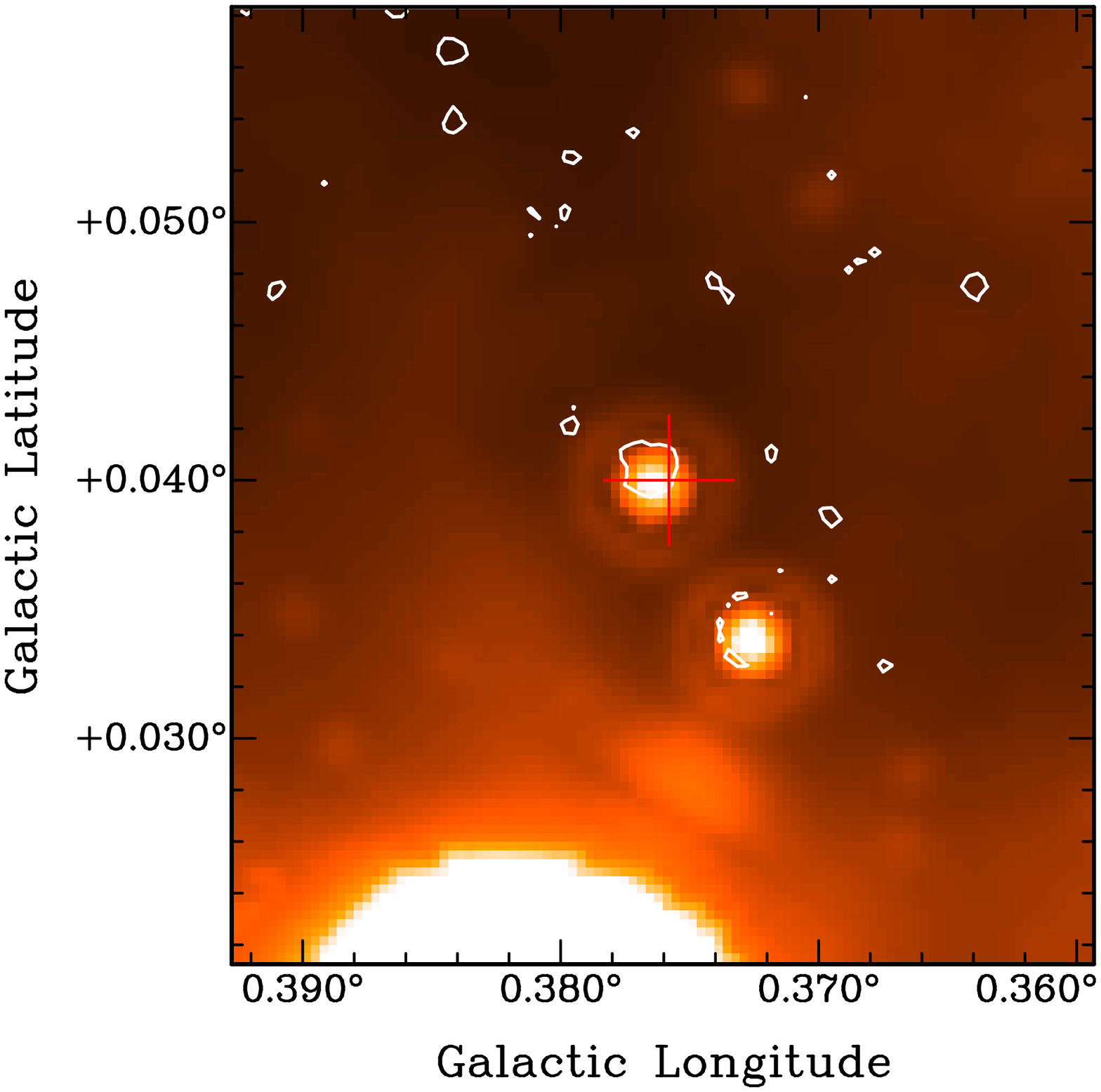}
  \hspace{-11.9cm}
  \includegraphics[width=0.4\textwidth,clip=true]{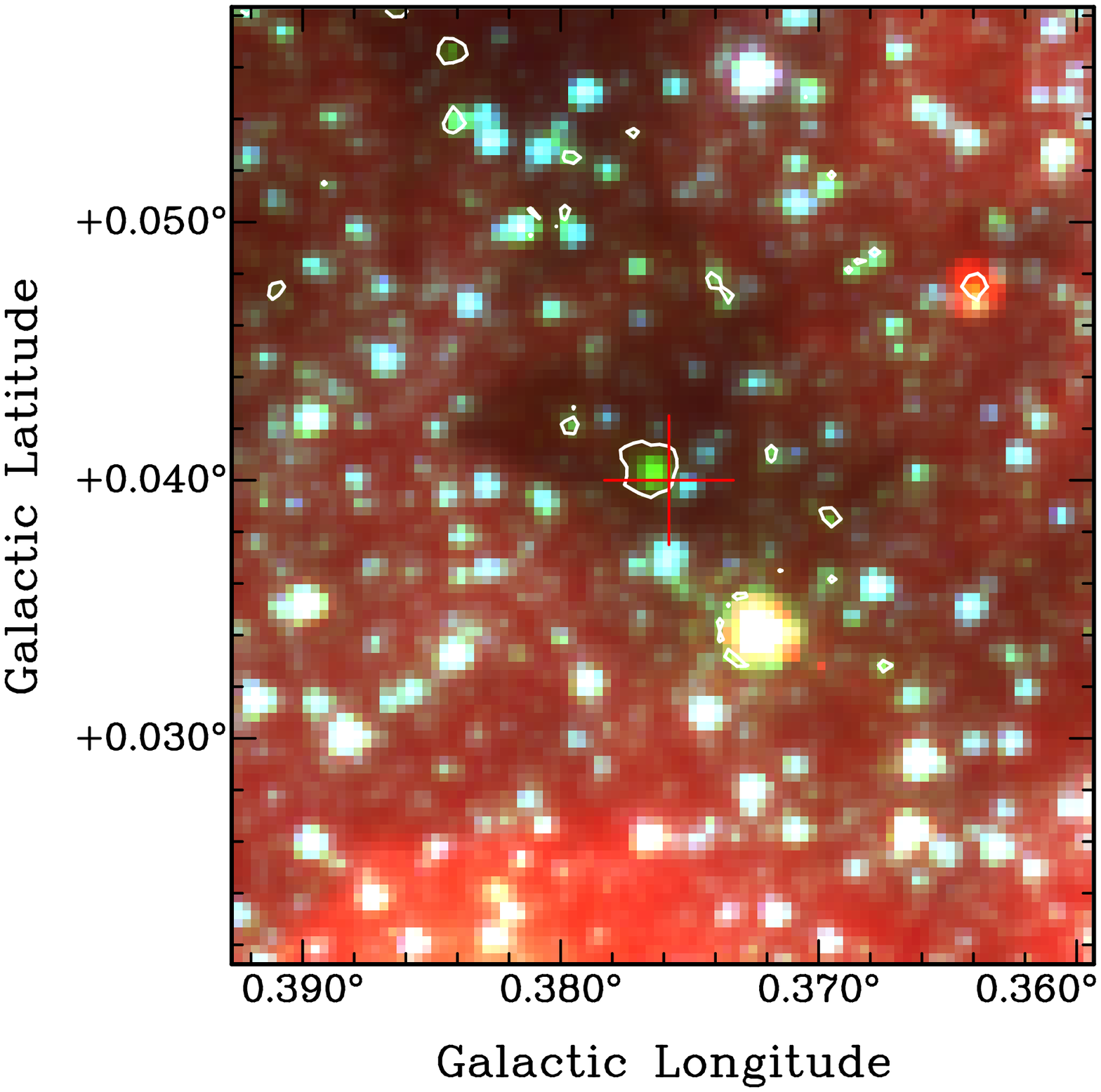}\\
  \caption{IRAC 3-color ({\it left}) and 24~\um\, ({\it right}) images of
    source g16.  The contours in both images are at a green ratio value of
    0.50.  The red plus sign ($+$) designates the position of 6.7~GHz maser
    emission detected by C10.\label{g16}}
\end{figure*}

\begin{figure*}[p]
  \centering
  G0.315$−$0.201\\
  \includegraphics[width=0.4\textwidth,clip=true]{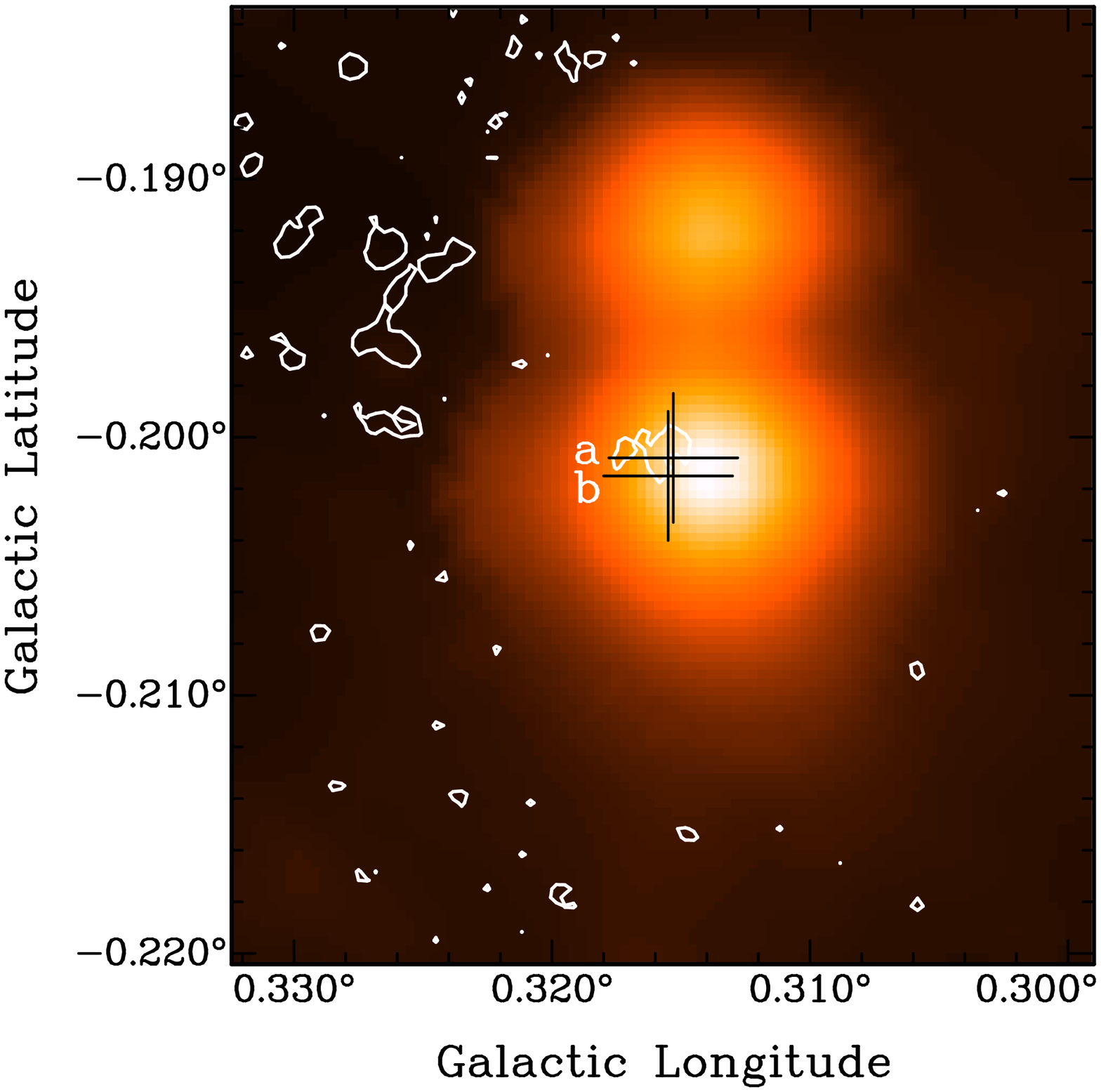}
  \hspace{-11.9cm}
  \includegraphics[width=0.4\textwidth,clip=true]{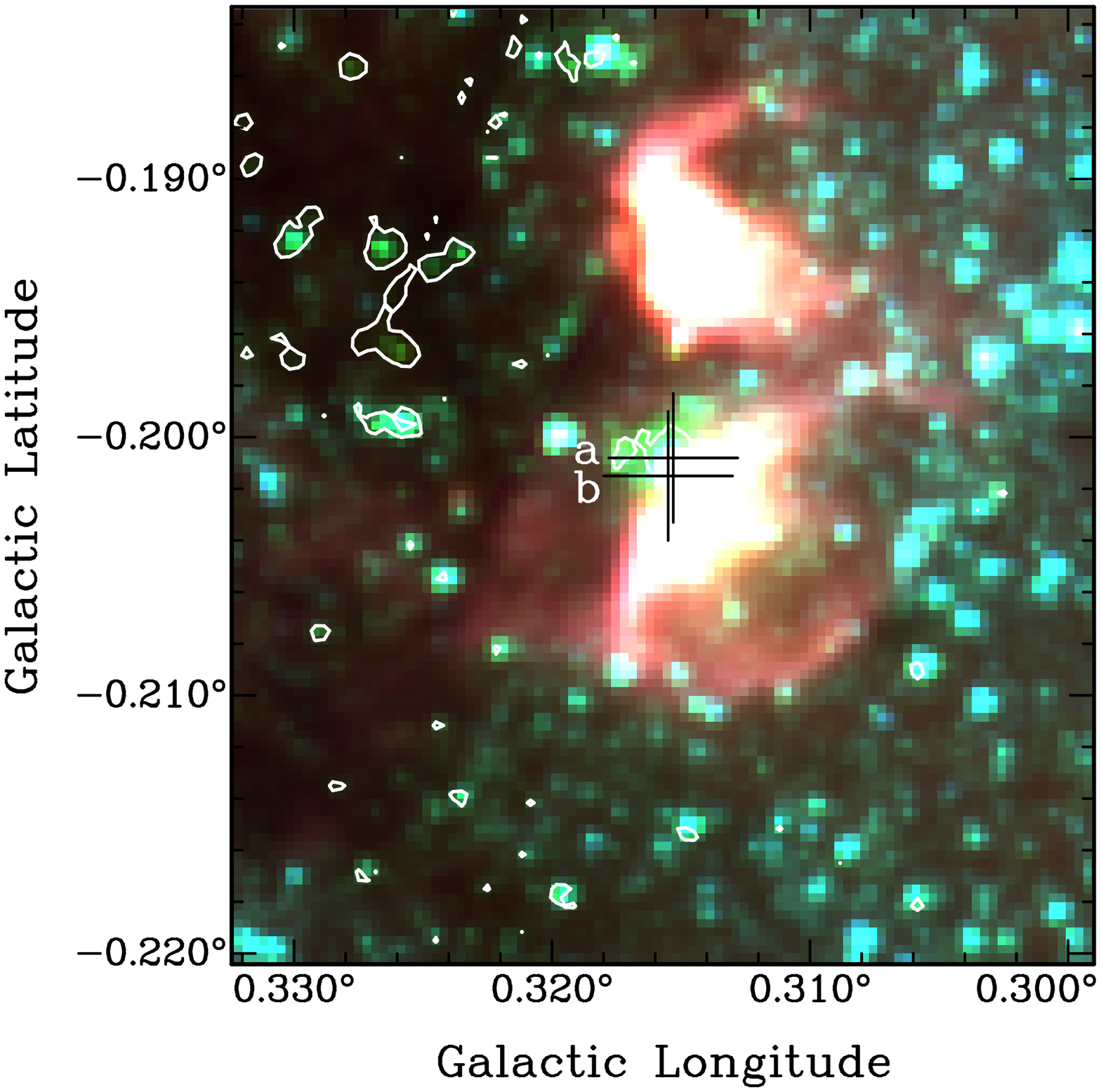}\\
  \includegraphics[angle=-90,width=0.4\textwidth,clip=true]{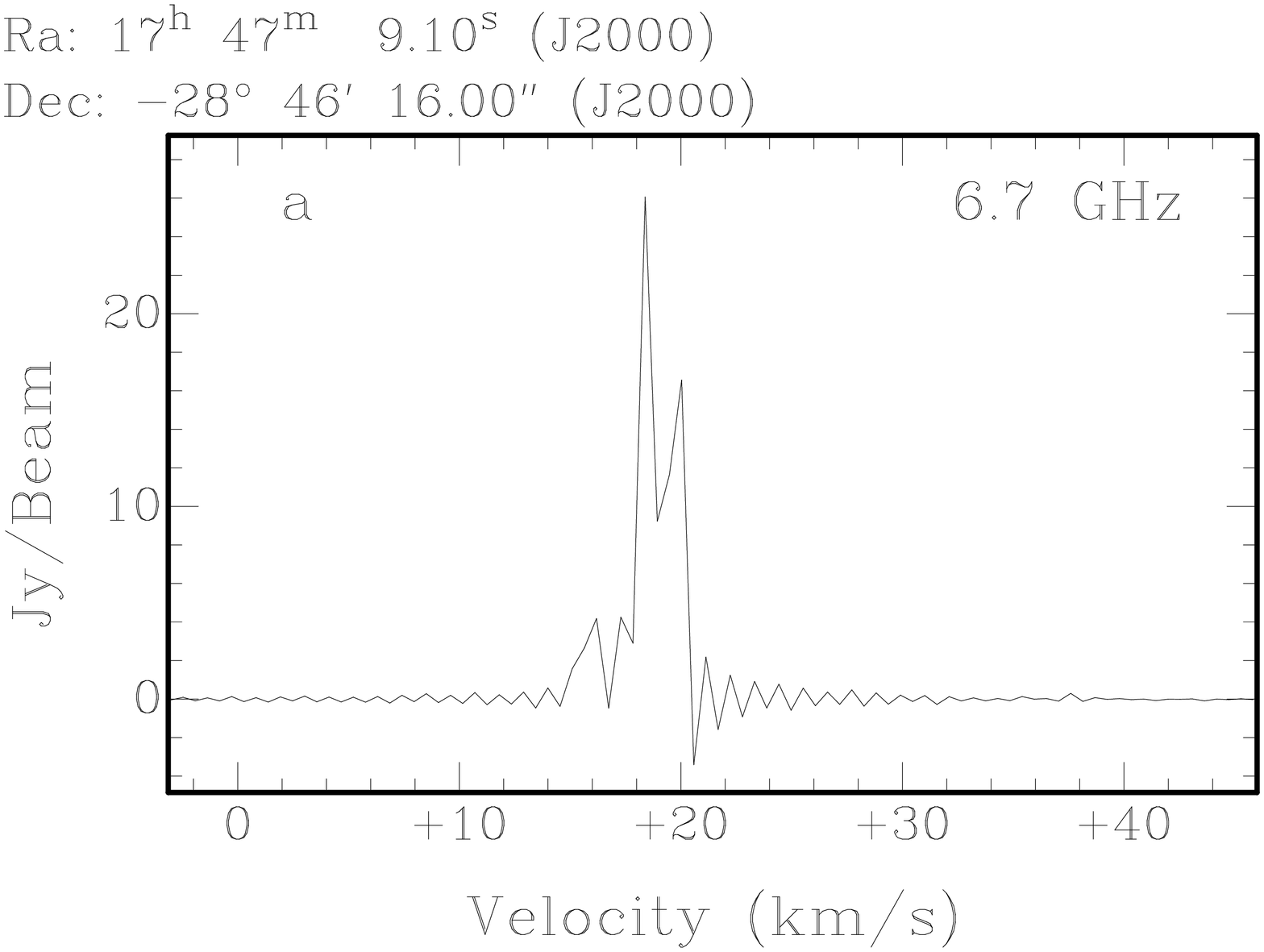}
  \includegraphics[angle=-90,width=0.4\textwidth,clip=true]{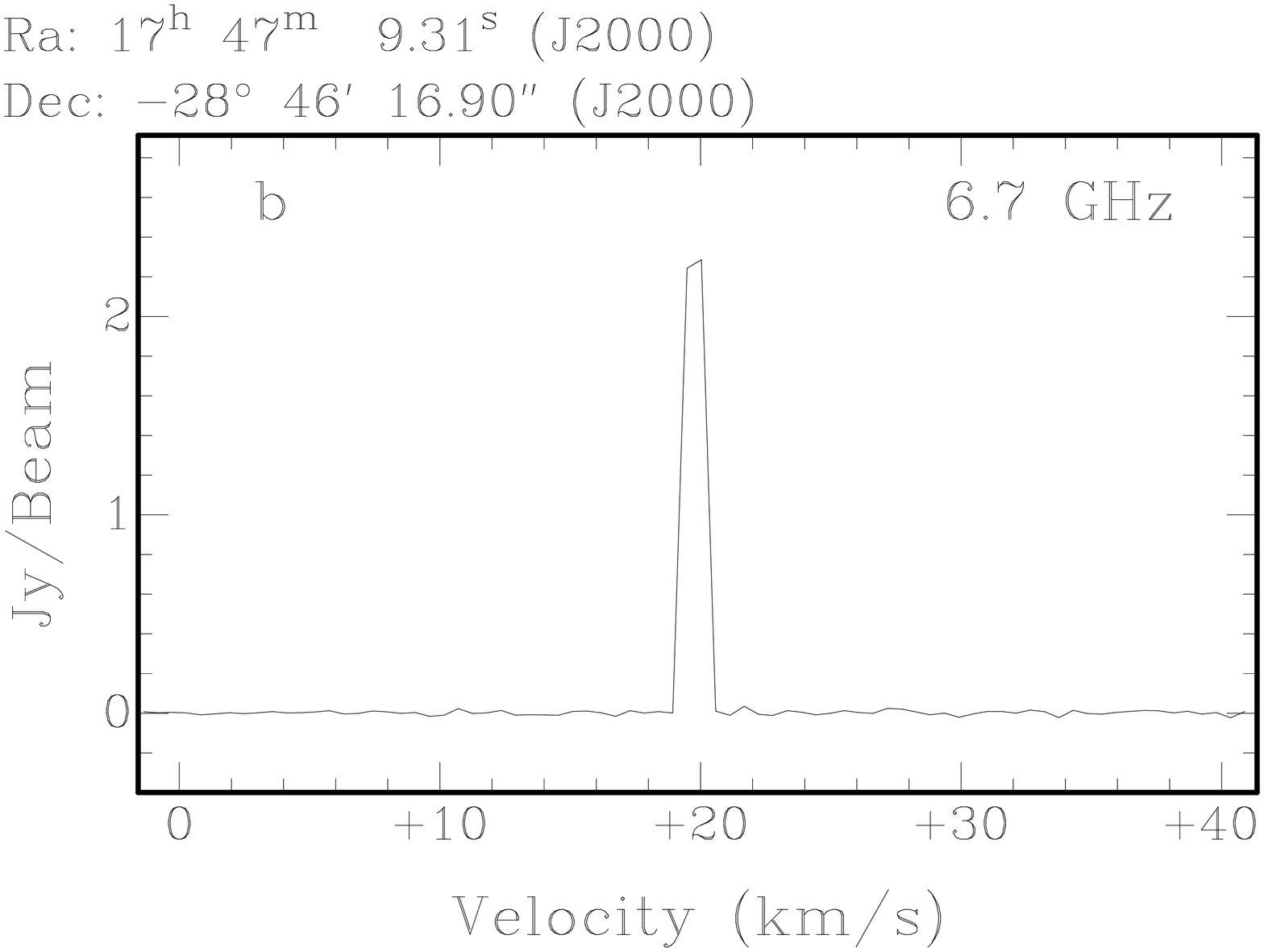}\\
  \caption{{\it Top}: IRAC 3-color ({\it left}) and 24~\um\, ({\it right})
    images of source g17.  The contours in both images are at a green
    ratio value of 0.40. The black plus signs ($+$) designate the positions of
    6.7~GHz maser emission detected with the EVLA.  {\it Bottom}: Spectra of
    6.7~GHz maser emission in the g17 field obtained with the
    EVLA.\label{g17}}
\end{figure*}

\begin{figure*}[p]
  \centering
  G0.167$-$0.445\\
  \includegraphics[width=0.4\textwidth,clip=true]{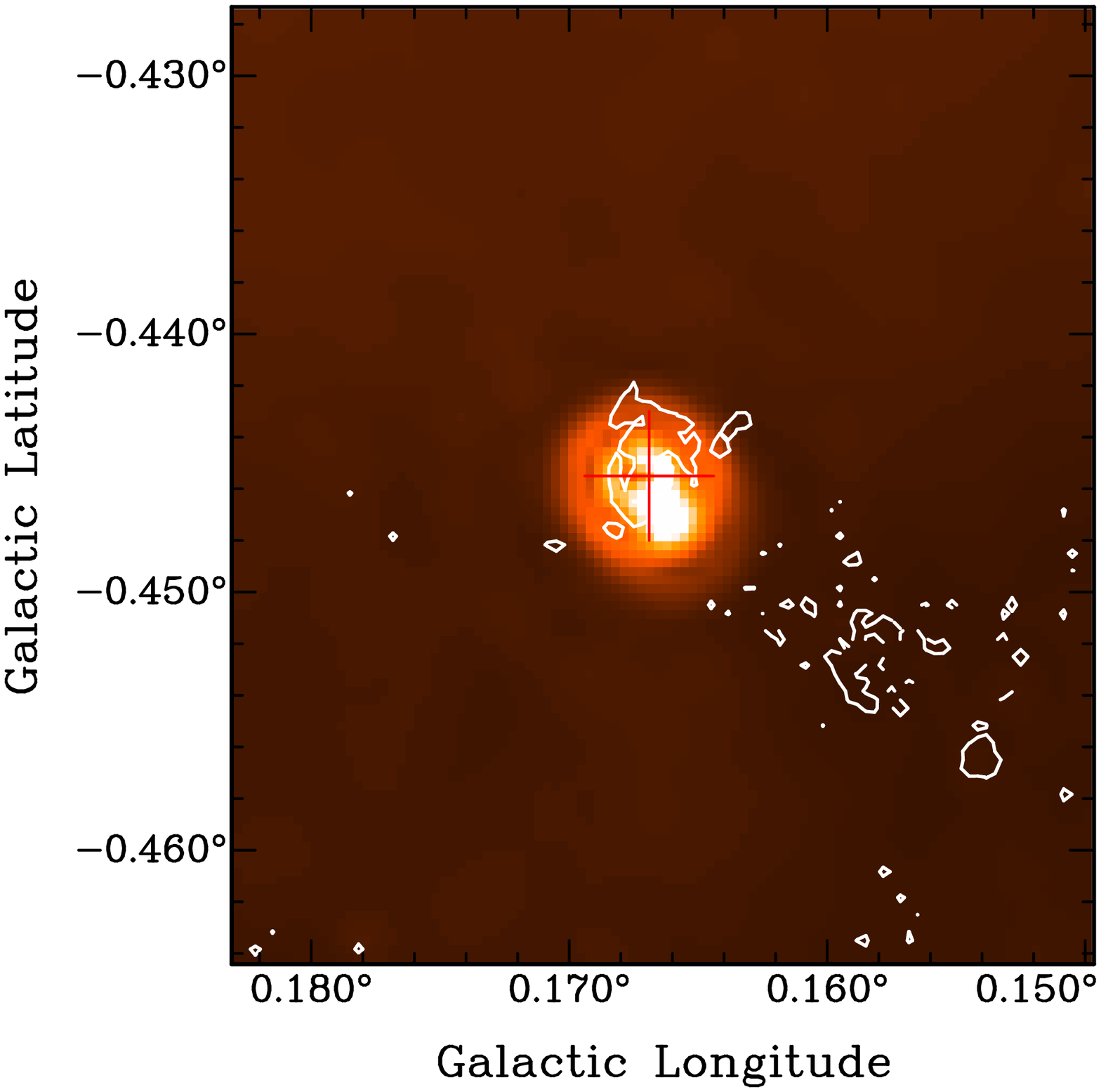}
  \hspace{-11.9cm}
  \includegraphics[width=0.4\textwidth,clip=true]{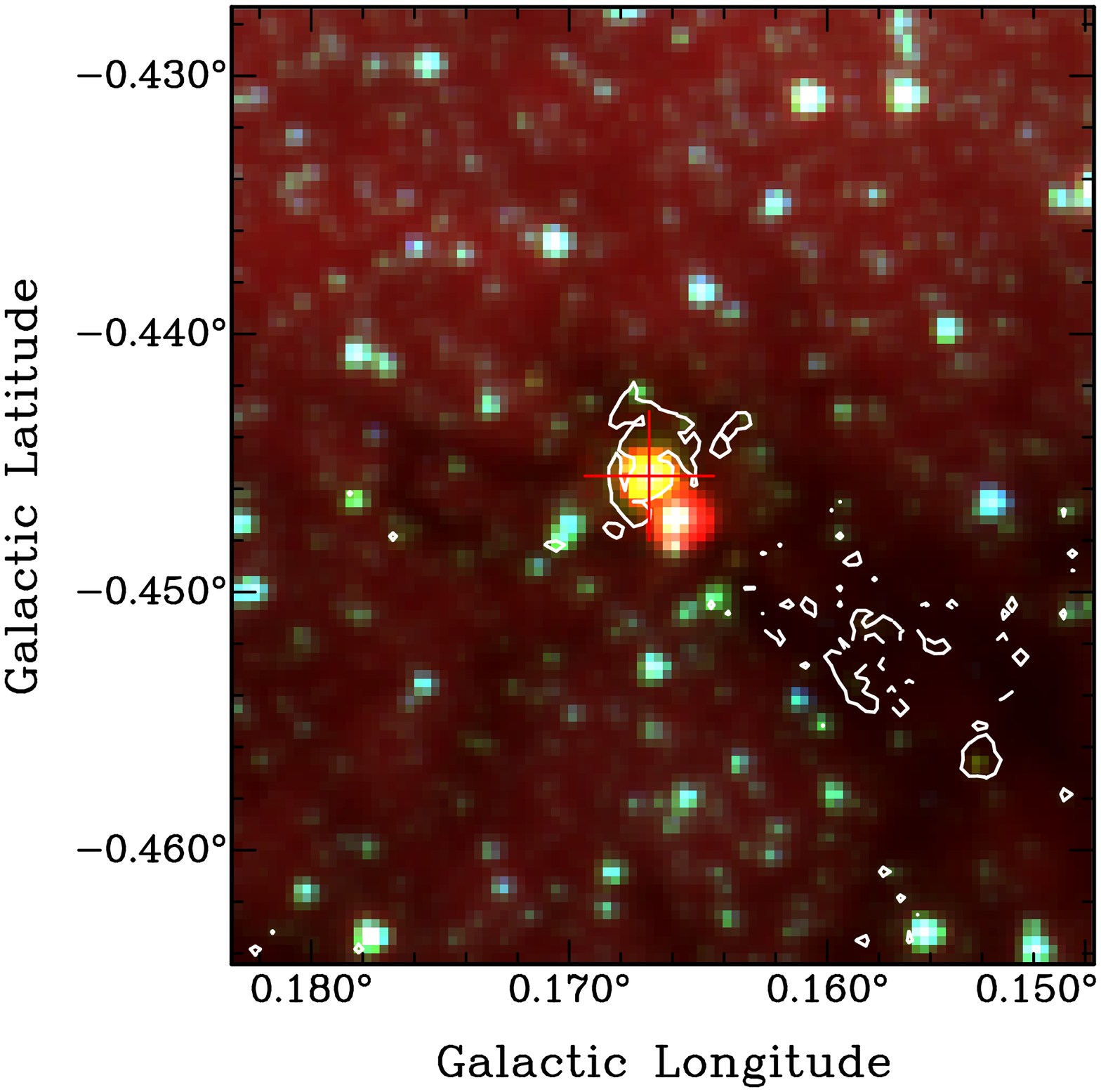}\\
  \caption{IRAC 3-color ({\it left}) and 24~\um\, ({\it right}) images of
    source g18.  The contours in both images are at a green ratio value of
    0.70.  The red plus sign ($+$) designates the position of 6.7~GHz maser
    emission detected by C10.\label{g18}}
\end{figure*}

\begin{figure*}[p]
  \centering
  G0.091$-$0.663\\
  \includegraphics[width=0.4\textwidth,clip=true]{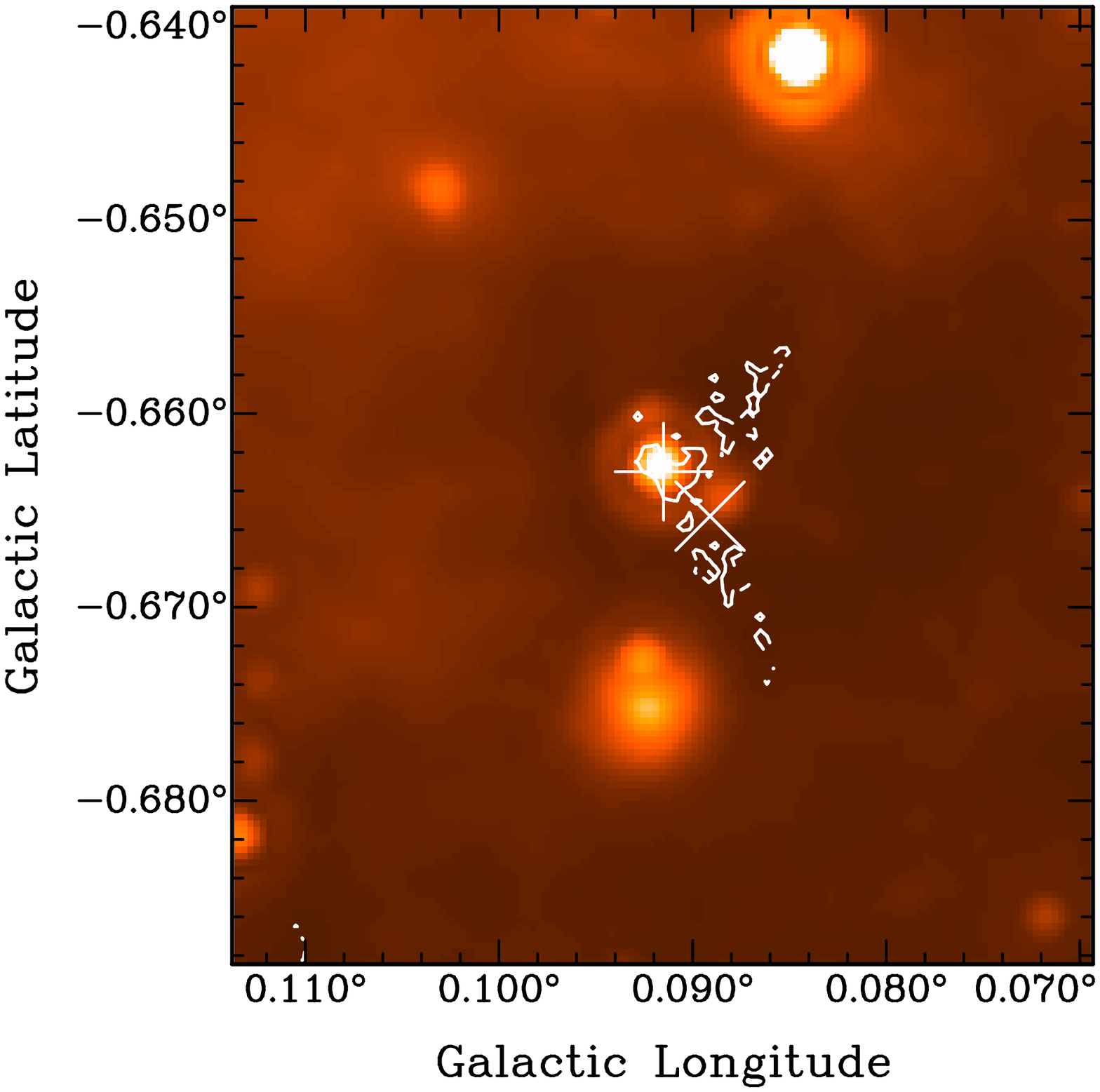}
  \hspace{-11.9cm}
  \includegraphics[width=0.4\textwidth,clip=true]{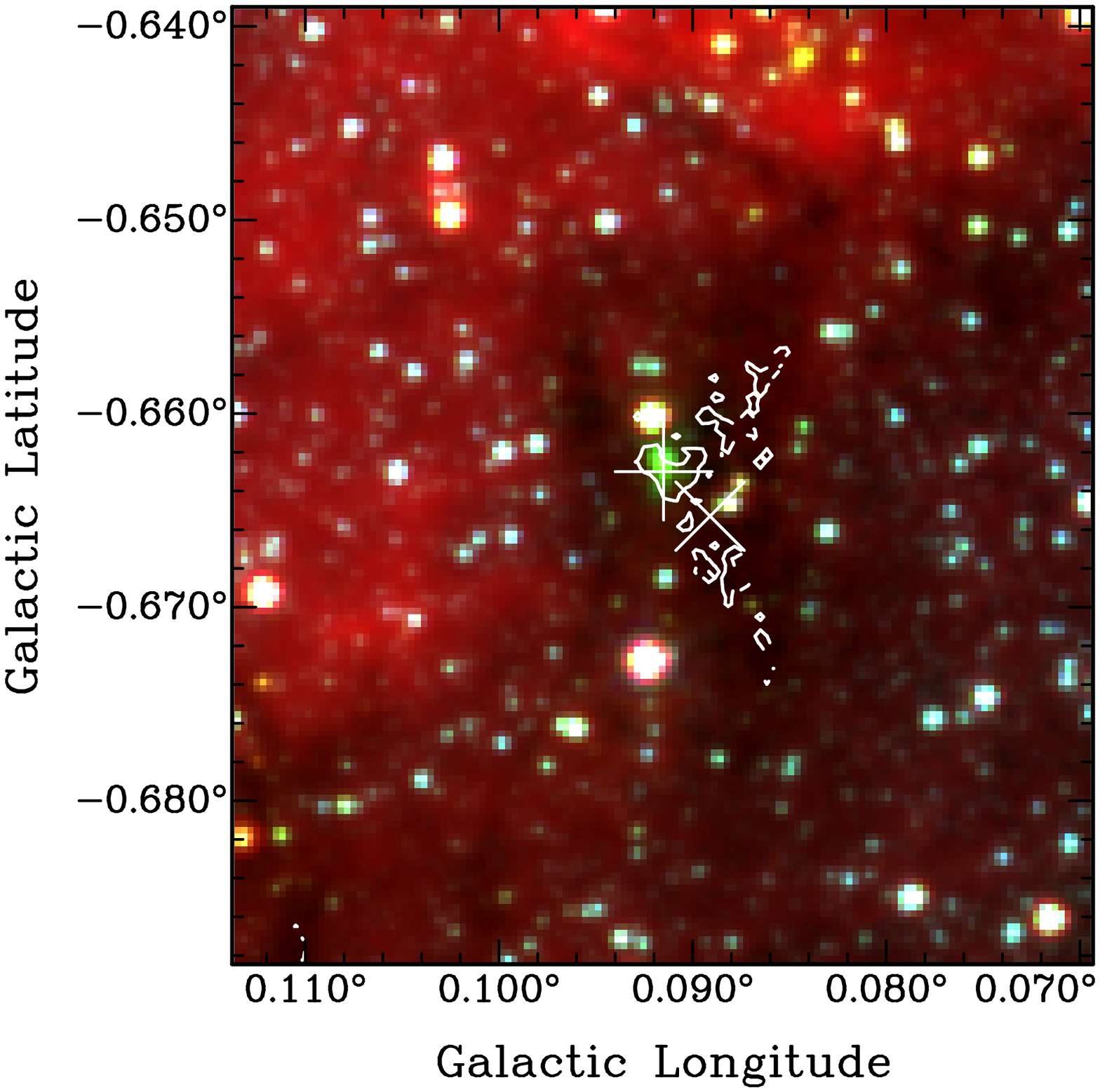}\\
  \includegraphics[angle=-90,width=0.4\textwidth,clip=true]{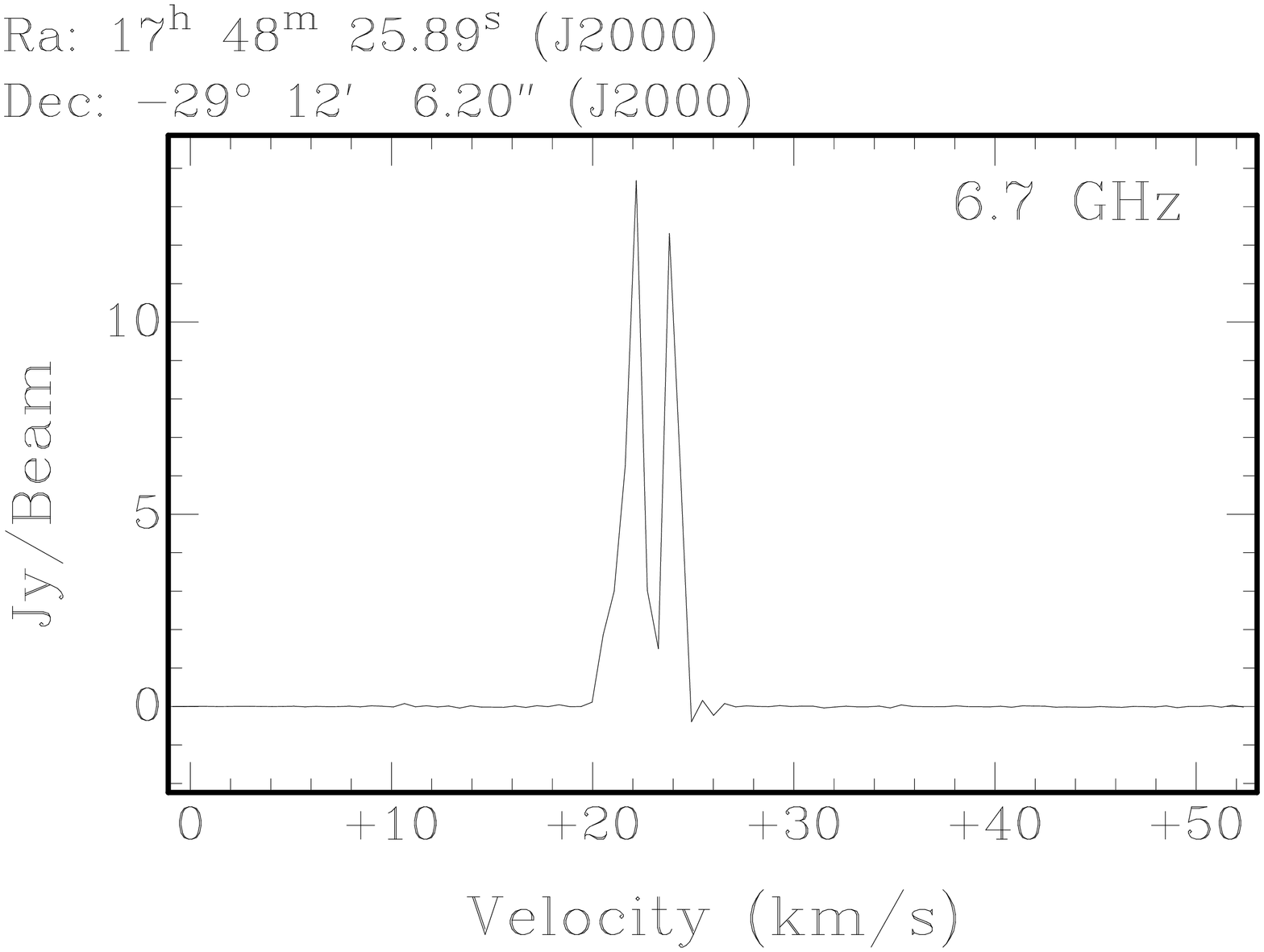}
  \includegraphics[angle=-90,width=0.4\textwidth,clip=true]{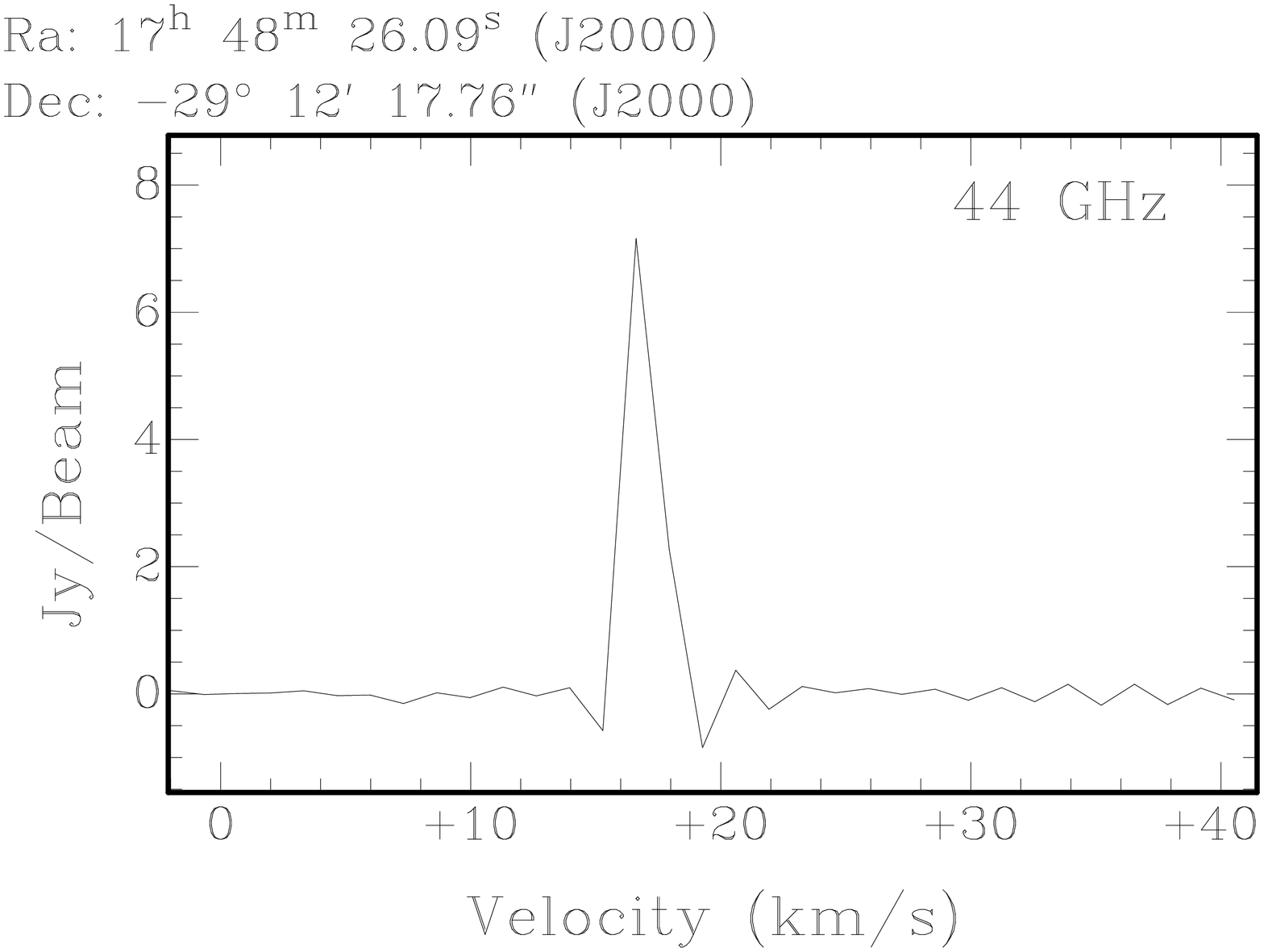}\\
  \caption{{\it Top}: IRAC 3-color ({\it left}) and 24~\um\, ({\it right})
    images of source g19.  The contours in both images are at a green ratio
    value of 1.00.  The white plus sign ($+$) designates the position of
    6.7~GHz maser emission detected with the EVLA, and the white cross sign
    ($\times$) designates the position of 44~GHz maser emission detected with
    the EVLA. {\it Bottom}: Spectra of 6.7~GHz ({\it left}) and 44~GHz ({\it
      right}) maser emission in the g19 field obtained with the
    EVLA.\label{g19}}
\end{figure*}

\begin{figure*}[p]
  \centering
  G0.084$-$0.642\\
  \includegraphics[width=0.4\textwidth,clip=true]{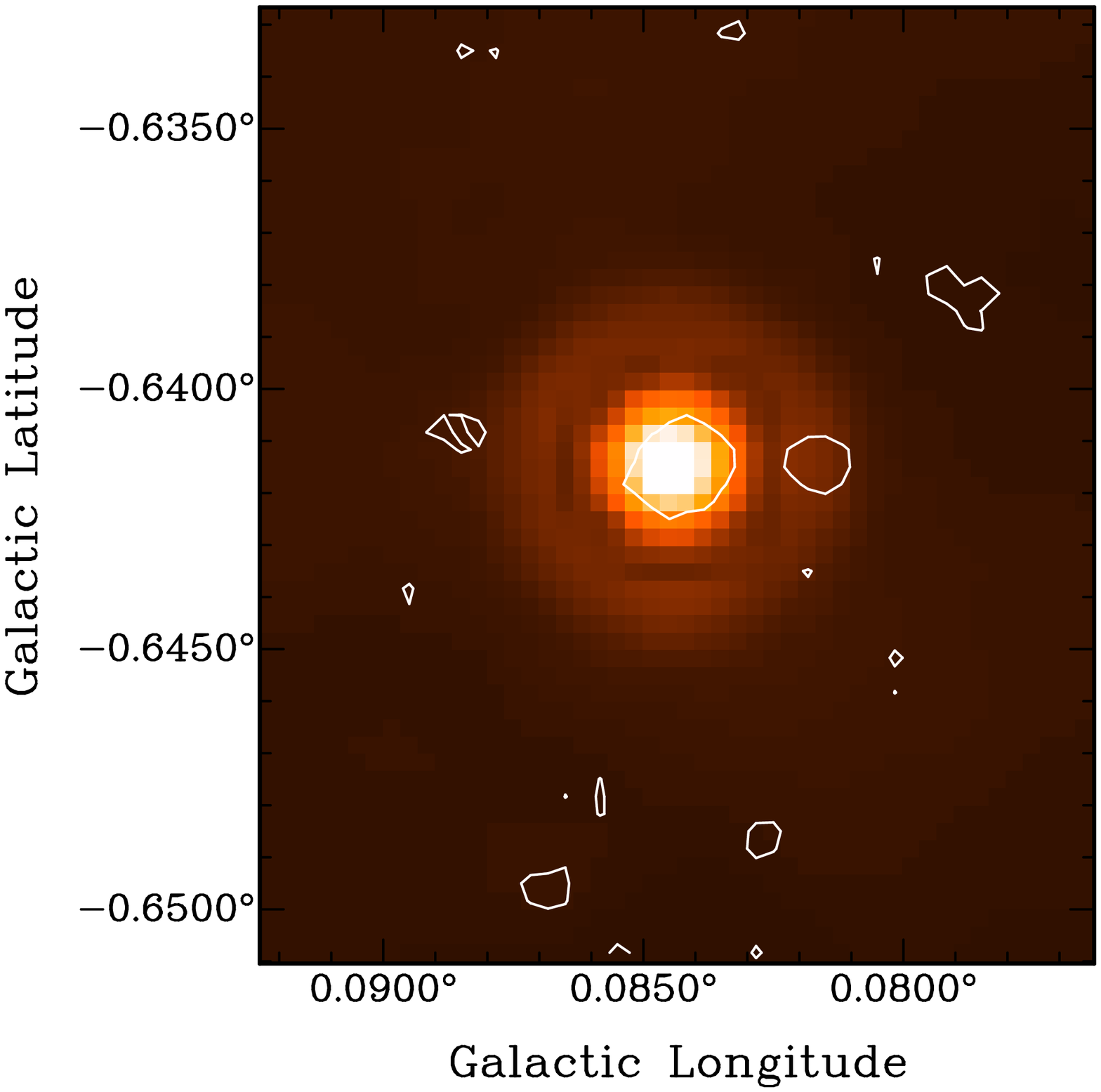}
  \hspace{-11.9cm}
  \includegraphics[width=0.4\textwidth,clip=true]{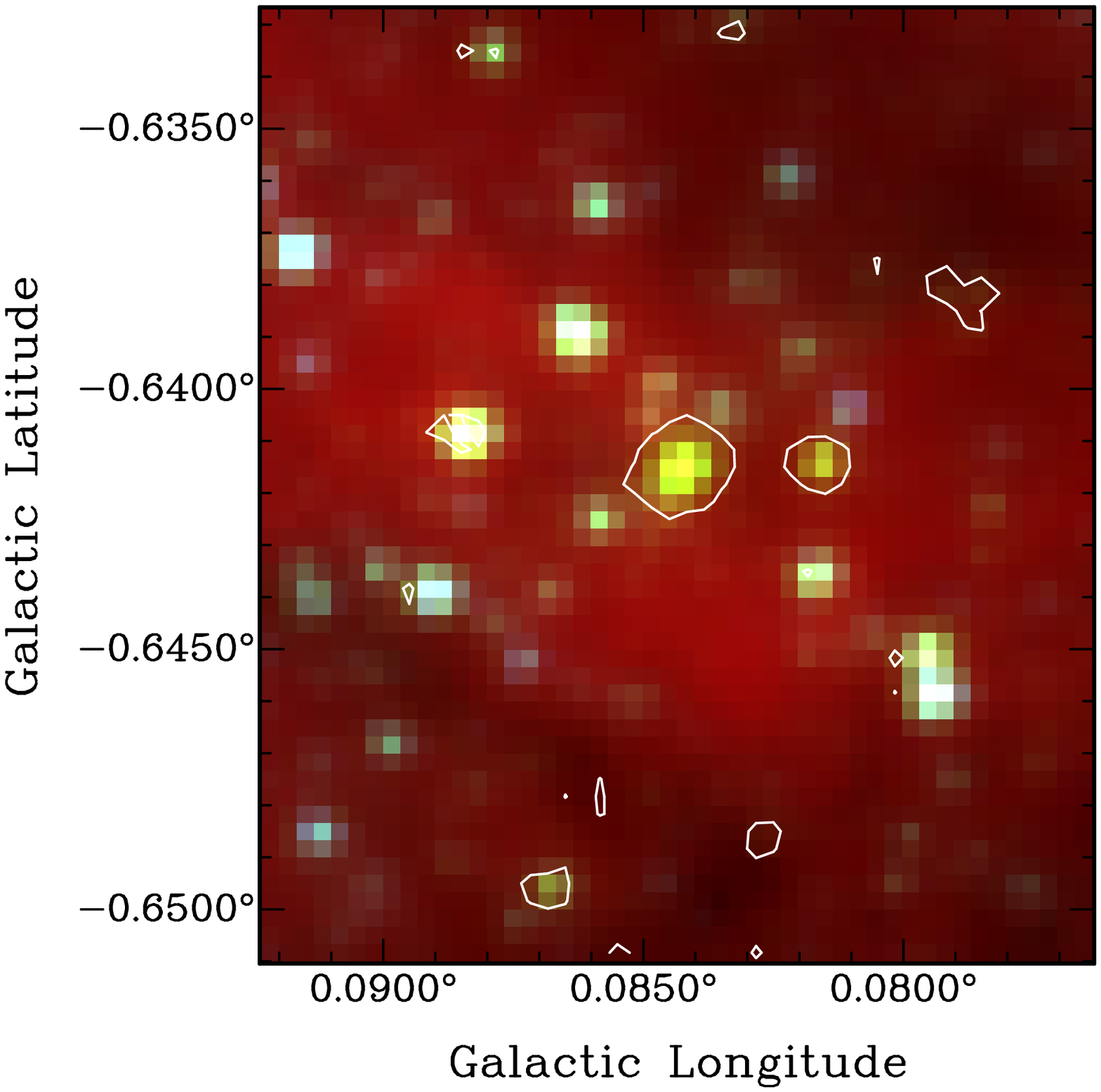}\\
  \caption{IRAC 3-color ({\it left}) and 24~\um\, ({\it right}) images of
    source g20.  The contours in both images are at a green ratio value of
    0.50. \label{g20}}
\end{figure*}

\begin{figure*}[p]
  \centering
  G359.972$-$0.459\\
  \includegraphics[width=0.4\textwidth,clip=true]{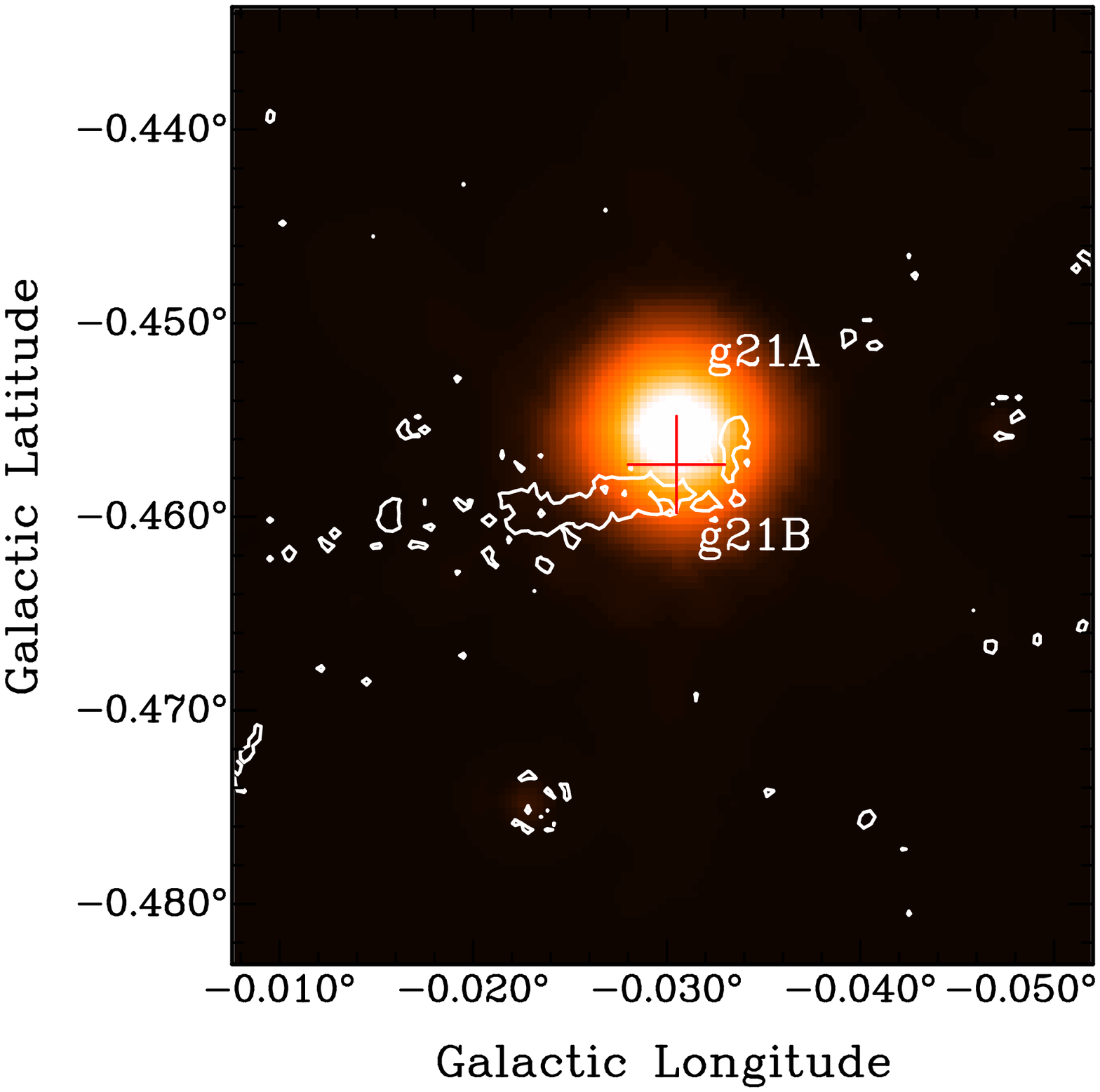}
  \hspace{-11.9cm}
  \includegraphics[width=0.4\textwidth,clip=true]{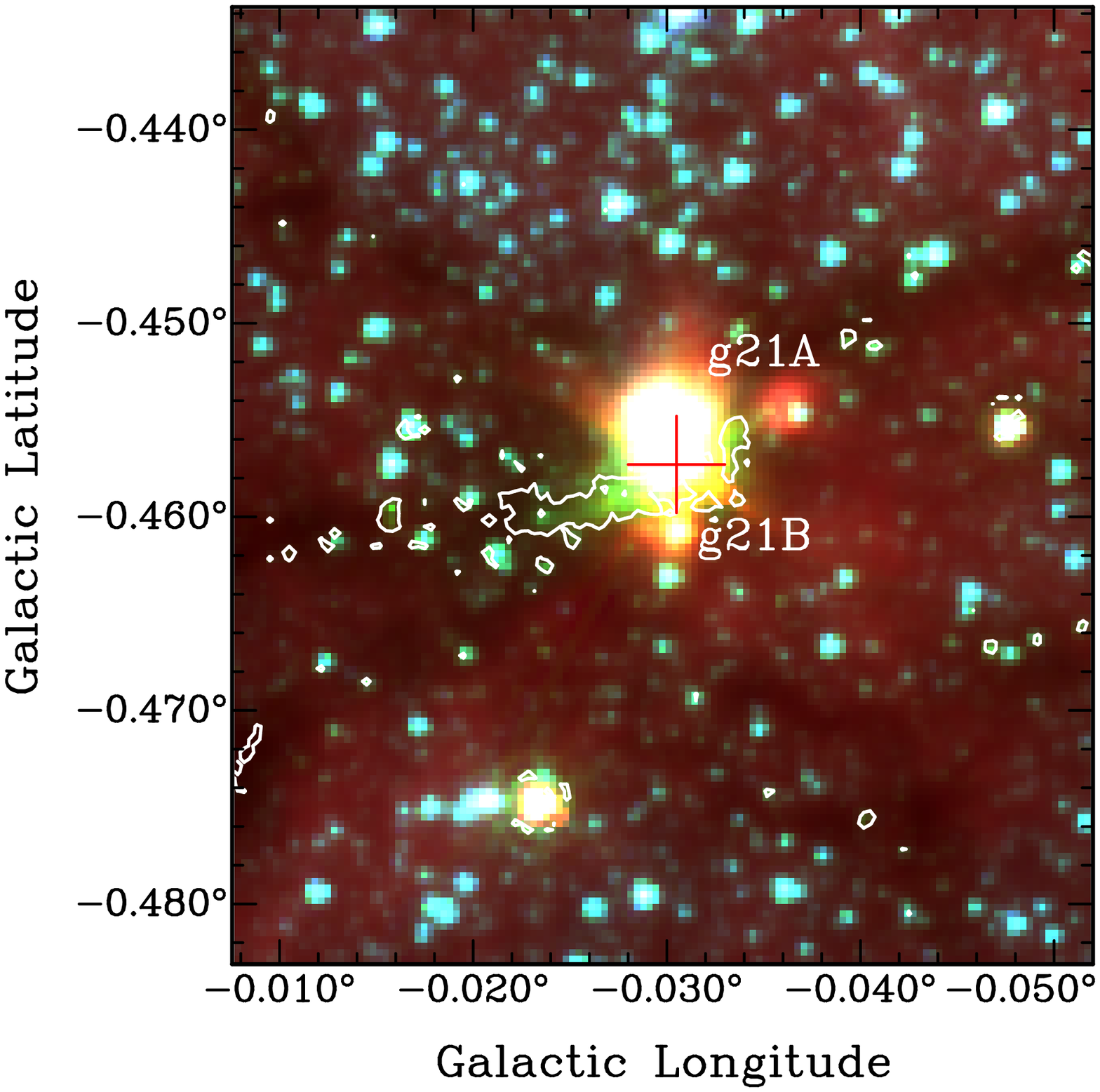}\\
  \caption{IRAC 3-color ({\it left}) and 24~\um\, ({\it right}) images of
    source g21.  The contours in both images are at a green ratio value of
    0.45.  The red plus sign ($+$) designates the position of 6.7~GHz maser
    emission detected by C10. \label{g21}}
\end{figure*}

\begin{figure*}[p]
  \centering
  G359.939$+$0.170\\
  \includegraphics[width=0.4\textwidth,clip=true]{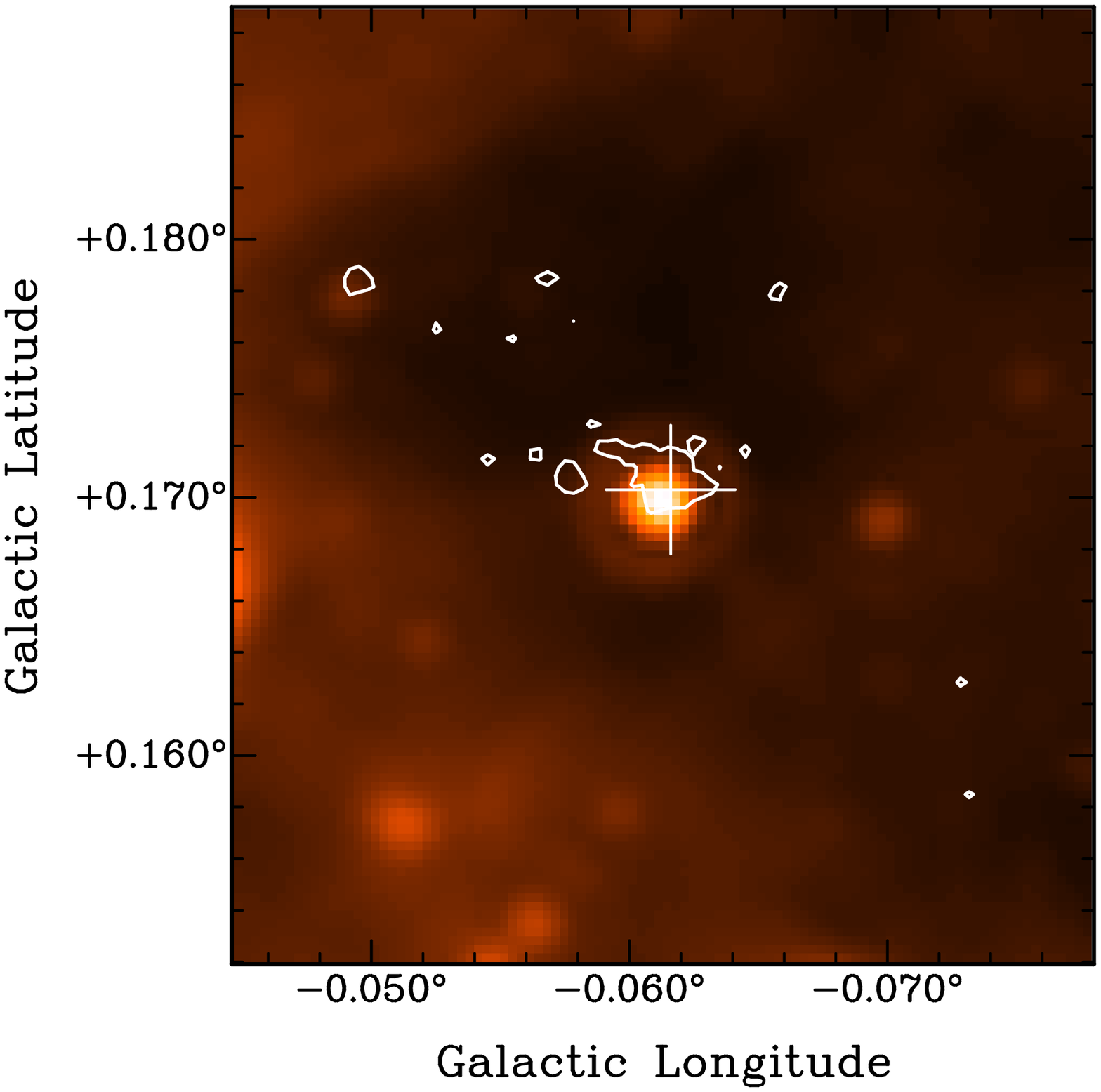}
  \hspace{-11.9cm}
  \includegraphics[width=0.4\textwidth,clip=true]{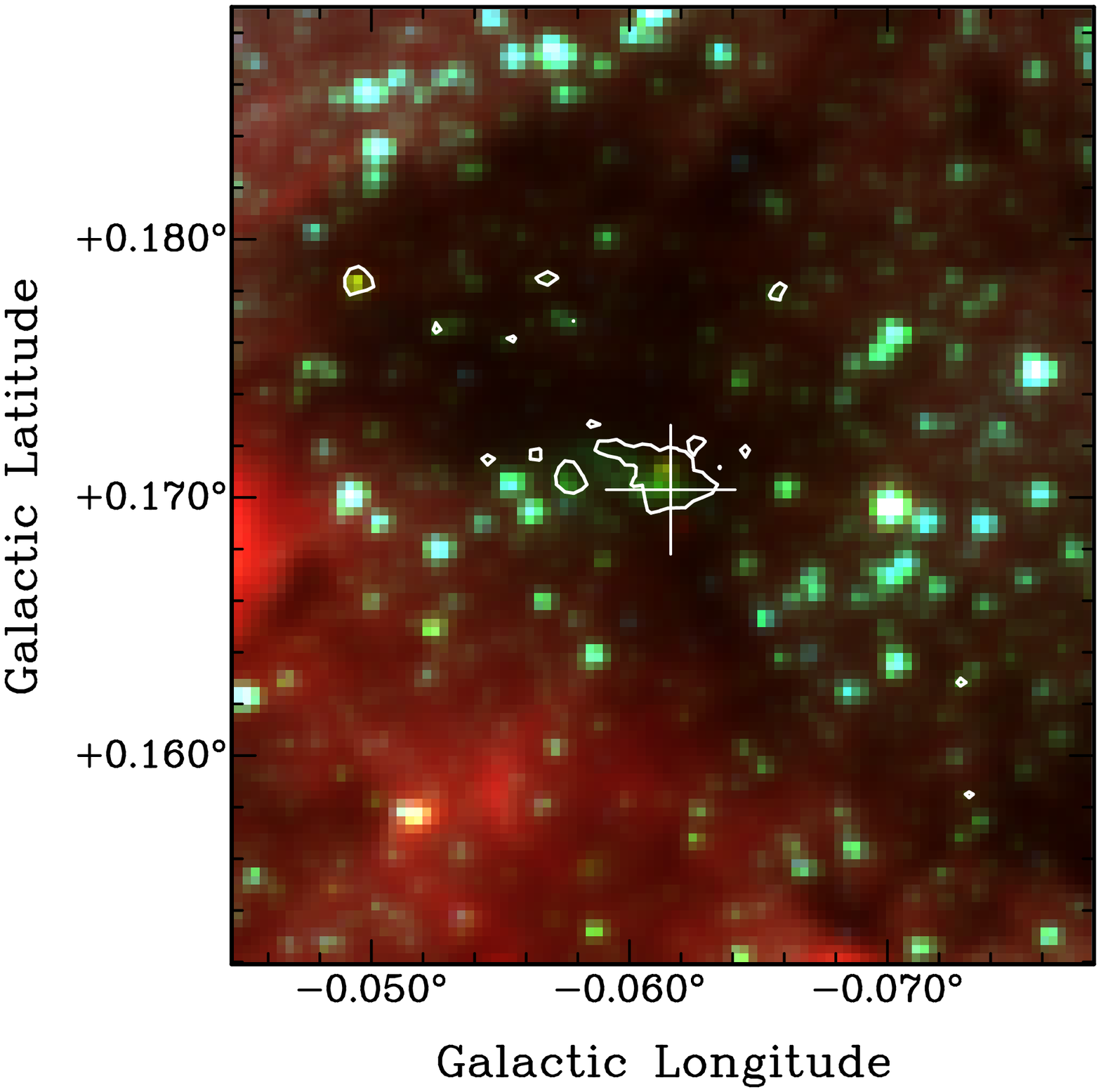}\\
  \includegraphics[angle=-90,width=0.4\textwidth,clip=true]{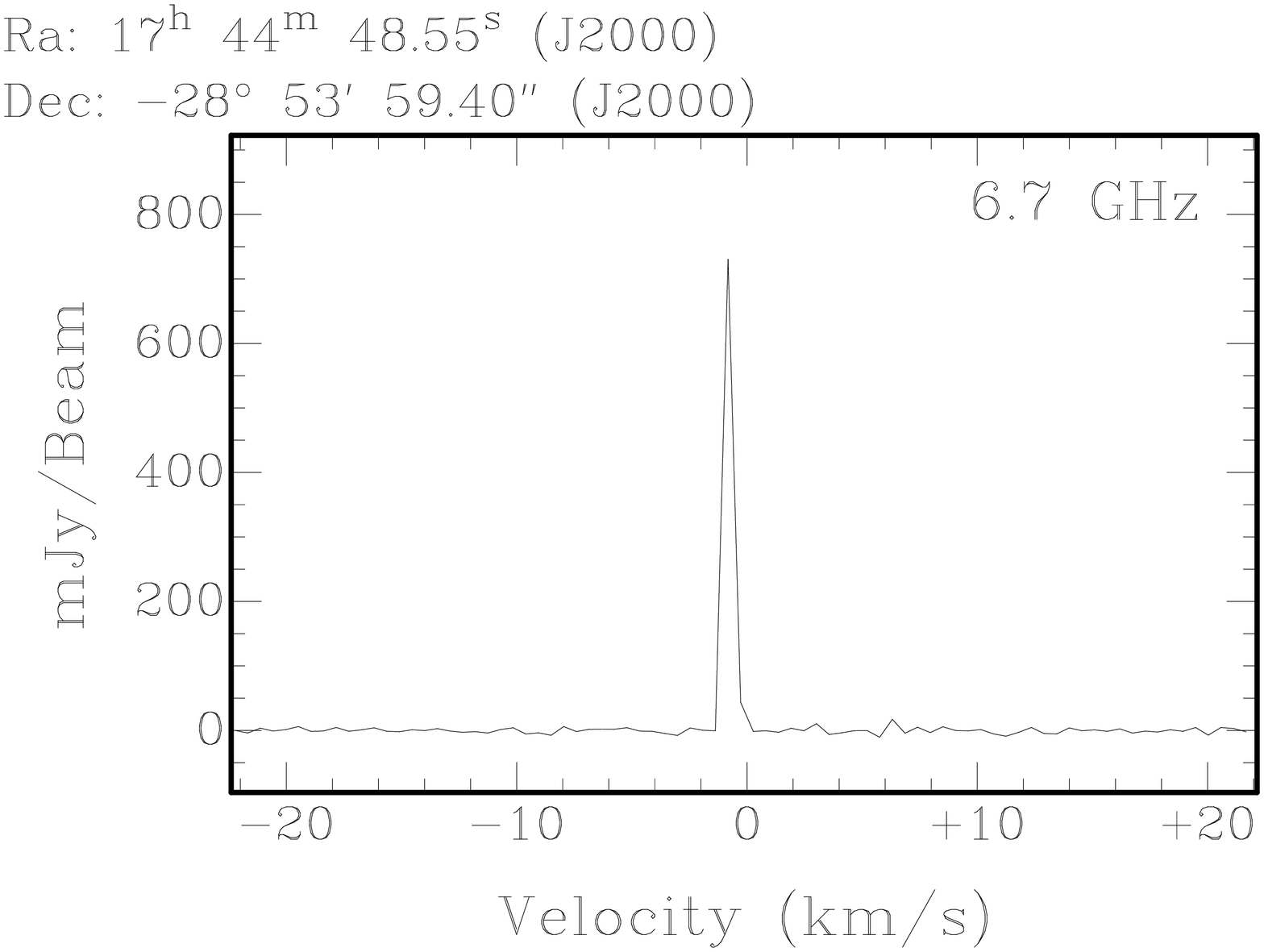}\\
  \caption{{\it Top}: IRAC 3-color ({\it left}) and 24~\um\, ({\it right})
    images of source g22.  The contours in both images are at a green ratio
    value of 0.80.  The white plus sign ($+$) designates the position of
    6.7~GHz maser emission detected with the EVLA. {\it Bottom}: Spectrum of
    6.7~GHz maser emission in the g22 field obtained with the
    EVLA.\label{g22}}
\end{figure*}

\begin{figure*}[p]
  \centering
  G359.932$-$0.063\\
  \includegraphics[width=0.4\textwidth,clip=true]{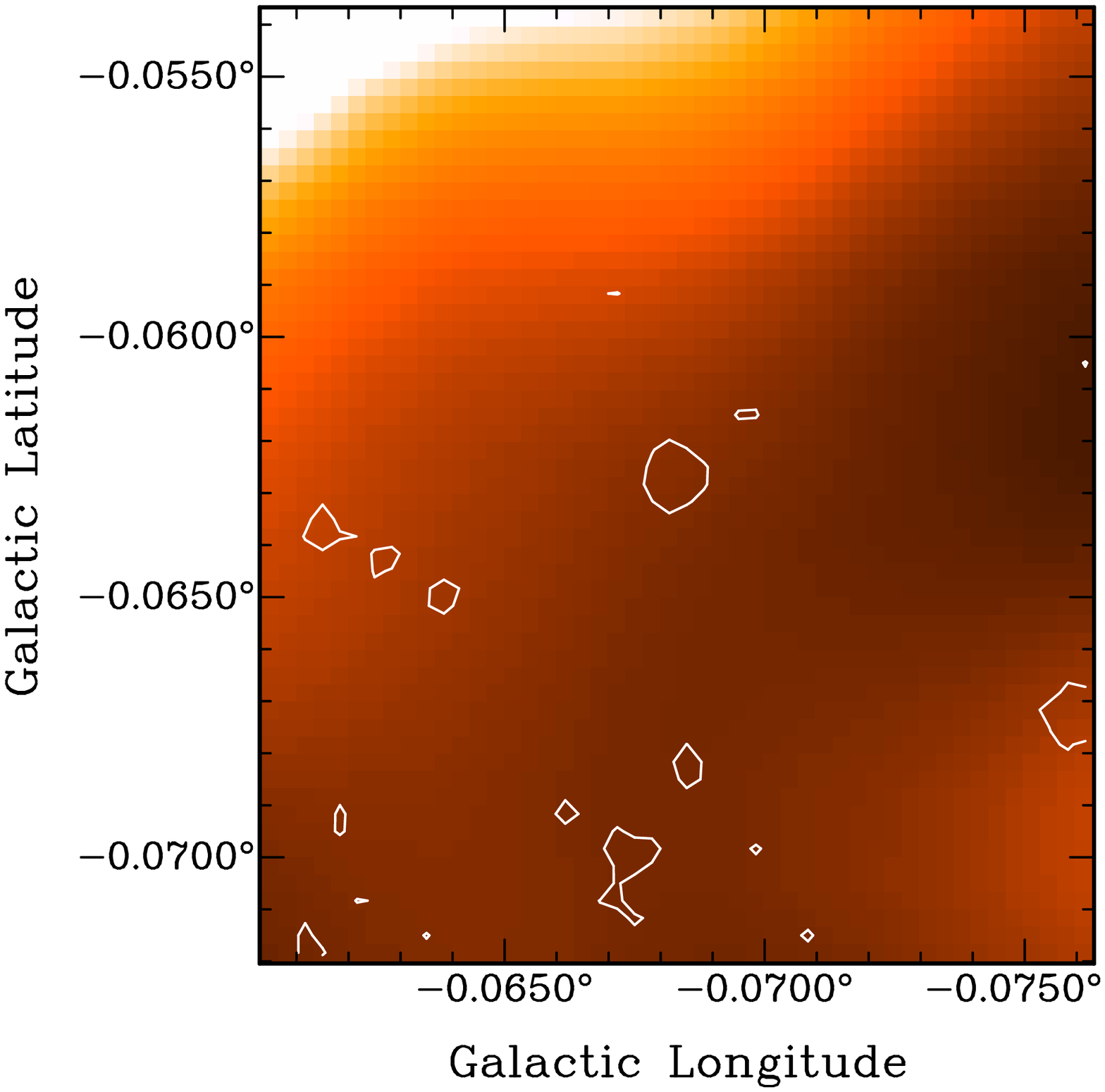}
  \hspace{-11.9cm}
  \includegraphics[width=0.4\textwidth,clip=true]{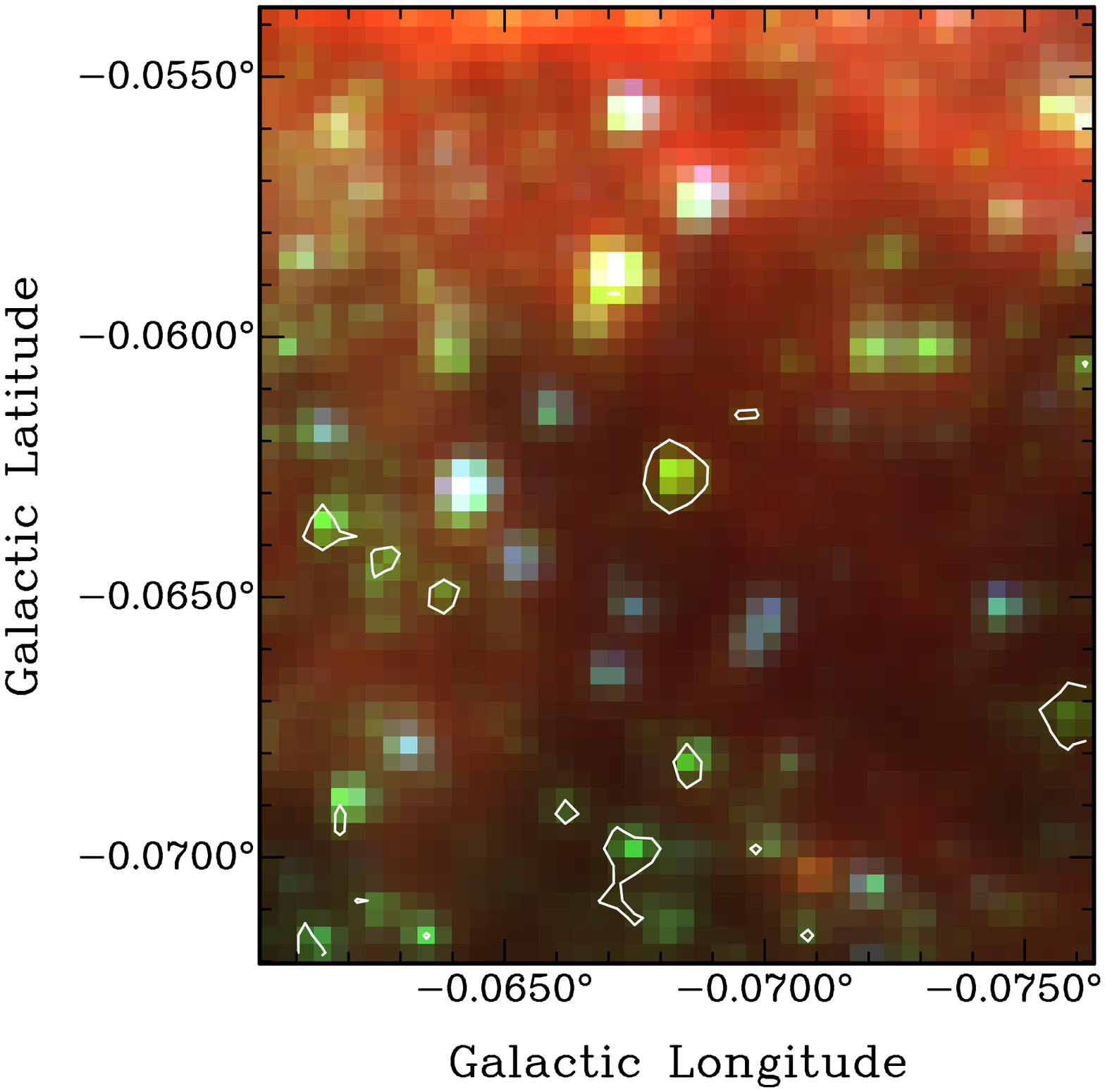}\\
  \caption{IRAC 3-color ({\it left}) and 24~\um\, ({\it right}) images of
    source g23.  The contours in both images are at a green ratio value of
    0.45. \label{g23}}
\end{figure*}

\begin{figure*}[p]
  \centering
  G359.907$-$0.303\\
  \includegraphics[width=0.4\textwidth,clip=true]{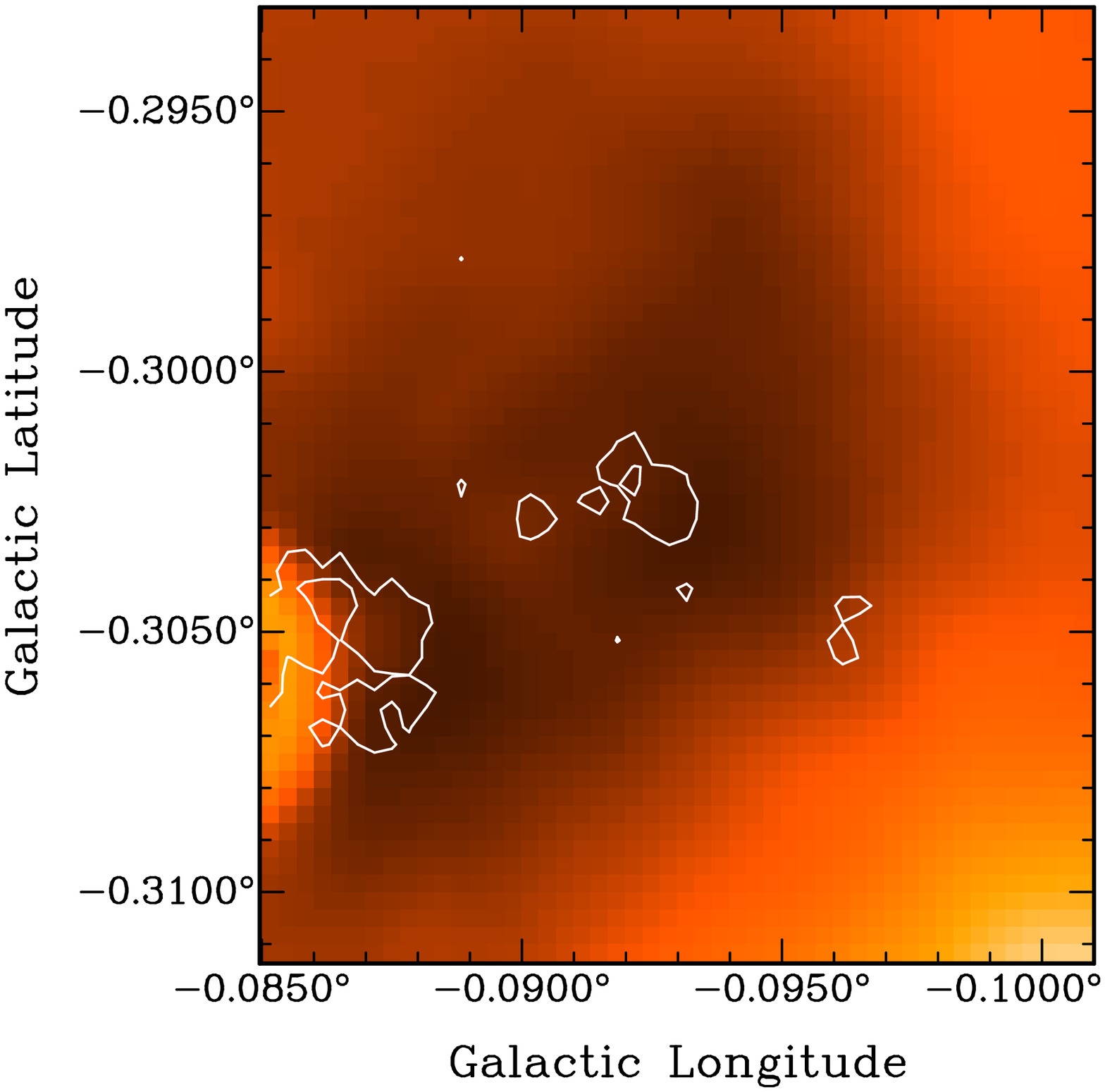}
  \hspace{-11.9cm}
  \includegraphics[width=0.4\textwidth,clip=true]{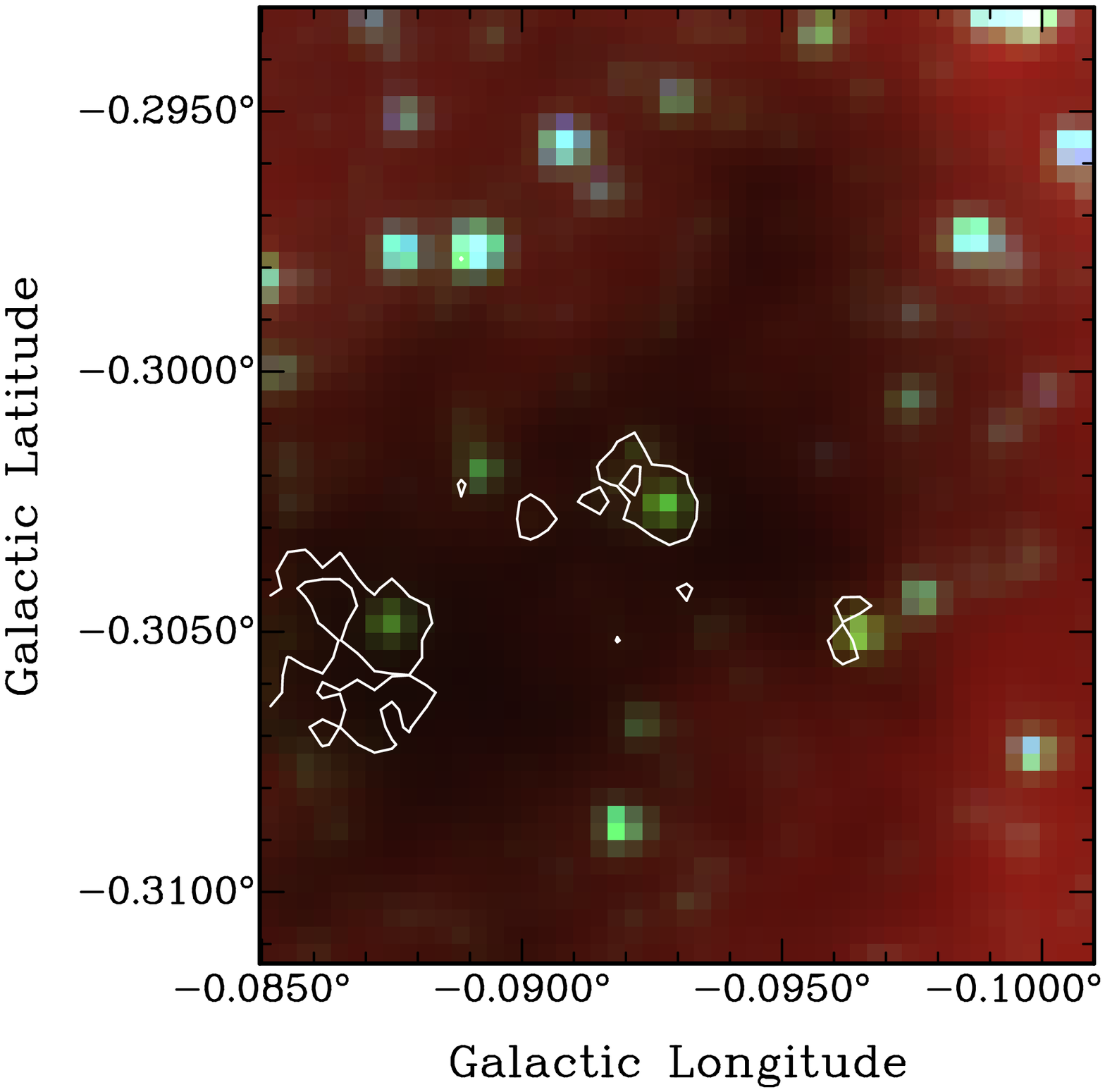}\\
  \caption{IRAC 3-color ({\it left}) and 24~\um\, ({\it right}) images of
    source g24.  The contours in both images are at a green ratio value of
    0.50. \label{g24}}
\end{figure*}

\begin{figure*}[p]
  \centering
  G359.841$-$0.080\\
  \includegraphics[width=0.4\textwidth,clip=true]{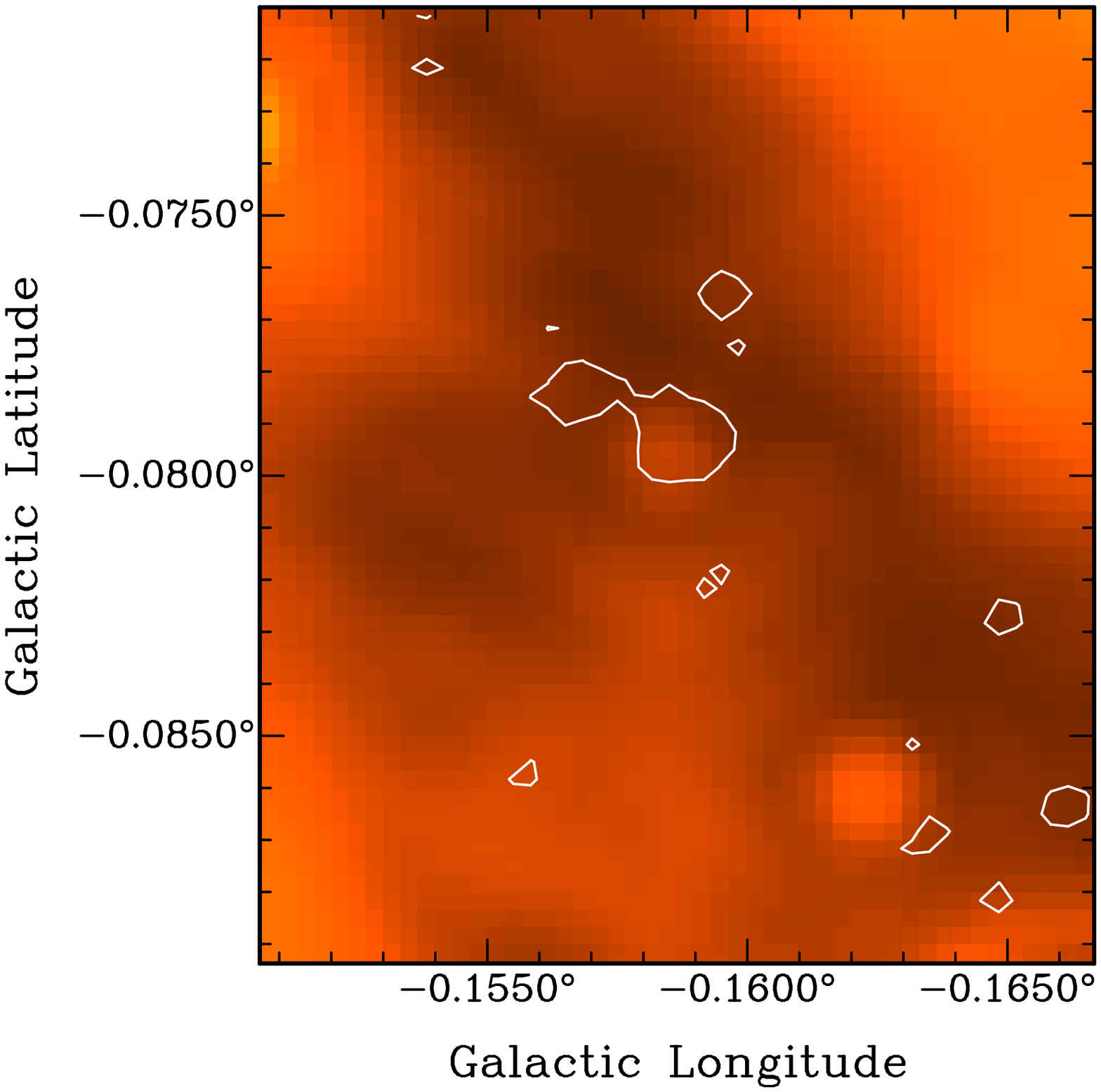}
  \hspace{-11.9cm}
  \includegraphics[width=0.4\textwidth,clip=true]{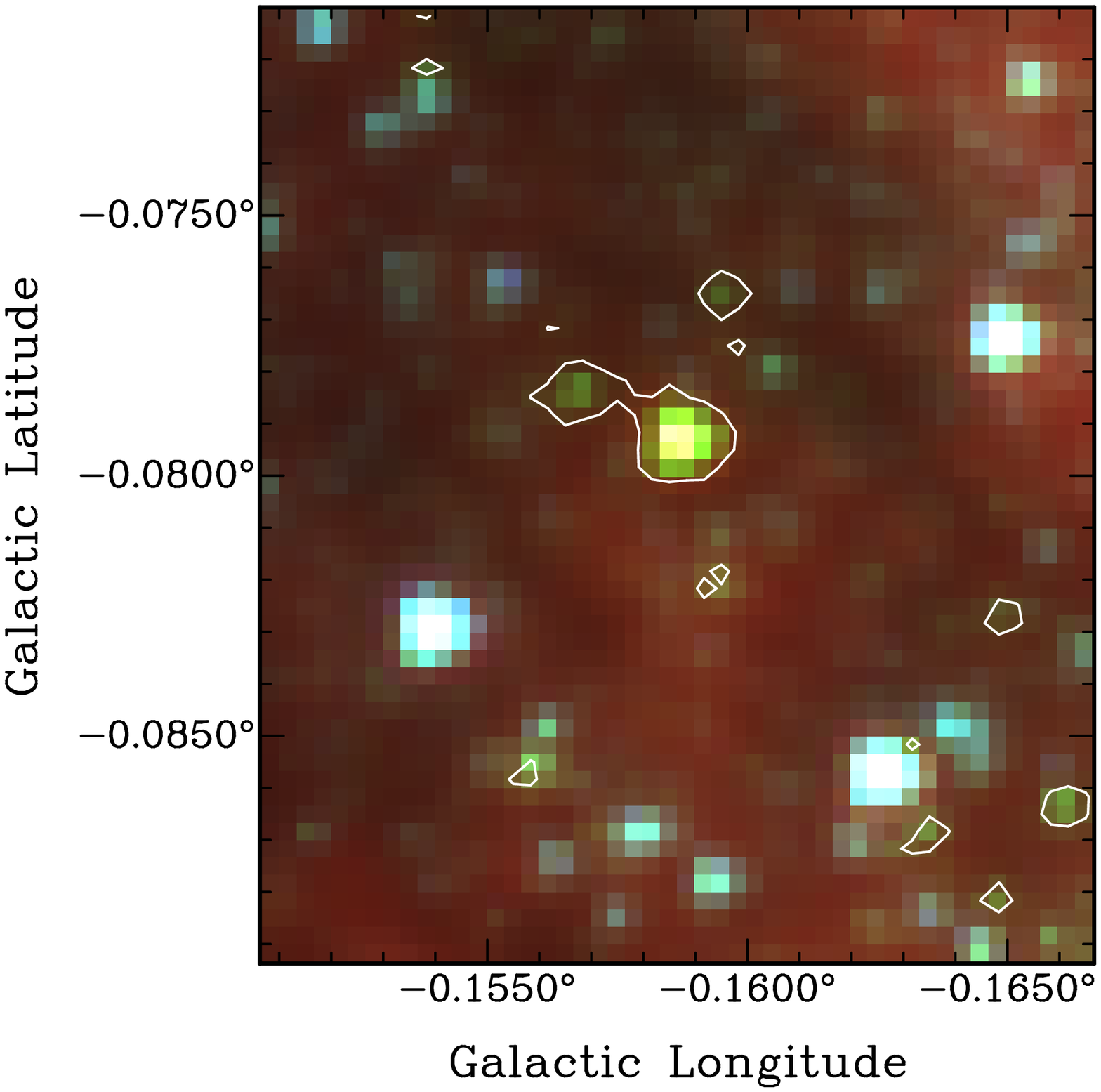}\\
  \caption{IRAC 3-color ({\it left}) and 24~\um\, ({\it right}) images of
    source g25.  The contours in both images are at a green ratio value of
    0.40. \label{g25}}
\end{figure*}

\begin{figure*}[p]
  \centering
  G359.618$-$0.245\\
  \includegraphics[width=0.4\textwidth,clip=true]{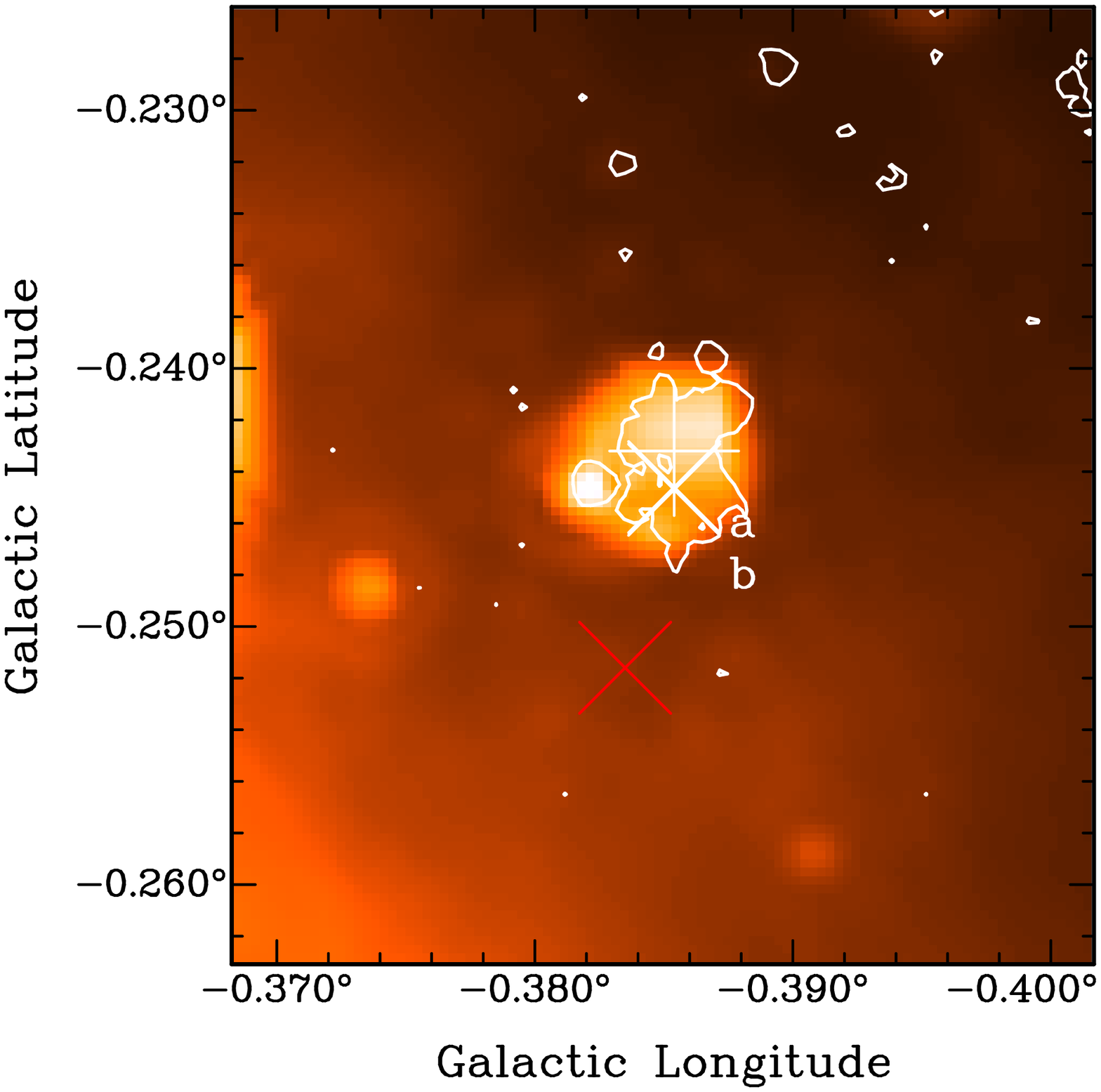}
  \hspace{-11.9cm}
  \includegraphics[width=0.4\textwidth,clip=true]{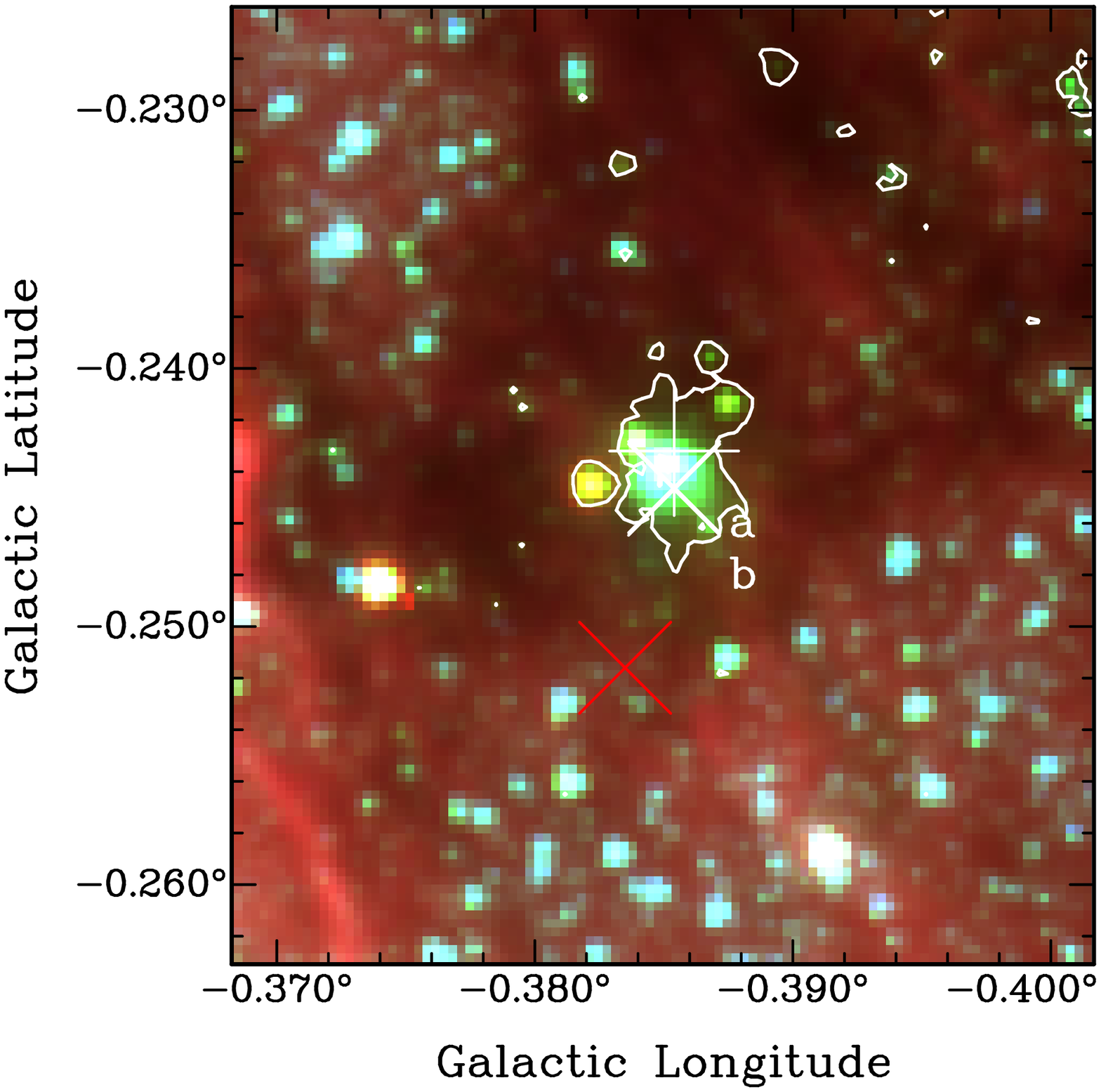}\\
  \includegraphics[angle=-90,width=.4\textwidth,clip=true]{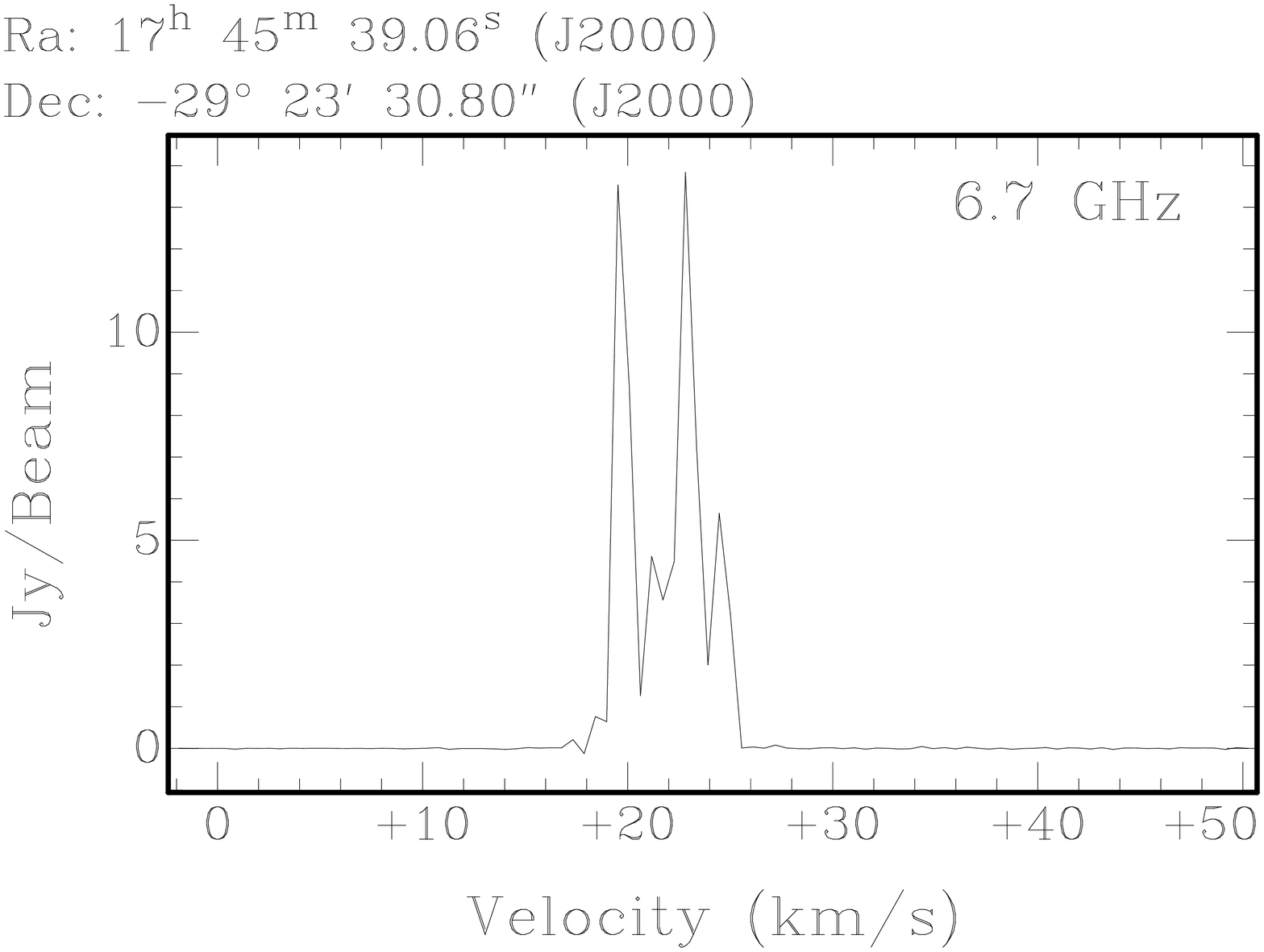}\\
  \includegraphics[angle=-90,width=.4\textwidth,clip=true]{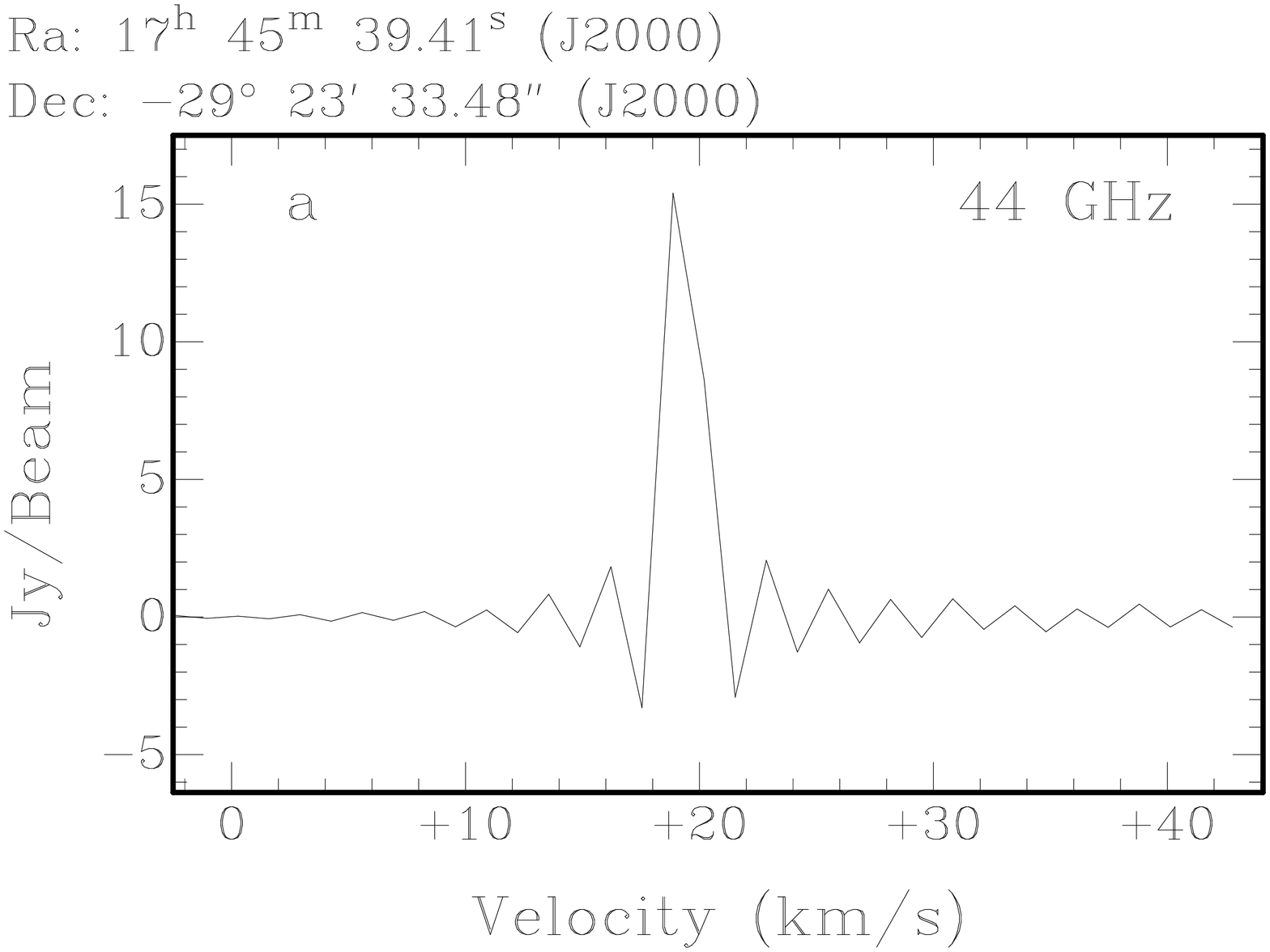}
  \includegraphics[angle=-90,width=.4\textwidth,clip=true]{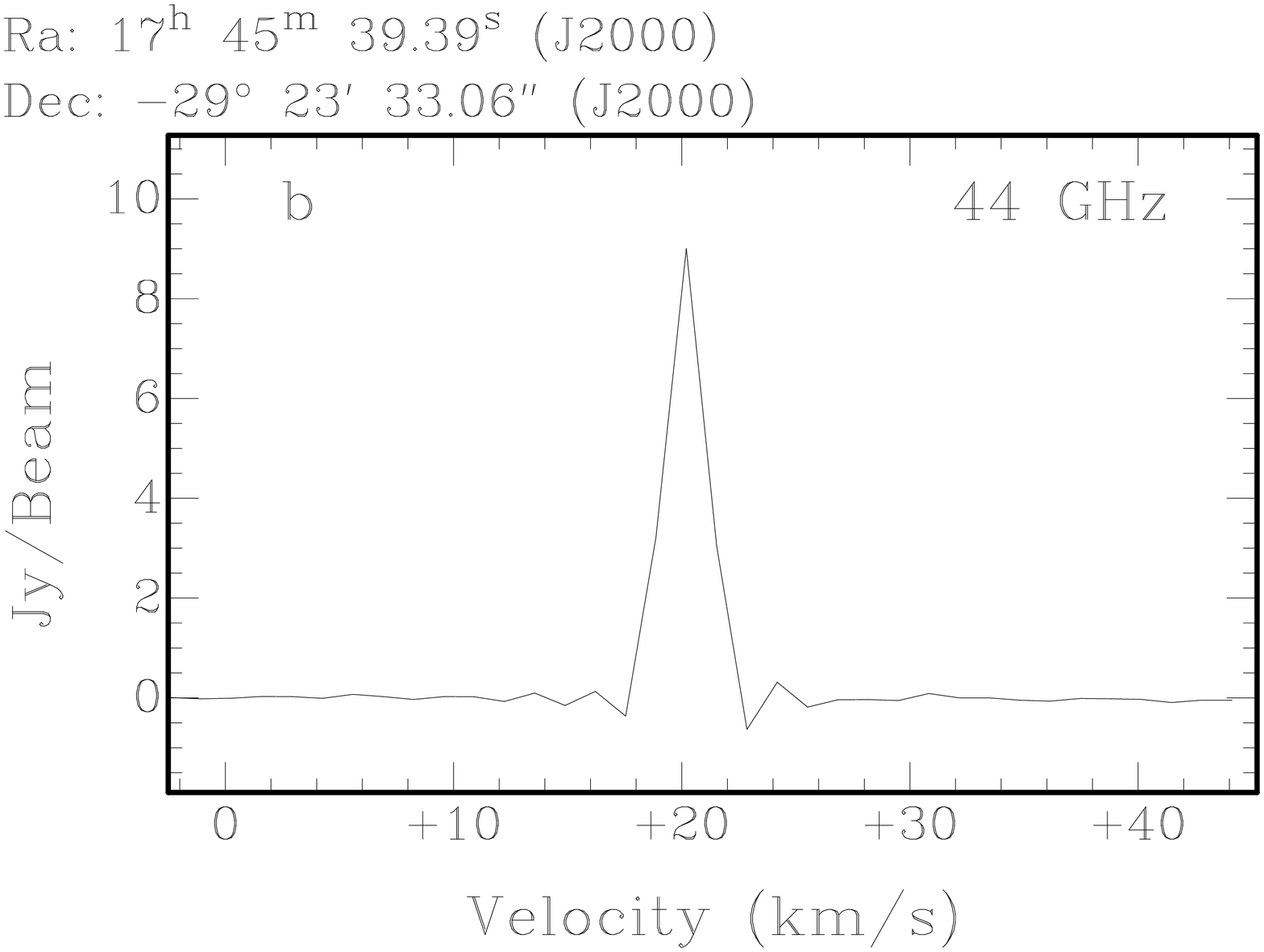}\\
  \caption{{\it Top}: IRAC 3-color ({\it left}) and 24~\um\, ({\it right})
    images of source g26.  The contours in both images are at a green ratio
    value of 0.50.  The white plus sign ($+$) designates the position of
    6.7~GHz maser emission detected with the EVLA, the white cross signs
    ($\times$) designate the position of 44~GHz masers detected with the EVLA,
    and the red cross sign disignates the position of 44~GHz maser emission
    from Y-Z09.  {\it Bottom}: Spectra of 6.7 and 44~GHz maser emission in the
    g26 field obtained with the EVLA. \label{g26}}
\end{figure*}

\begin{figure*}[p]
  \centering
  G359.599$-$0.032\\
  \includegraphics[width=0.4\textwidth,clip=true]{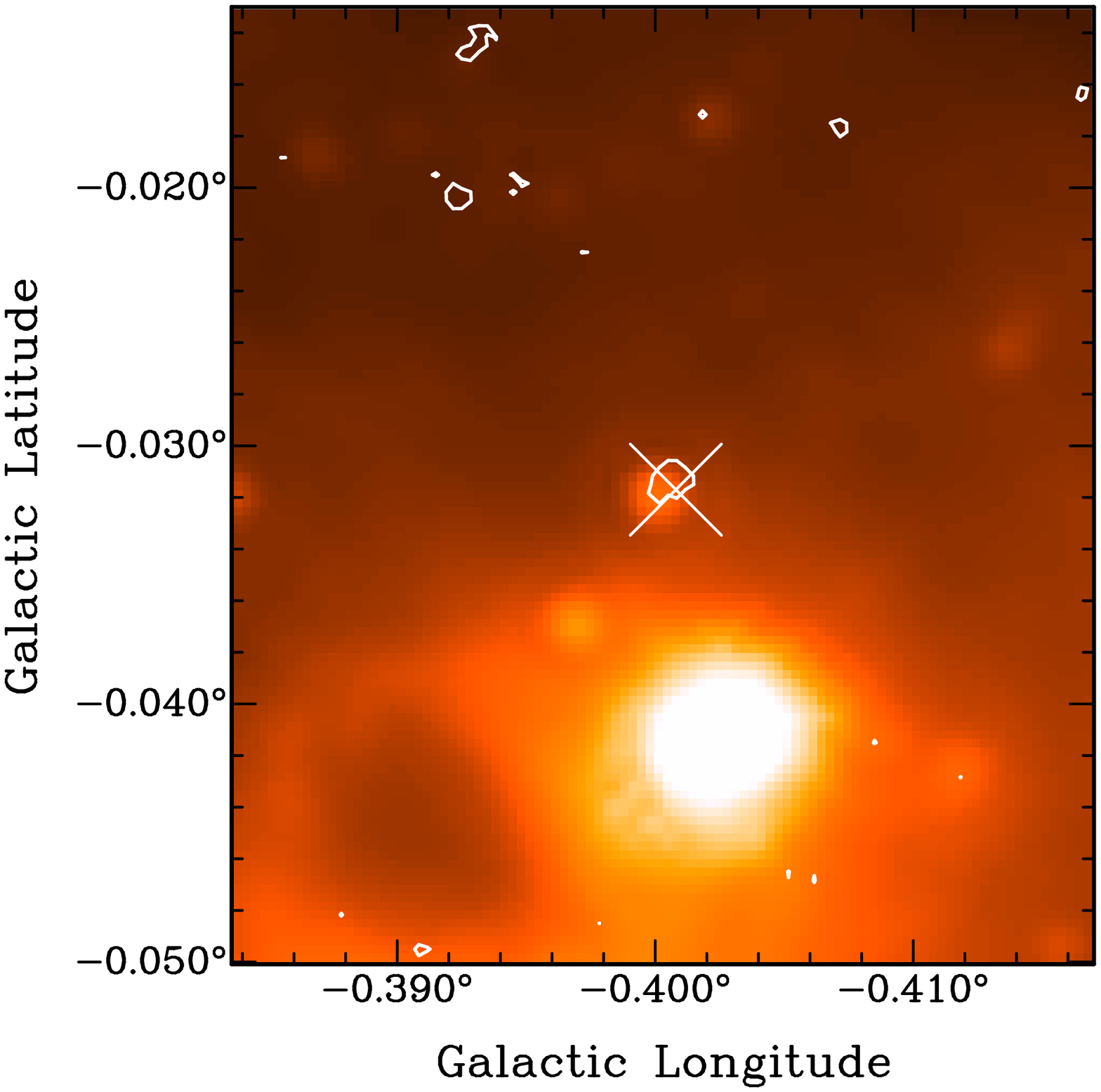}
  \hspace{-11.9cm}
  \includegraphics[width=0.4\textwidth,clip=true]{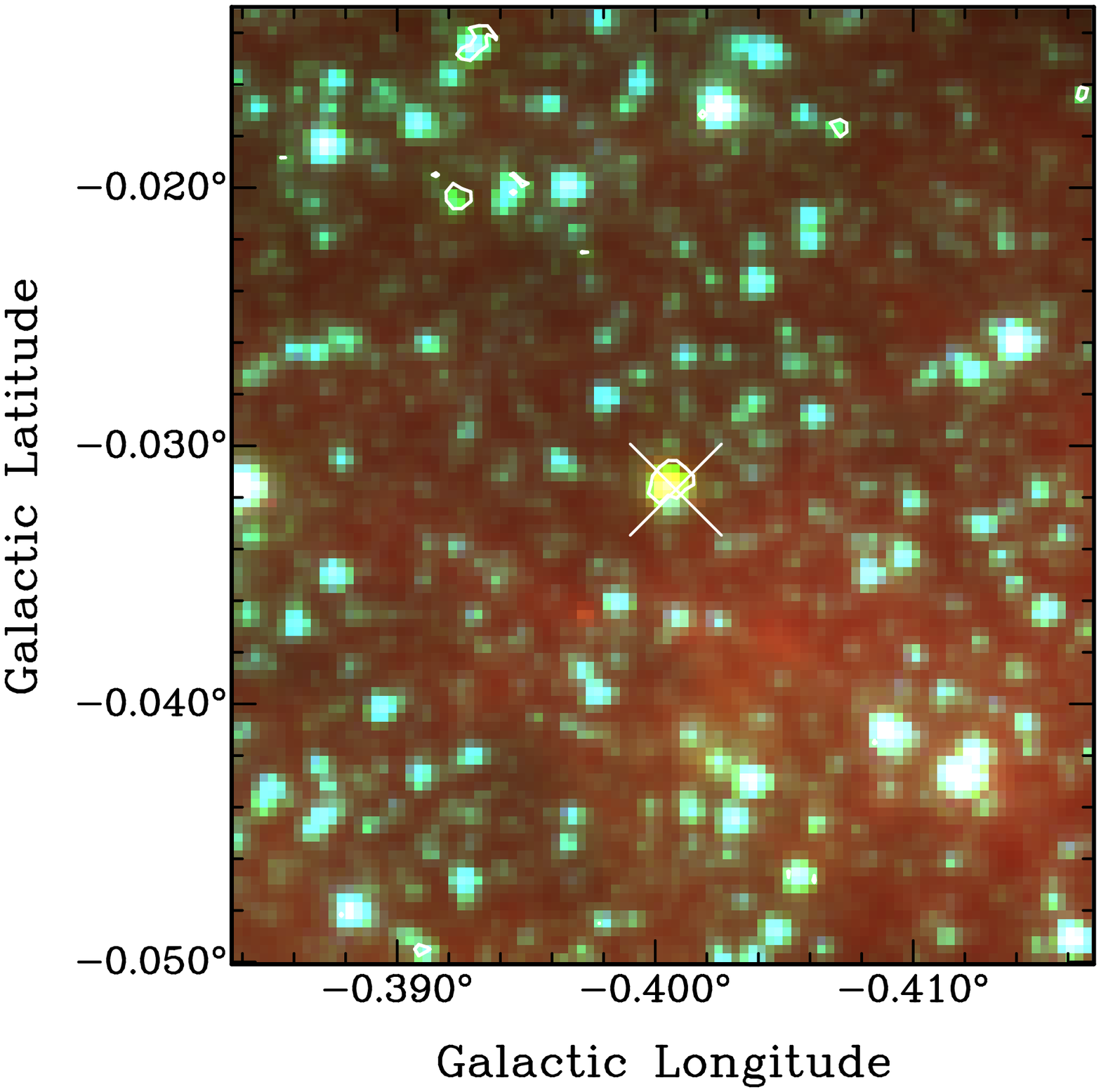}\\
  \includegraphics[angle=-90,width=.4\textwidth,clip=true]{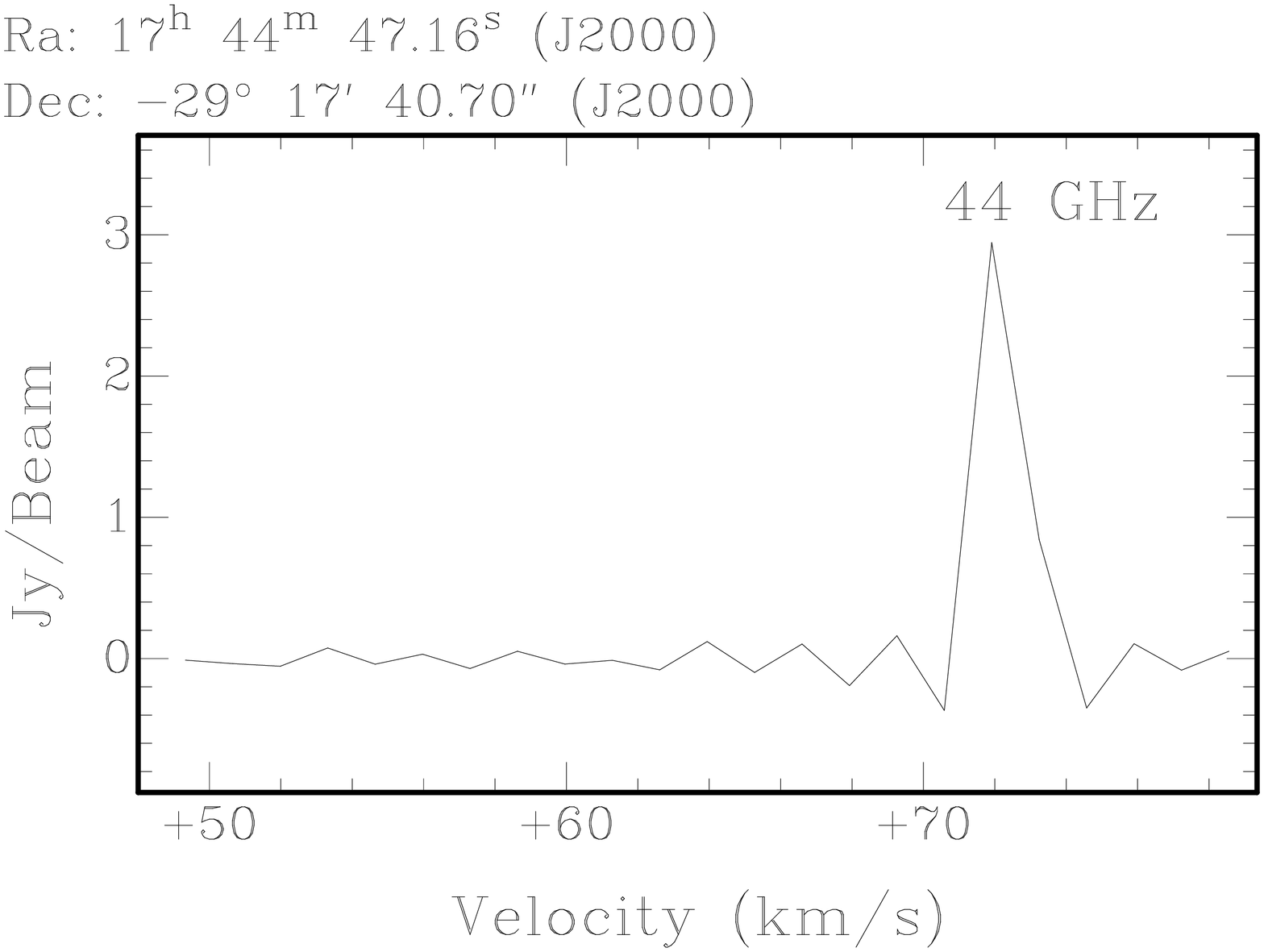}\\
  \caption{{\it Top}: IRAC 3-color ({\it left}) and 24~\um\, ({\it right})
    images of source g27.  The contours in both images are at a green ratio
    value of 0.35.  The white cross sign ($\times$) designates the position of
    44~GHz maser emission detected with the EVLA.  {\it Bottom}: Spectrum of
    44~GHz maser emission in the g27 field obtained with the EVLA.\label{g27}}
\end{figure*}

\begin{figure*}[p]
  \centering
  G359.57$-$0.270\\
  \includegraphics[width=0.4\textwidth,clip=true]{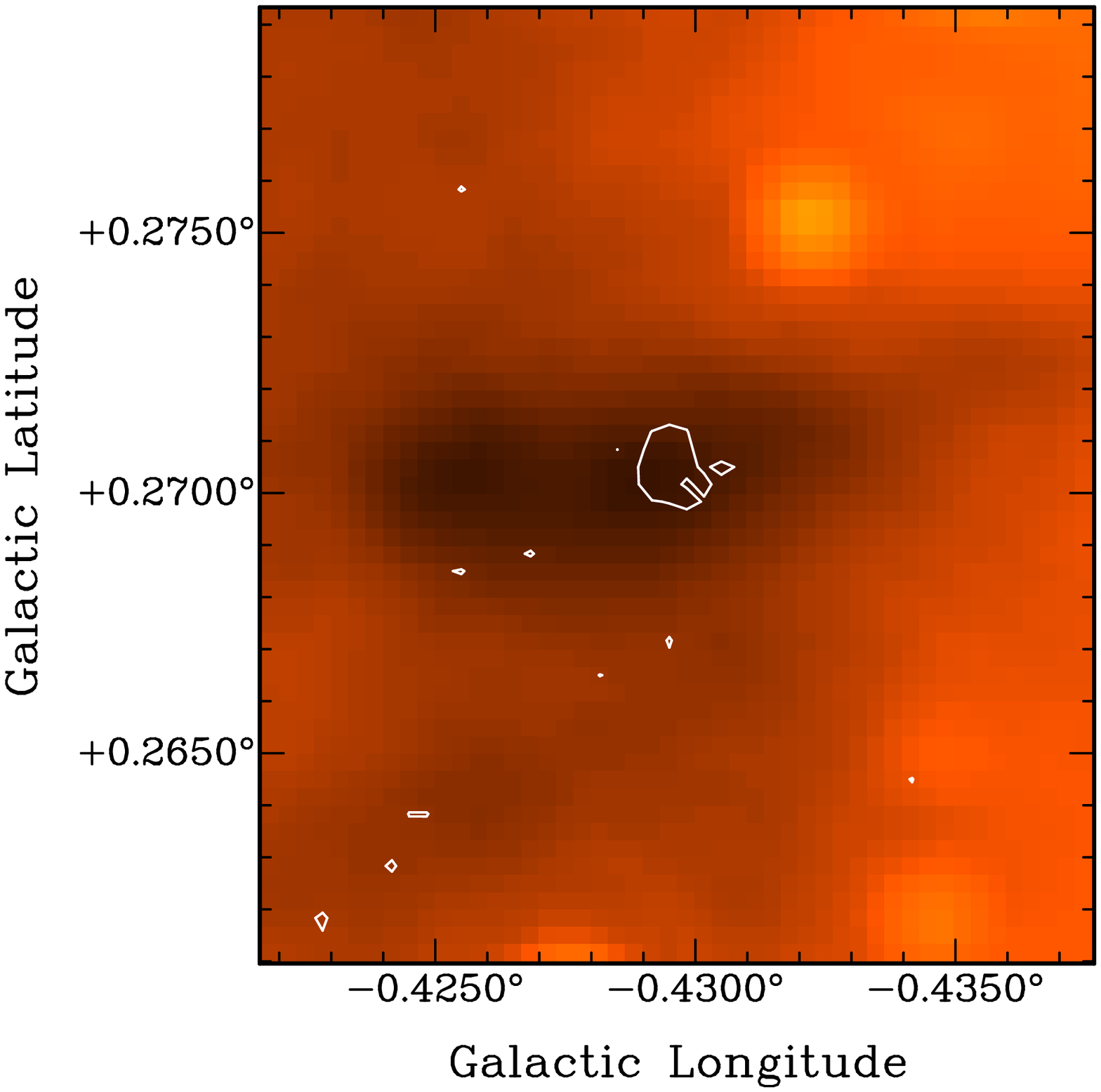}
  \hspace{-11.9cm}
  \includegraphics[width=0.4\textwidth,clip=true]{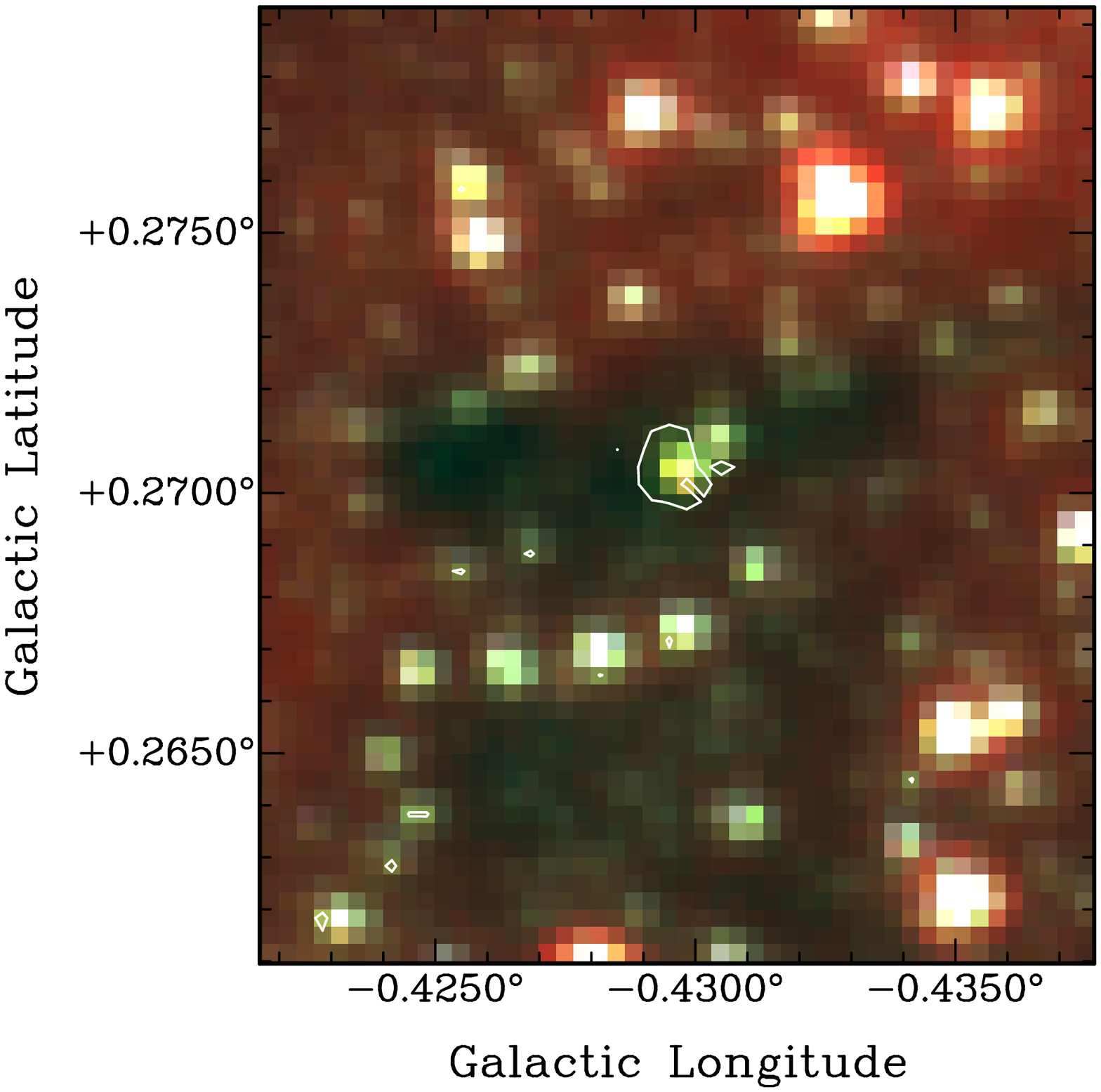}\\
  \caption{IRAC 3-color ({\it left}) and 24~\um\, ({\it right})
    images of source g28.  The contours in both images are at a green ratio
    value of 0.45.\label{g28}}
\end{figure*}

\begin{figure*}[p]
  \centering
  G359.437$-$0.102\\
  \includegraphics[width=0.4\textwidth,clip=true]{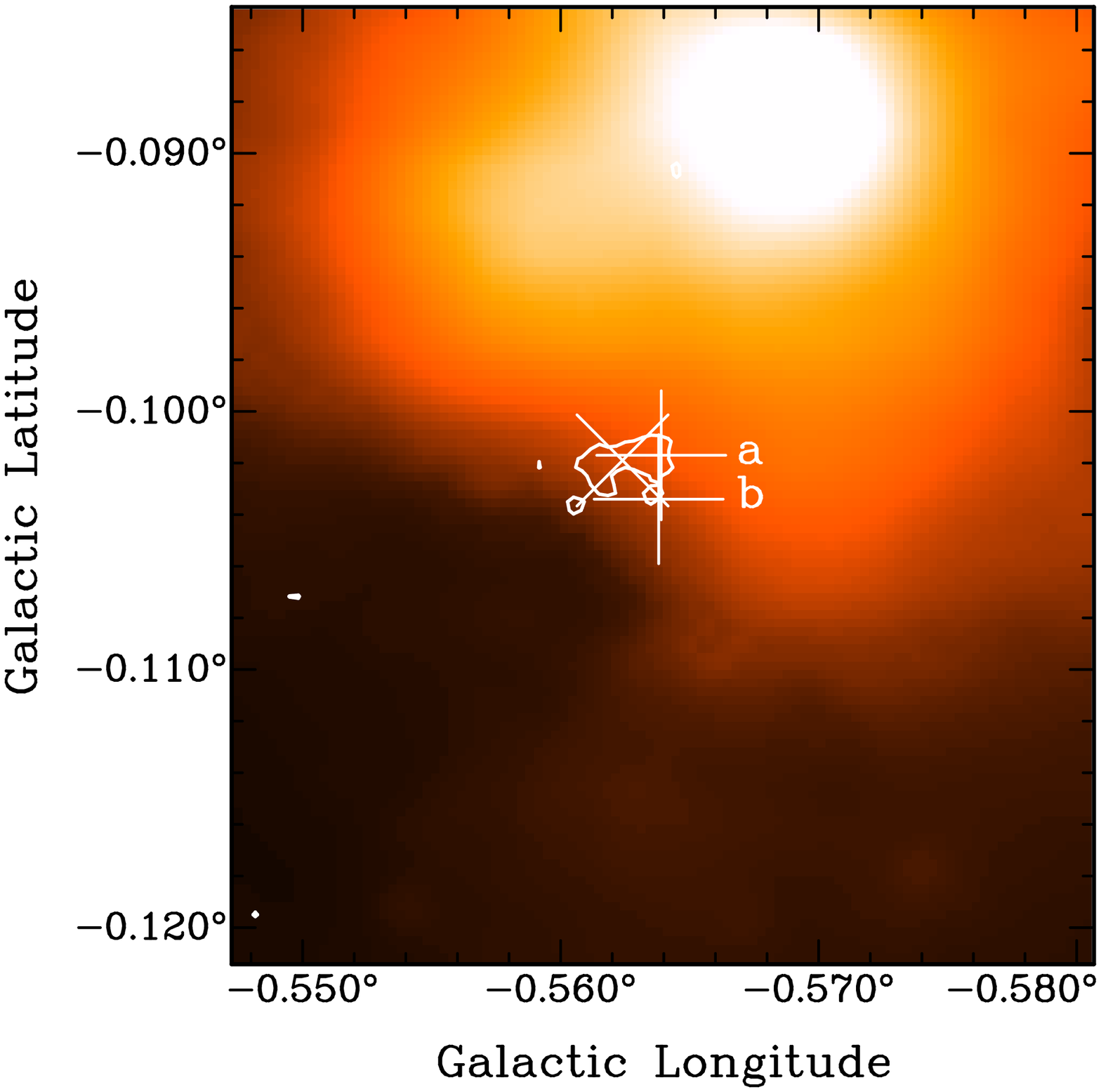}
  \hspace{-11.9cm}
  \includegraphics[width=0.4\textwidth,clip=true]{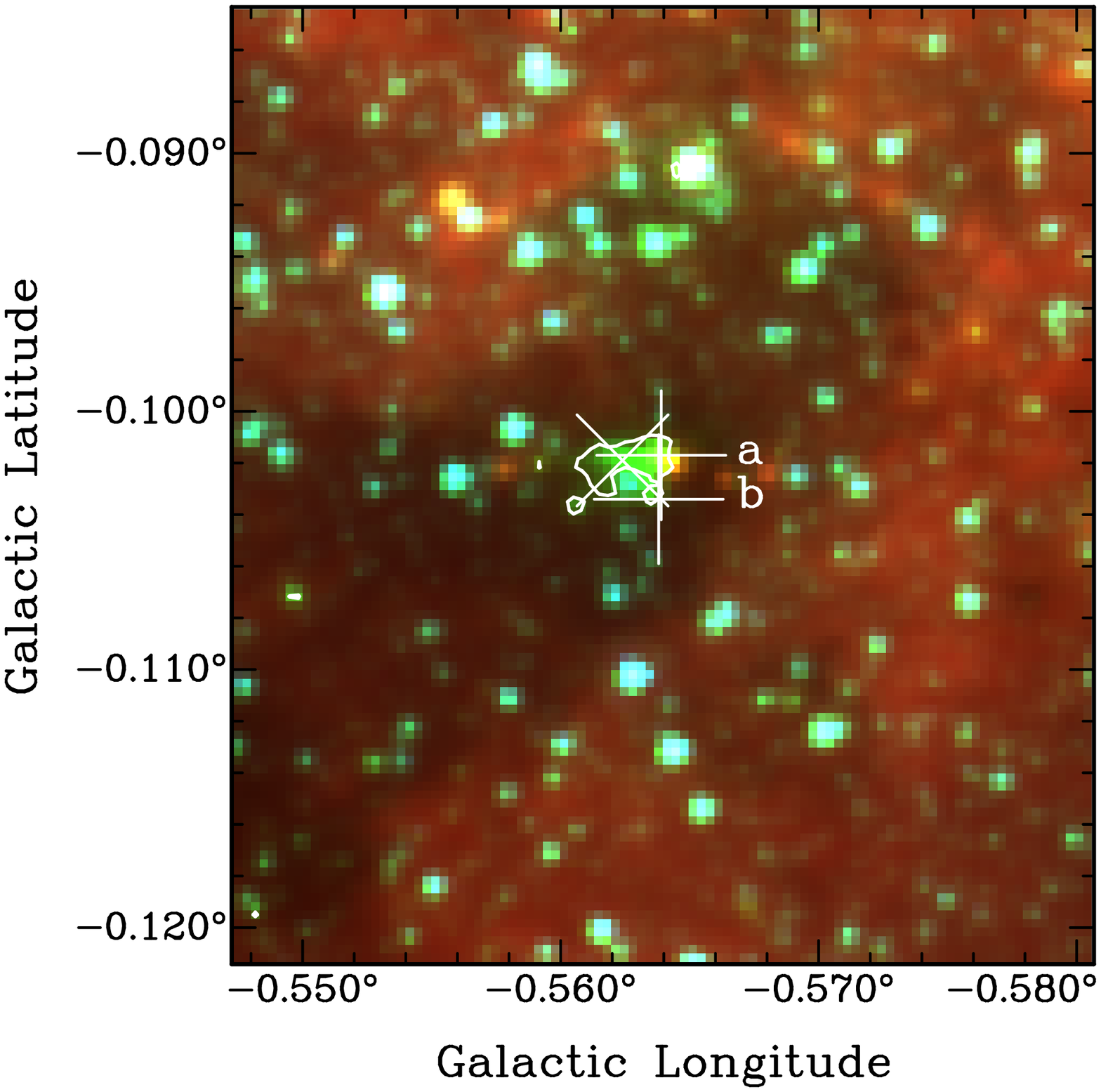}\\
  \includegraphics[angle=-90,width=.4\textwidth,clip=true]{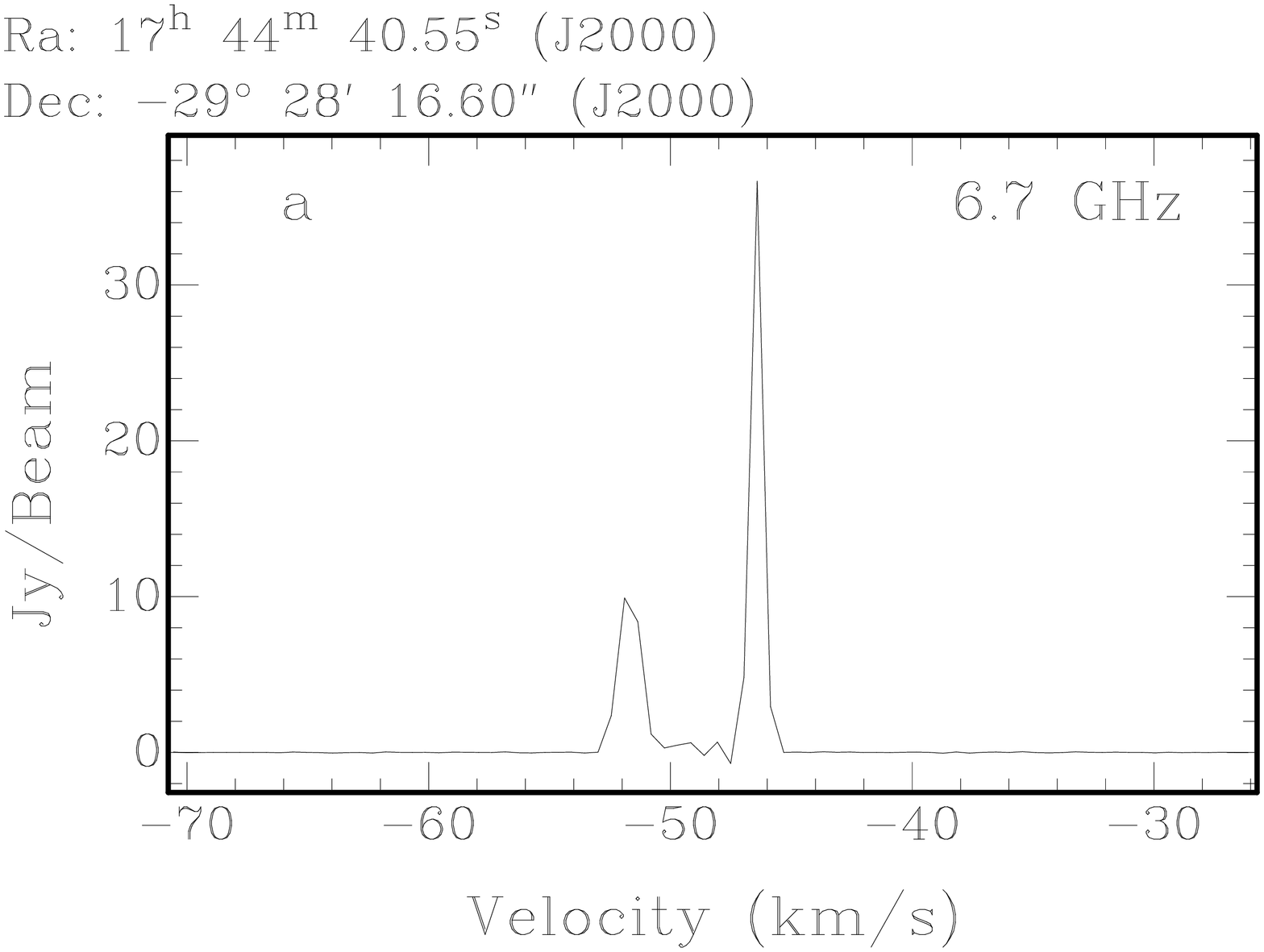}
  \includegraphics[angle=-90,width=.4\textwidth,clip=true]{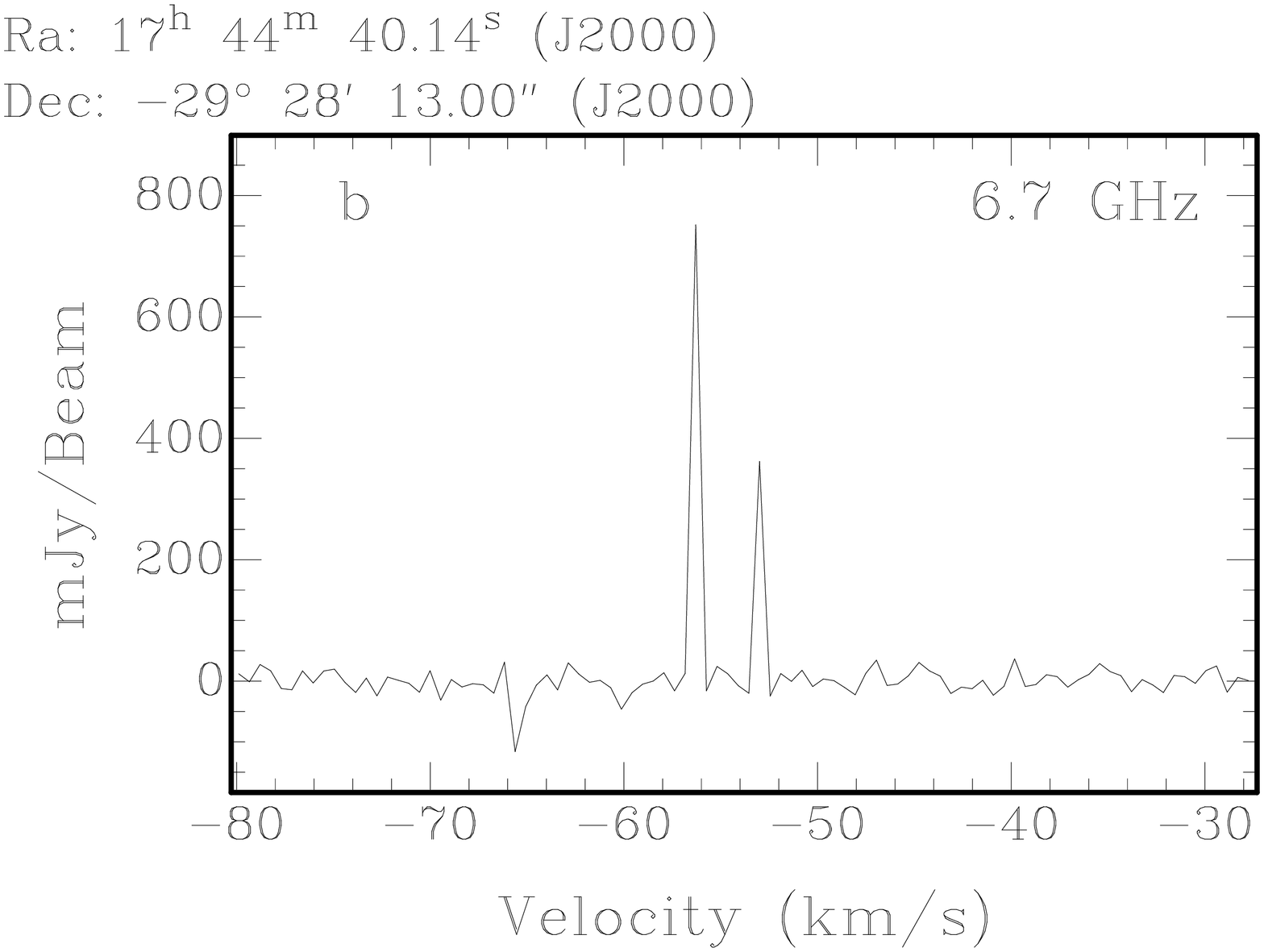}\\
  \includegraphics[angle=-90,width=.4\textwidth,clip=true]{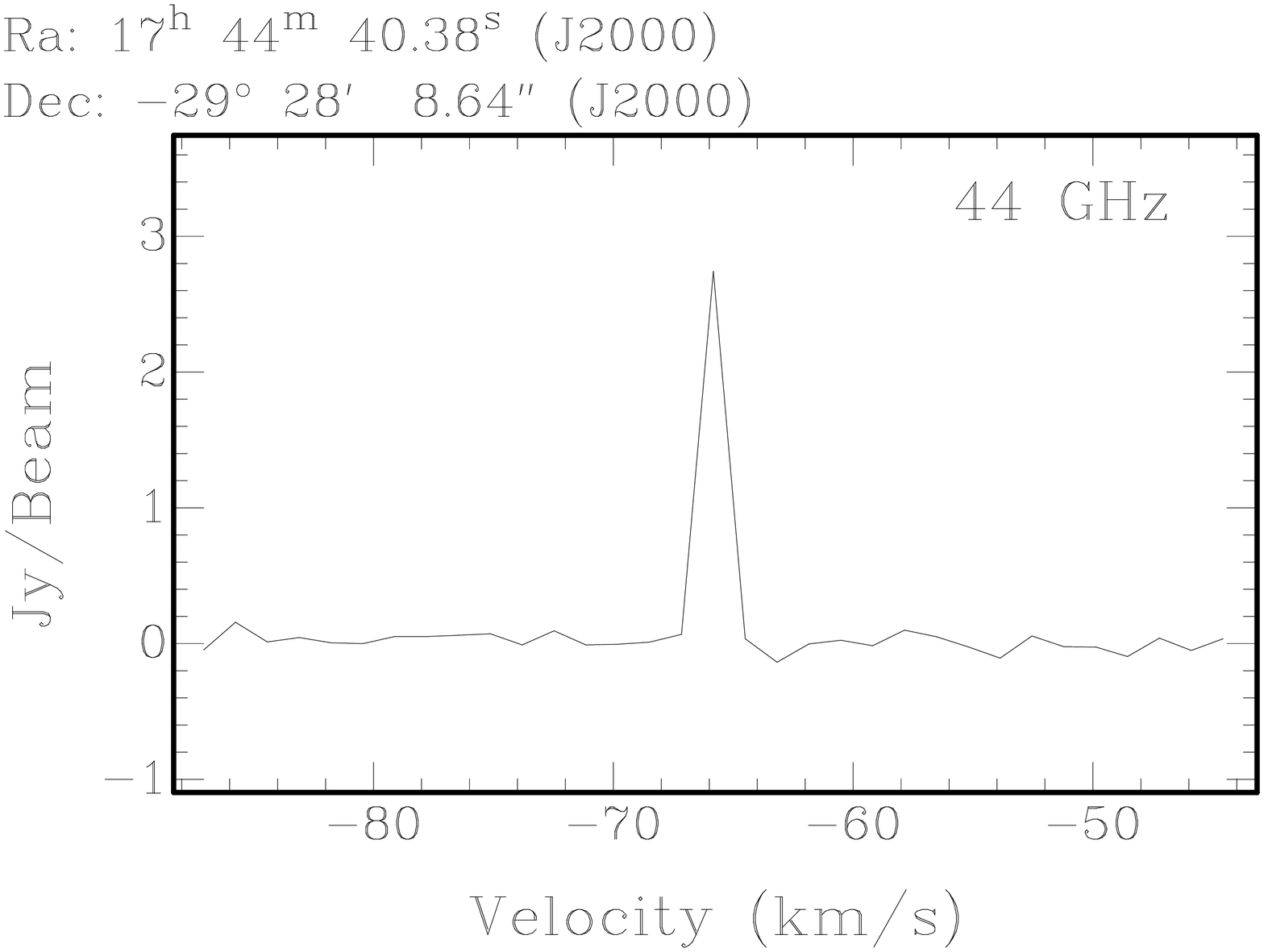}\\
  \caption{{\it Top}: IRAC 3-color ({\it left}) and 24~\um\, ({\it right})
    images of source g29.  The contours in both images are at a green ratio
    value of 0.60.  The white plus signs ($+$) designate the position of
    6.7~GHz masers detected with the EVLA, and the white cross sign ($\times$)
    designates the position of 44~GHz maser emission detected with the EVLA.
    {\it Bottom}: Spectra of 6.7 and 44~GHz maser emission in the g29 field
    obtained with the EVLA.\label{g29}}
\end{figure*}

\begin{figure*}[p]
  \centering
  G359.30$+$0.033\\
  \includegraphics[width=0.4\textwidth,clip=true]{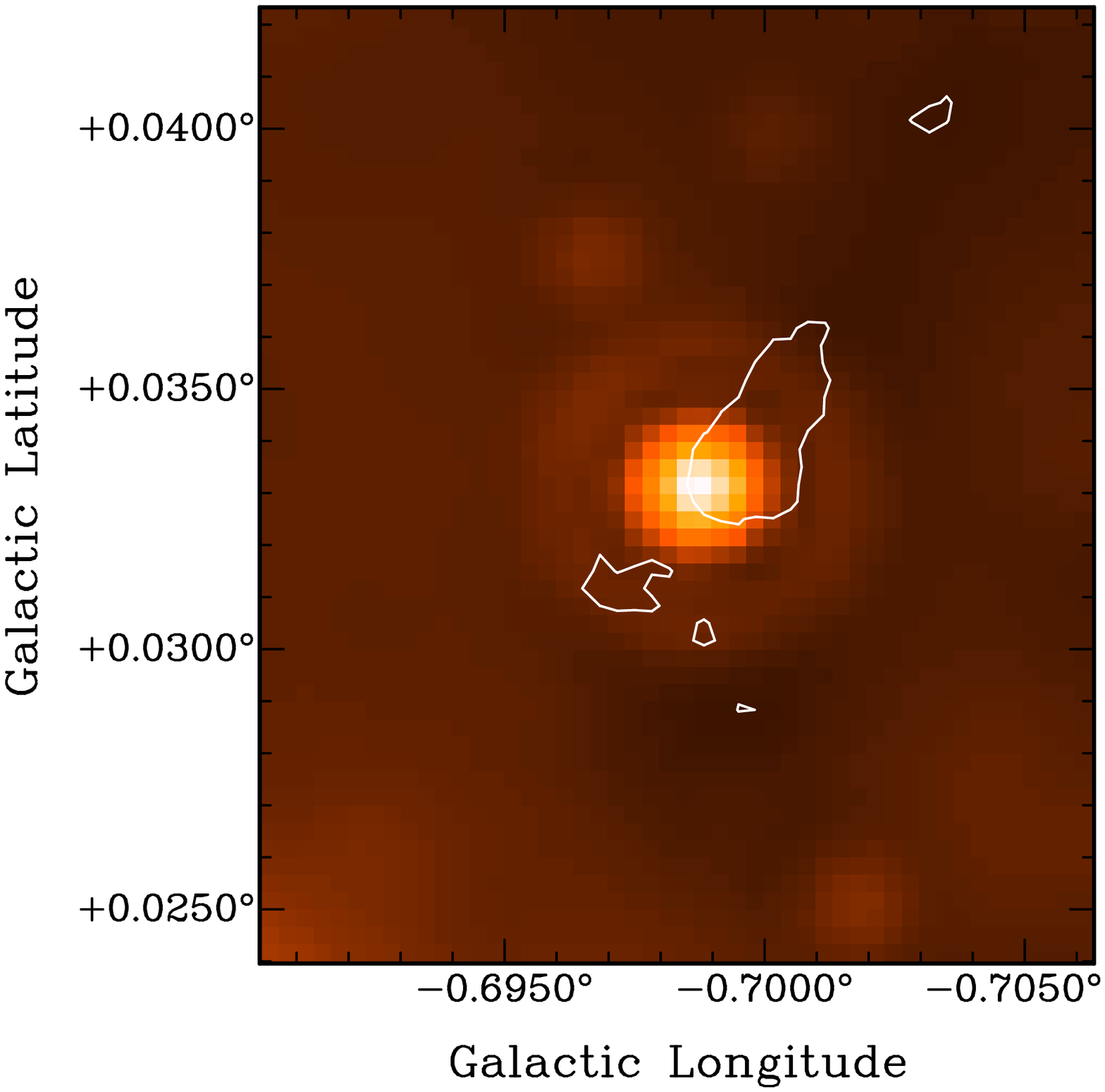}
  \hspace{-11.9cm}
  \includegraphics[width=0.4\textwidth,clip=true]{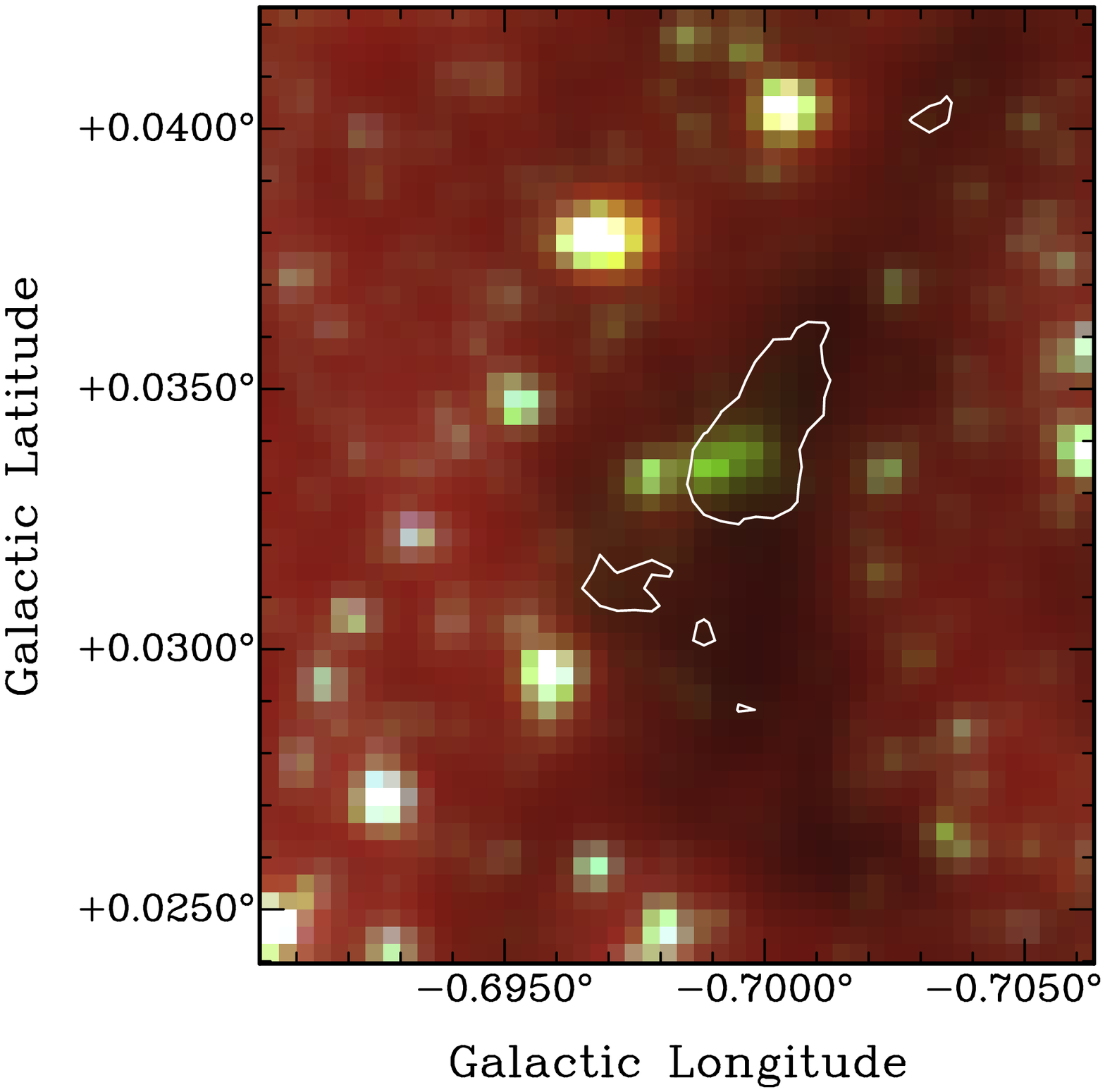}\\
  \caption{IRAC 3-color ({\it left}) and 24~\um\, ({\it right}) images of
    source g30.  The contours in both images are at a green ratio value of
    0.65. \label{g30}}
\end{figure*}

\begin{figure*}[p]
  \centering
  G359.199$+$0.041\\
  \includegraphics[width=0.4\textwidth,clip=true]{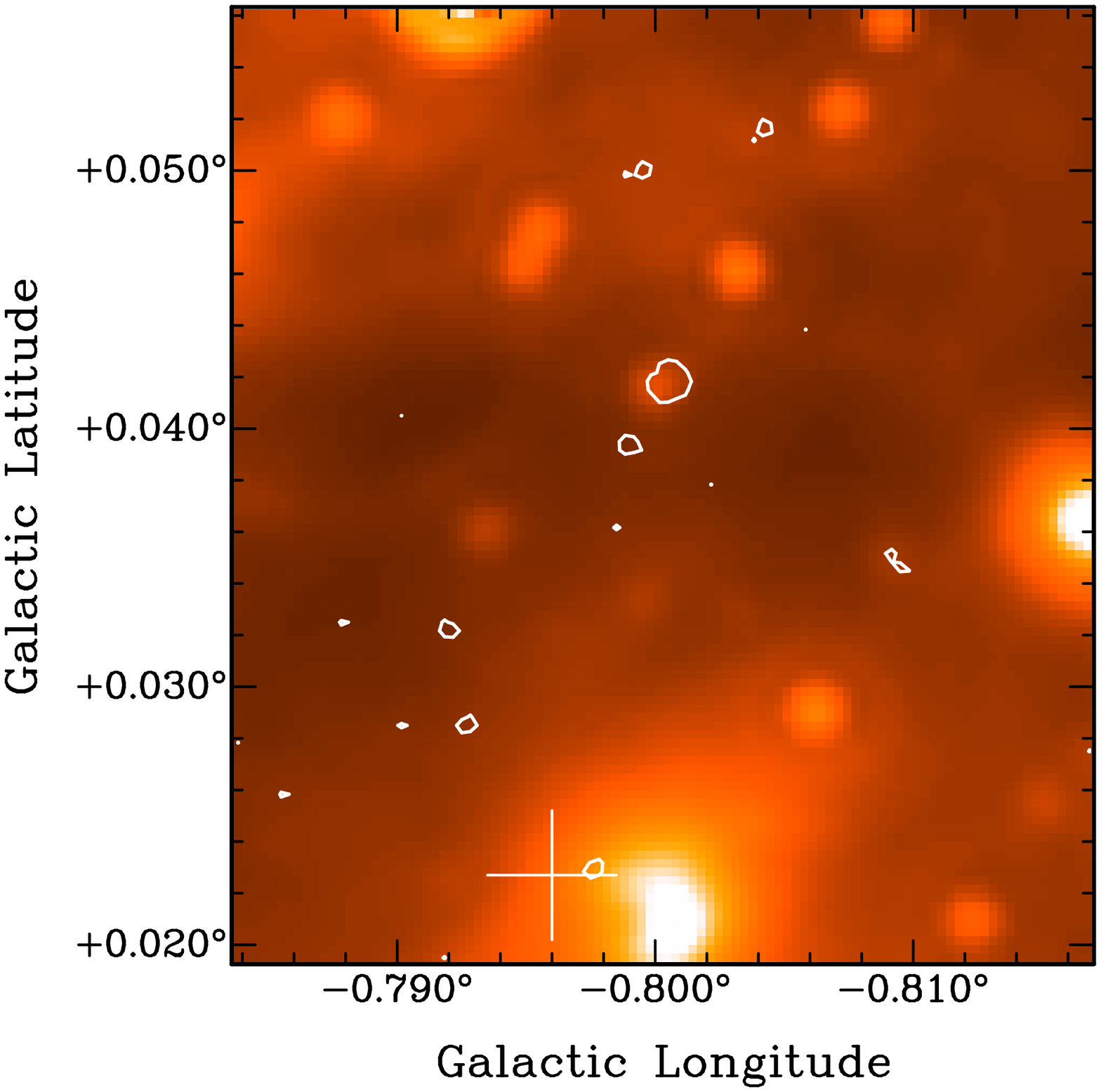}
  \hspace{-11.9cm}
  \includegraphics[width=0.4\textwidth,clip=true]{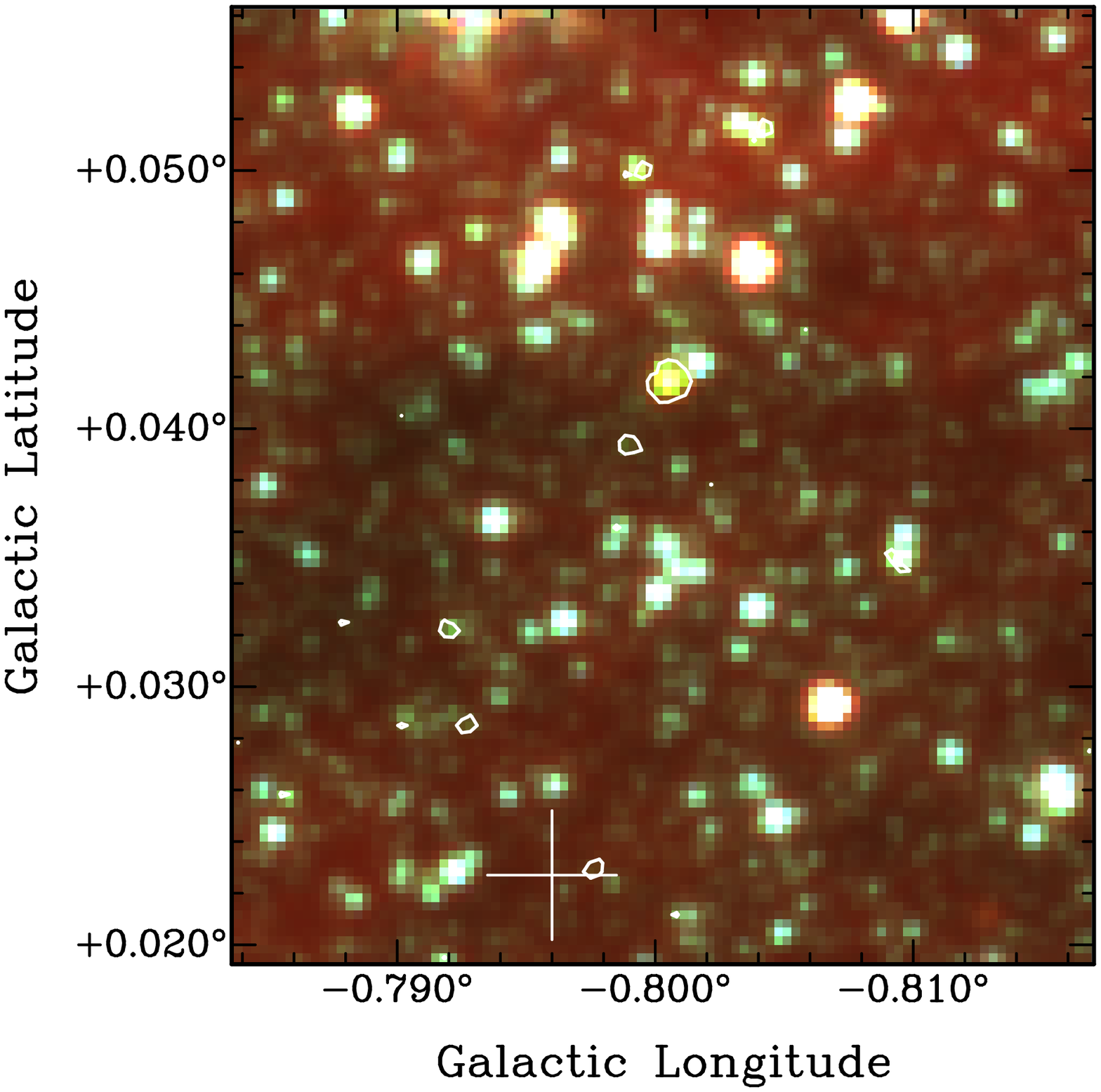}\\
\includegraphics[angle=-90,width=.4\textwidth,clip=true]{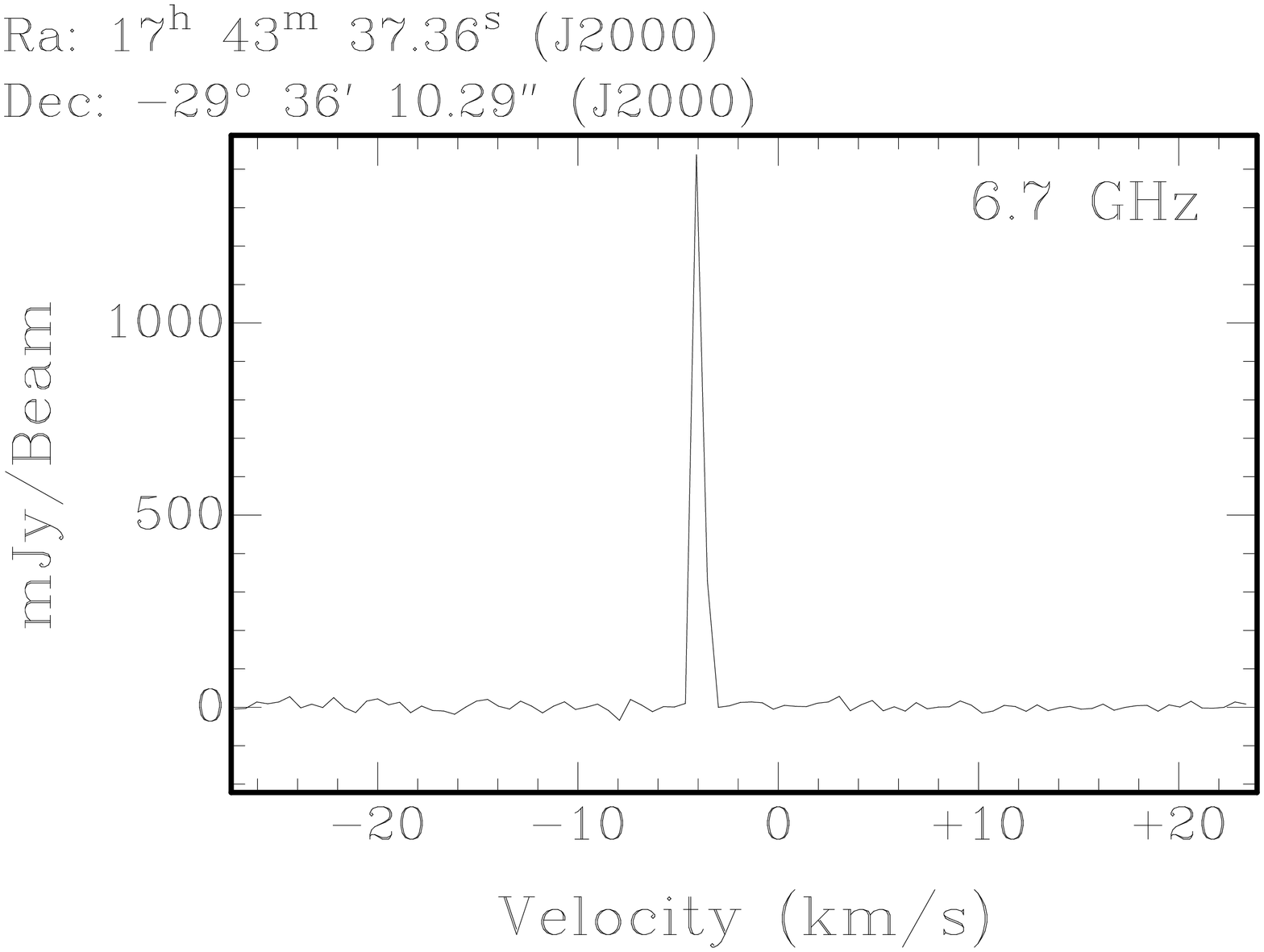}\\
  \caption{{\it Top}: IRAC 3-color ({\it left}) and 24~\um\, ({\it right})
    images of source g31.  The contours in both images are at a green ratio
    value of 0.40.  The white plus sign ($+$) designates the position of
    6.7~GHz maser emission detected with the EVLA. Because the location of the
    maser emission is $>$10\arcsec\, from the green source, we do not classify
    it as an association.  {\it Bottom}: Spectrum of 6.7~GHz maser emission in
    the g31 field obtained with the EVLA. \label{g31}}
\end{figure*}

\begin{figure*}[p]
  \centering
  G358.980$+$0.084\\
  \includegraphics[width=0.4\textwidth,clip=true]{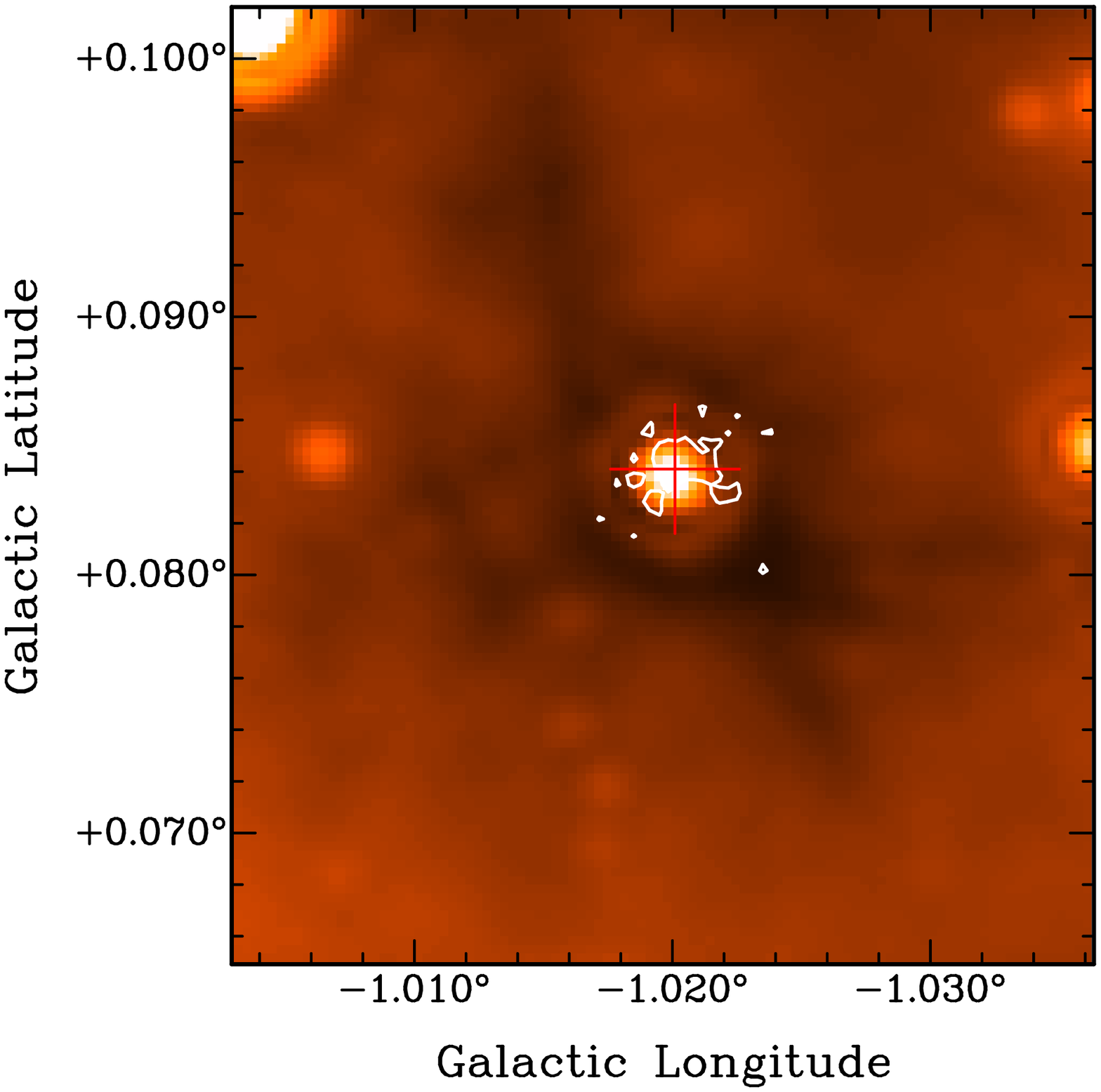}
  \hspace{-11.9cm}
  \includegraphics[width=0.4\textwidth,clip=true]{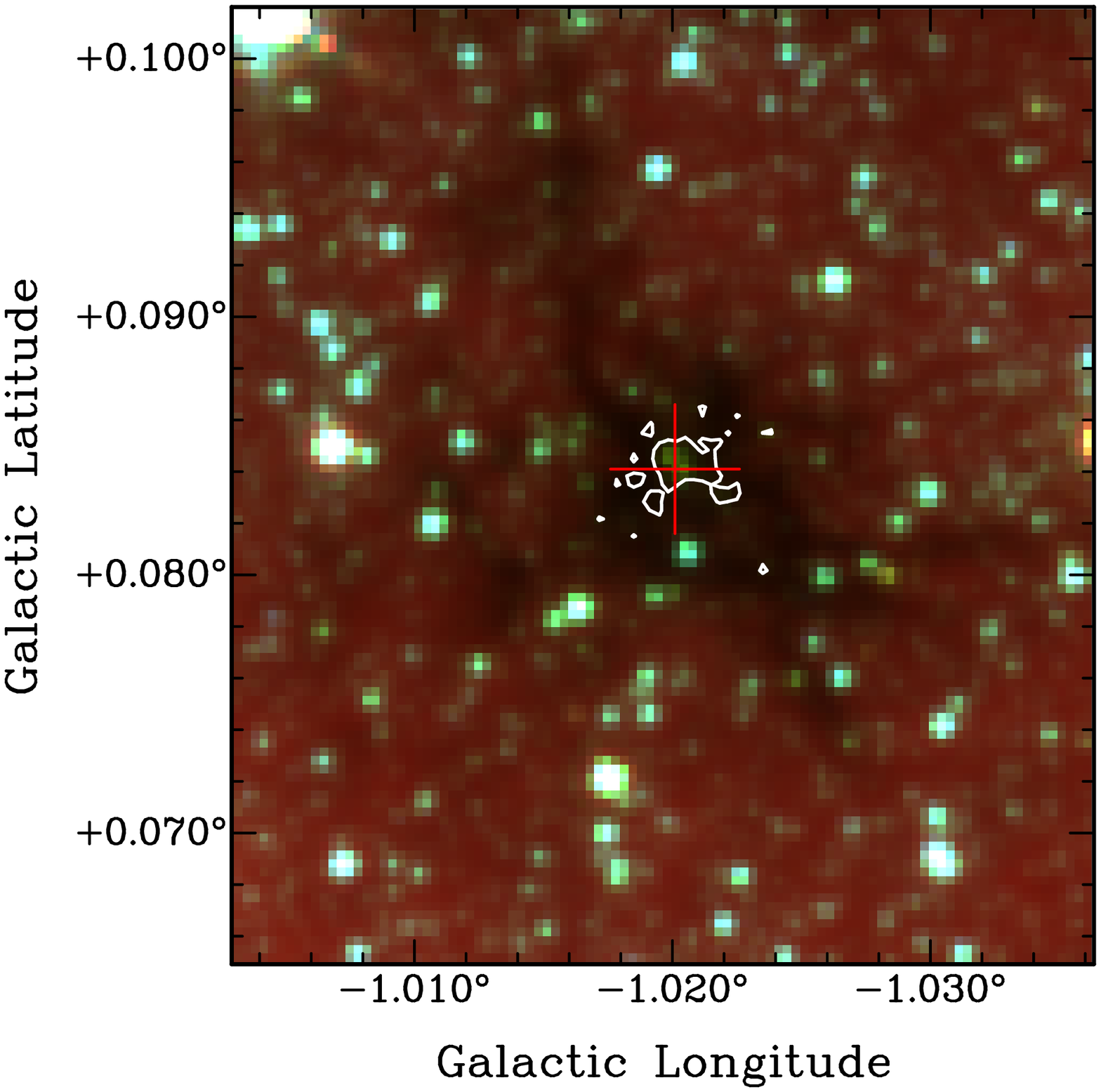}\\
  \caption{IRAC 3-color ({\it left}) and 24~\um\, ({\it right})
    images of source g32.  The contours in both images are at a green ratio
    value of 0.90.  The white red sign ($+$) designates the position of
    6.7~GHz maser emission detected by C10. \label{g32}}
\end{figure*}

Here we describe each of the green sources in our sample.  {\it Spitzer}/IRAC
3-color images of each source, using data obtained as part of GLIMPSE
\citep{benj03} and another IRAC survey of the Galactic center region
\citep{stol06, aren08, rami08}, are contained in Figures~\ref{g0}-\ref{g32}.
In addition, Figures~\ref{g0}-\ref{g32} also contain {\it Spitzer}/MIPS
24~\um\, images of each source using data obtained by Y-Z09.  For the fields
toward which we detect \meth\, maser emission, the maser positions are
overlaid on the images, and the spectra are included in the figures.  The
masses of the sources given in the following sections are from Y-Z09, and are
calculated as described in Section~\ref{masses}. In general, the 6.7~GHz
masers that are associated with our green sources are located on the enhanced
4.5~\um\, emission that define the sources.  The average separation between
the center of the green sources and their associated 6.7~GHz masers is
2.3$''$. Only 2 of the 13 associated 6.7~GHz masers are $>$~5$''$ from the
center of the green sources, supporting the idea that these masers are formed
close to the central protostellar object.  44~GHz masers are found at an
average angular separation of 10.6$''$ from the center of their associated
green sources.  The 44~GHz masers masers are farther away from the center of
the green sources than the 6.7~GHz masers, near the edges of the enhanced
4.5~\um\, emission and consistent with their creation in outflows.

In 6 fields (g6-g10, g31), we detect 6.7~GHz masers in our observations but
do not associate them green sources.  We include the positions of these masers
in their appropriate figures to show where they reside in relation to the
green sources in our sample.  In general, the masers that are not associated
with the green sources are not associated with any strong IRAC or MIPS
emission.  These masers may be associated with star formation along the line
of sight to the green sources, but at an evolutionary state during which
no significant IR emission is detectable.  Alternatively, they may reside on
the far side of the Galaxy, making the detection of infrared emission toward
them difficult.


\subsubsection{g0 (G1.041$-$0.072)}
Green excess source g0 consists of two knots of enhanced 4.5~\um\, emission
(Fig.~\ref{g0}), and is located to the west of Sgr~D. The mass of g0
(16.1$\pm$3.5~\Msun) indicates that it is a site of high-mass star formation.
Source g0 is located within an IRDC that is at the eastern edge of the
prominent dust ridge seen toward the Galactic center \citep{lis94,lis98,
  lis01}.  Because it is within 10$'$ of the Galactic plane, we assume that
this source is at the distance of the Galactic center.  We do not detect any
\meth\, maser emission toward this source.

\subsubsection{g1 (G0.955$-$0.786), g2 (G0.868$-$0.697), g4 (G0.955$-$0.786)}
Sources g1, g2, and g4 (Figs.~\ref{g1}, \ref{g2}, and \ref{g4}) are all
coincident with the positions of known planetary nebulae
\citep{vand01,jaco04}, are in close proximity to one another, and are isolated
from other star formation regions.  All three sources are associated with
24~\um\, emission, but not show no correlation with other observational
signatures of star formation, such as IRDCs or \meth\, maser emission.
Because they are likely to be a planetary nebulae rather than protostars, we
exclude them from our statistical analysis.

\subsubsection{g3 (G0.826$-$0.211)}
Similar to g0, green source g3 is embedded within an IRDC that may be part of
the prominent dust ridge at positive Galactic longitude near the Galactic
center.  As seen in Figure~\ref{g3}, g3 appears more orange than green in the
IRAC 3-color image, but, as the contours on the image show, it does have an
enhancement at 4.5~\um.  Its Galactic latitude places it at the distance of
the Galactic center, and its possible association with the dust ridge and its
proximity to Sgr~B2 support this claim.  Source g3 has a mass of
10.6$\pm$0.0~\Msun, and is not associated with any \meth\, maser emission.

\subsubsection{g5 (G0.708+0.408)}
Green source g5 has the highest Galactic latitude of all the sources in our
sample, making it likely to be foreground to the Galactic center.  It is
embedded within a small IRDC, and displays faint 24~\um\, emission
(Fig.~\ref{g5}).  This source, which displays no \meth\, maser emission,
has a mass of 6.0$\pm$3.0~\Msun.

\subsubsection{g6 (G0.693$-$0.045), g7 (G0.679$-$0.037), g8 (G0.667$-$0.037), g9
  (G0.667$-$0.035), g10 (G0.665$-$0.053)} Green excess sources g6 through g10
are located in close proximity to one another in a large IRDC complex.  This
IRDC is part of the dust ridge, and is adjacent to Sgr~B2, one of the most
massive star-forming regions in the Galaxy.  The green sources are near the
southern edge of the IRDC, just north of the bright Sgr~B2 \hii\, regions,
suggestive that star formation is progressing from south to north in this
cloud (Y-Z09).  We assign the Galactic center distance to each of these green
sources.  Source g6, at 29.9$\pm$8.2 the most massive of our sources, appears
reddish in Figure~\ref{g6}, and is coincident with bright 24~\um\, emission.
We do not associate g6 with any \meth\, maser emission, but we do detect a
6.7~GHz maser emission to its north.  This maser site, which displays multiple
velocity features, is not associated with any obvious IRAC emission, but is
coincident with faint 24~\um\, emission. Sources g7, g8, and g9 are shown in
Figure~\ref{g789}.  Bright green in color in the IRAC image, g7 also displays
bright bright 24~\um.  It has a mass of 9.4$\pm$1.2, and is not associated
with any \meth\, masers.  Sources g8 and g9 are reddish in color, and the the
region immediately surrounding them harbors seven 6.7~GHz masers (within
$\sim$~1\arcmin), along with several sites of 44~GHz maser emission (within
$\sim$~1\arcmin).  The velocities of the 6.7~GHz masers fall in the range of
$\sim$~48-73~\kms, and the 44~GHz masers fall in the range of
$\sim$~46-78~\kms, both of which are consistent with the measured velocity of
ionized gas seen toward this region.  The large number of maser sites, along
with their spread in velocity, indicates a high rate of star formation in this
area.  Source g9 has a mass of 21~$\pm$~7~\Msun, but due to a poorly
constrained fit, g8 has no derived mass.  Both g8 and g9 are located in a
region of bright 24~\um\, emission.  Source g10 (9.5$\pm$0.7) displays
extended 4.5~\um\, emission, and is not coincident with 24~\um\, emission
(Fig.~\ref{g10}).  We detect three 6.7~GHz masers in the g10 field, but none
are close enough to be associated with the green source.  Two of these masers
display a single velocity feature, while the other shows two components.  None
of the maser sites are coincident with bright IRAC or 24~\um\, emission.

\subsubsection{g11 (G0.542$-$0.476), g12 (G0.517$-$0.657),
  g13(G0.483$-$0.701),g14 (G0.477$-$0.727), g15 (G0.408$-$0.504)}
Foreground green excess sources g11-g15 are associated with the Sharpless 20
\citep[Sh20;][]{shar59,dutr03} star formation region, which is centered at
$l=0.5$\degreesym, b$=-0.3$\degreesym\, \citep{mars74}.  Source g11
(5.2$\pm$2.5~\Msun) displays bright, compact 4.5~\um\, emission, but no
correlated 24~\um\, emission (Fig.~\ref{g11}).  It is located within an IRDC,
adjacent to a bright loop of 8~\um\, emission that is likely part of Sh20.  We
did not detect any \meth\, masers toward g11.  Green source g12, which has a
mass of 2.1~$\pm$~1.2~\Msun, shows faint, extended 4.5~\um\, emission along
with coincident 24~\um\, emission (Fig.~\ref{g12}).  The IRDC that harbors g12
is adjacent to Sh20.  We detect two 44~GHz \meth\, masers associated with g12,
with single velocity features at 15 and 17~\kms.  One of the maser sites
resides within the extent of the 4.5~\um\, emission, while the other is
$\sim$~10\arcsec\, away.  The 4.5~\um\, and 24~\um\, emission from g13 are
bright and extended (Fig.~\ref{g13}).  This source is locate just south of
Sh20, embedded within an IRDC.  It has a mass of 9.9~$\pm$~1.7~\Msun, and we
detect a single 44~GHz \meth\, maser toward it.  This maser has a single
velocity feature at 13~\kms\, that is located on a knot of 4.5~\um\, emission
19\arcsec\, away from the center of the green source.  Located just south of
g13, g14 displays two knots of 4.5~\um\, emission, both of which are
coincident with 24~\um\, emission (Fig.~\ref{g14}).  This 7.4$\pm$1.7~\Msun,
source displays no \meth\, maser emission, and is found within an IRDC.
Green source g15 (Fig.~\ref{g15}) shows extended 4.5~\um\, emission coincident
with 24~\um\, emission.  This source, with a mass of 14.3~$\pm$~4.2~\Msun,
is not associated with a 44~GHz maser.  The 4.5~\um\, emission from
g15 is positionally coincident with a 6.7~GHz maser that has a single velocity
feature at 26~\kms.  It is also surrounded by a bright ring of 8 and 24~\um\,
emission, so it is unclear if g15 is embedded in an IRDC, or if the region in
which it sits is only dark at 8~\um\, relative to the bright ring of emission.

\subsubsection{g16 (G0.376+0.040)}
Located at the distance of the Galactic center, the extended green source g16
(Fig.~\ref{g16}) coincides with one of the string of submillimeter continuum
emitting clouds that comprise the dust ridge.  In addition to being located
within an IRDC, g16 also displays bright 24~\um\, emission.  This
source, with a mass of 10.1~$\pm$~0.6~\Msun, is coincident with a 6.7~GHz
maser, but no 44~GHz maser.  The 6.7~GHz maser has a single velocity feature
at 37~\kms, and is located 2\arcsec\, away (detected by C10 only) from the
center of the green source, but coincident with the enhanced 4.5~\um\,
emission.

\subsubsection{g17 (G0.315$-$0.201)}
The foreground source g17 (Fig.~\ref{g17}) lies in the vicinity of two stellar
cluster candidates that are located within 1\arcmin\, of each other and within
the Sharpless 20 region. This region also has variable X-ray emission,
indicating the presence of very young stars \citep{law04}.  Source g17 is
associated with an IRDC, and it displays no 24~\um\, emission (one of only two
with a maser and no 24~\um\, emission).  A possible reason for its lack of
24~\um\, emission is that it lies next to a region of bright 24~\um\,
emission, which may be overwhelming any emission coincident with the 4.5~\um\,
emission.  Green source g17, with a mass of 12.7~$\pm$~2.6~\Msun, is
associated with pair of 6.7~GHz masers that are $<$~5\arcsec\, from the source.
One of the maser sites displays a single velocity feature at 20~\kms, and the
other displays several features from $\sim$~16--20~\kms.  As seen in
Figure~\ref{g17}, both masers lie at an interface where bright 4.5~\um\,
emission transitions into bright 8~\um\, emission.  No 44~GHz maser is
detected toward g17.

\subsubsection{g18 (G0.167$-$0.445)}
Another foreground source, g18 is found toward the \hii\,
region RCW~141.  This enhanced 4.5~\um\, source appears orange in
Figure~\ref{g18}, and has a mass of 14.0~$\pm$~4.3~\Msun.  It is
found within a filamentary IRDC and is associated with 24~\um\, emission.
While not detected in our data, a 6.7~GHz maser was detected by C10
positionally coincident with g18.  The 6.7~GHz spectrum of this source
displays several velocity features from 9--17~\kms. No 44~GHz maser is
detected toward g18.

\subsubsection{g19 (G0.091$-$0.663), g20 (G0.084$-$0.642)}
Green sources g19 (Fig.~\ref{g19}) and g20 (Fig.~\ref{g20}), are located in
close proximity to one another in the same IRDC complex toward the \hii\,
region RCW~141.  Both sources are likely foreground to the Galactic center,
and are coincident with bright 24~\um\, emission.  While g20
(7.3~$\pm$~1.3~\Msun) is not associated with any \meth\, masers, g19
(5.5~$\pm$~1.3~\Msun) is associated with a 6.7~GHz and a 44~GHz maser.  The
6.7~GHz maser spectrum displays two bright features between 20 and 25~\kms.
It is located within the confines of the green source, $\sim$~2\arcsec\, from
the center of its extended emission. The 44~GHz maser emission was detected
$\sim$~14\arcsec\, from the green source, not directly on the 4.5~\um\,
emission.  This 44~GHz maser displays a single emission feature at 17~\kms.

\subsubsection{g21A (G359.972$-$0.459A) and g21B (G359.972$-$0.459B)}
Single SED fits were performed for most green sources (Y-Z09), but g21 was
best fit by two sources, designated g21A (11.8~$\pm$~2.5~\Msun) and g21B
(23.8~$\pm$~6.6~\Msun).  These sources are seen as distinct lobes of 4.5~\um\,
emission in Figure~\ref{g21}, and appear to be in the same star forming
region, which is foreground to the Galactic center and in the vicinity
of RCW~137.  Like many of the other green sources, g21A and B are associated
with an IRDC and are coincident with 24~\um\, emission.  There is one 6.7~GHz
maser that is located $<$~10\arcsec\, from g21A and g21B (7\arcsec\, and
1\arcsec, respectively).  Because it is closer to g21B, we associate the maser
with that source.  This maser consists of a single, bright emission feature at
23~\kms\, in the C10 data.  No 44~GHz masers are detected toward either of the
g21 sources.

\subsubsection{g22 (G359.939+0.170)}
Another example of an extended green source embedded within an IRDC, g22 is
likely located foreground to the Galactic center near an \hii\, complex.  The
enhanced 4.5~\um\, emission is coincident with both 24~\um\, emission and a
6.7~GHz maser (Fig.~\ref{g22}).  The maser has a single, bright emission
feature at -0.8~\kms, and is located directly on the green region that defines
the source, close to the peak of the 24~\um\, emission.  Source g22 has a mass
of 4.9~$\pm$~1.8~\Msun, and is not associated with any 44~GHz maser emission.

\subsubsection{g23 (G359.932$-$0.063)}
Of all our sources, g23 is the closest (in projection) to the Galactic
center.  It is embedded in an IRDC that runs parallel to the Galactic plane,
just south of Sgr~A$^*$.  It is a compact 4.5~\um\, source (Fig.~\ref{g23})
with no 24~\um\, counterpart.  Source g23 has a mass of 12.0~$\pm$~3.2~\Msun,
is located at the distance of the Galactic center, and is not associated with
any \meth\, masers.

\subsubsection{g24 (G359.907$-$0.303)}
Green source g24 shows a clear enhancement at 4.5~\um, but its emission at this
wavelength is relatively faint (Fig.~\ref{g24}).  It displays no corresponding
24~\um\, emission, and no \meth\, maser emission.  This source is located
within an IRDC, adjacent to a region of bright 8~\um\, emission.  It has a
mass of 6.3~$\pm$~2.5~\Msun, and is likely located foreground to the Galactic
center.

\subsubsection{g25 (G359.841$-$0.080)}
Much like g23, green source g25 is found within an IRDC that is south of
Sgr~A$^*$ and parallel to the Galactic plane.  This source is comprised of two
knots of 4.5~\um\, emission (Fig.~\ref{g25}), the brighter of which is
correlated with 24~\um\, emission.    This source, which displays no \meth\,
maser emission, is located at the distance of the Galactic center and has a
mass of 10.3~$\pm$~1.2~\Msun.

\subsubsection{g26 (G359.618$-$0.245)}
Similarly to sources g21A and g21B, green source g26 is located foreground to
the Galactic center, and is in the vicinity of RCW~137.  This bright, extended
green source, which has a derived mass of 7.1~$\pm$~0.9~\Msun, is located
within a plume-shaped IRDC and is coincident with 24~\um\, emission
(Fig.~\ref{g26}).  6.7~GHz maser emission is detected toward this source, and
its spectrum displays multiple velocity features between 19 and 25~\kms.  The
6.7~GHz maser site is within the confines of the green source.  In
addition to the 6.7~GHz maser emission, we also detect two sites of 44~GHz
maser emission $\sim$~3\arcsec\, away, with velocities of 19 and
20~\kms.  The multiple velocity features of 6.7~GHz maser emission, along with
the multiple sites of 44~GHz maser emission, indicate that a cluster of stars
may be forming in the IRDC that contains g26.

\subsubsection{g27 (G359.599$-$0.032)}
What stands out about g27 (Fig.~\ref{g27}) is that it is the only green source
not clearly associated with an IRDC.  Its compact 4.5~\um\, emission is,
however, coincident with 24~\um\, emission, indicating that it could still be
a region of high-mass star formation.  Indeed, it has a mass of
17.5~$\pm$~3.0~\Msun.  This Galactic center green source is not associated
with any 6.7~GHz maser emission, but is associated with a 44~GHz maser.  This
44~GHz maser is $<$~2\arcsec\, from the center of the source, and has a
velocity of 72~\kms.

\subsubsection{g28 (G359.57+0.270)}
Source g28 is a faint, slightly extended 4.5~\um\, source that is not
coincident with 24~\um\, emission (Fig.~\ref{g28}).  Its height above the
Galactic plane results in a distance assignment that is foreground to the
Galactic center.  This source (4.0~$\pm$~1.3~\Msun) is found within a small
IRDC that has a relatively low 8~\um\, flux decrement relative to its
surroundings.  We do not detect any \meth\, masers toward g28.

\subsubsection{g29 (G359.437$-$0.102)}
Located in the Sgr~C region at the distance of the Galactic center, green
source g29 is embedded within an IRDC and displays extended 4.5~\um\, emission
(Fig.~\ref{g29}).  This 14.1~$\pm$~2.4~\Msun\, source is not associated with
24~\um\, emission, making it one of only two sources that has maser emission
but no 24~\um\, emission.  Despite the lack of this particular star-forming
indicator, the association of g29 with both 6.7 and 44~GHz maser emission
indicates that star formation is indeed underway toward this source.  The
6.7~GHz maser emission arises in two locations that are near the edge of the
4.5~\um\, emission and are $\sim$~6\arcsec\, apart.  Each of these maser sites
displays two velocity components ($-$53 to $-$45~\kms, and $-$57 to
$-$53~\kms).  The single 44~GHz maser is 5\arcsec\, from the center of green
source, but still within the extent of the 4.5~\um\, emission.  This maser has
a velocity of $-$66~\kms.

\subsubsection{g30 (G359.30+0.033)}
Green excess source g30 displays extended 4.5~\um\, and is found within an IRDC that
is long, filamentary, and perpendicular to the Galactic plane (Fig.~\ref{g30}).
We find bright 24~\um\, emission associated with this green source, but no
\meth\, masers.  This source is likely to be at the distance of the Galactic
center, and it has a mass of 8.9~$\pm$~0.7~\Msun.

\subsubsection{g31 (G359.199+0.041)}
A compact, bright 4.5~\um\, emission source, g31 is located at the distance of
the Galactic center in an IRDC.  It has a mass of 14.4~$\pm$~4.6~\Msun, and is
coincident with 24~\um\, emission.  We detect a 6.7~GHz maser detected in the
field, but it is $\sim$~70\arcsec\, away from g31, and thus not associated
with the green source.  The maser site is $\sim$~15\arcsec\, from a small region of
bright 24~\um\, emission, but is not near any bright IRAC emission.  This
44~GHz maser displays a single velocity feature at $-$4~\kms.

\subsubsection{g32 (G358.980+0.084)}
Green source g32 shows faint, extended 4.5~\um\, emission (Fig.~\ref{g32}), is
embedded within an IRDC, and is likely located at the distance of the Galactic
center. This source is associated with bright 24~\um\, emission and 6.7~GHz
maser emission, making it very likely that this is a star-forming source.  The
6.7~GHz maser emission is positionally coincident with the extended, enhanced
4.5~\um\, emission, and displays a single velocity feature at $\sim$~6~\kms\,
(in the C10 data only).  The SED fit for this source indicates that it has a
mass of 10.5~$\pm$~6.0~\Msun.  No 44~GHz maser emission is detected toward
g32.

\section{Discussion}

\subsection{Comparison of Galactic Center and Foreground Sources}

The CMZ contains $\sim$~5~$\times$~10$^{7}$~\Msun\, of molecular gas
\citep{pier00}, including many prominent IRDCs.  How exactly star formation
proceeds in these clouds, however, remains unclear.  Based on molecular line
emission seen throughout the CMZ, such as SiO and HCO$^+$
\citep[e.g.,][]{mart00,riqu10}, as well as several H$_3^+$ lines of sight
toward the CMZ \citep{oka05}, the chemistry of molecular gas in the Galactic
center is likely to be different from the gas in the Galactic disk.  In
addition, a NH$_3$ study of of Galactic center clouds distributed between
l=$-1$\degreesym\, and 3\degreesym\, \citep{huet93} shows a two-temperature
distribution of molecular gas at T$_{kin}~\sim$200K and T$_{kin}\sim$25K,
while the dust temperature of the clouds in the CMZ remains low
\citep[$\leq$~30~K; e.g.,][]{oden84,cox89,pier00}.  Stronger turbulence
\citep[cf.][]{morr96} also differentiates CMZ molecular clouds from their
counterparts in the Galactic disk.  Thus, it is possible that the unique
environment in CMZ molecular clouds results in different initial conditions
for star formation.  The recent results of Y-Z09, however, show that the
Kennicut law \citep{kenn98} holds in the Galactic center, so the relationship
between star formation rate per unit area and surface mass density in the
Galactic center is similar to that of the Galactic disk, at least to first
order.  Nevertheless, it is possible that some important differences exist
between Galactic center and disk star formation.  To test this possibility, we
compare the 6.7 (radiatively excited) and 44~GHz (collisionally excited)
\meth\, maser detection rates toward foreground and Galactic center green
sources.  The foreground sources have 6.7 and 44~GHz \meth\, maser detection
rates of 44$\pm$17\% and 25$\pm$13\%, respectively (errors for detection rates
are calculated using $\sqrt{N}$ counting statistics).  Galactic center green
sources have 6.7~GHz \meth\, maser detection rate of 33$\pm$15\%, and a 44~GHz
maser detection rate of 27$\pm$13\%.

\begin{figure}
\centering
\includegraphics[angle=-90,width=0.49\textwidth,clip=true]{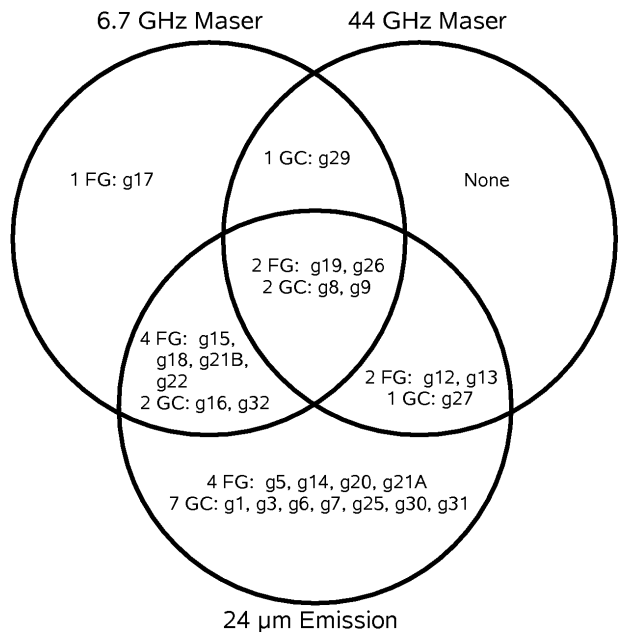}
\caption{Diagram showing the number of sources that harbor 6.7~GHz masers,
  44~GHz masers, and 24~\um\, emission.  `FG' stands for sources foreground to
  the Galactic center, and `GC' stands for Galactic center sources.  Sources
  in regions that overlap display at least two of the star formation
  indicators.  Two Galactic center sources (g10 and g23) and three foreground
  sources (g11, g24, and g28) display no maser emission and no 24~\um\,
  emission.  These five sources are excluded from this diagram.\label{venn}}
\end{figure}

The overall detection rate of 44~GHz masers is roughly the same for foreground
and Galactic center sources.  The detection rate of 6.7~GHz masers may be
higher for foreground sources than Galactic center sources, but taking into
account their large error bars, they are also consistent with being the same
(Fig.~\ref{venn}; see also Tables~\ref{maser-summary}\, and
\ref{fg-gc-summary}).  The possible difference in the foreground and Galactic
center 6.7~GHz maser detection rates could be due to small-number statistics
(i.e., the relatively low number of sources resulting in large errors).  If
the $\sqrt{N}$ statistics errors are taken as true 1$\sigma$\, errors, then
the difference in detection rates is roughly a 1$\sigma$\, result.  Thus, it
is possible that the detection rates for 6.7~GHz \meth\, masers (in addition
to the 44~GHz \meth\, masers) are the same for both foreground and Galactic
center green sources.  If this is the case, then we find no obvious difference
between star formation in these two regions, as traced by \meth\, maser
detections.

Despite the small number of statistics, we do, however, find that 2/7
(29$\pm$20\%) of foreground sources with 6.7~GHz masers have 44~GHz masers and
that 3/5 (60$\pm$35\%) of Galactic center sources with 6.7~GHz masers have
44~GHz masers.  Thus, the Galactic center sources with 6.7~GHz masers may have
a relative overabundance of 44~GHz masers (again, a roughly 1$\sigma$\,
result).  One possible explanation for this could be the properties of the
molecular clouds that harbor the green sources.  \citet{prat08}\, show that
44~GHz maser emission is enhanced at high densities
(n(H$_{2})$~$\sim$~10$^5-10^6$ cm$^{-3}$) and warm temperatures (80 to 200~K).
Thus, the environment in clouds within the Galactic center may give rise to
more favorable conditions for the creation of the 44~GHz \meth\, masers.

Although not firmly established, some recent work \citep{elli07,bree10} has
proposed a sequence of \meth\, maser evolution in star forming regions.  In
this evolutionary sequence, Class~I \meth\, masers, generated by protostellar
outflows, are formed before Class~II \meth\, masers, with an overlap period
lasting $\sim~1.5~\times~10^{4}$~years.  Subsequent work by \citet{voro10}\,
shows that Class~I masers are also formed at later evolutionary states, when
expanding \hii\, regions collide with neighboring molecular clouds, possibly
accounting for the exceptions noted by \citet{bree10} to their proposed
sequence.  If this sequence is correct, then a relative over-abundance of
44~GHz masers in Galactic center sources, along with a relative
under-abundance of 6.7~GHz masers, suggests that the Galactic center sources
are, on average, younger than the foreground sources.  According to Y-Z09,
there was a burst in the Galactic center star formation rate about 10$^5$
years ago (rising to 1.4~\Msun~yr$^{-1}$), which may have given rise to an
over-abundance of Stage~I young stellar objects in the Galactic center,
supporting this result.  Other recent work \citep[e.g.,][]{font10}, however,
suggests that there is no correlation between evolutionary state and \meth\,
maser class, so further study is needed to test these results.

We also compare our maser detection results with other studies of \meth\,
maser emission toward 4.5~\um\, emission sources.  In one study,
\citet{cham09}\, studied 25~GHz Class~I \meth\, maser emission toward a sample
of 47 green fuzzies found within IRDCs.  These green fuzzies were located
using the Green Fuzzy Finder (GFF), which identifies contiguous pixels with an
enhancement at 4.5~\um.  In another study, \citet{cyga09} studied 6.7 and
44~GHz \meth\, maser emission toward a sample of $\sim$~20 EGOs.
\citet{cyga09} identified EGOs by eye using 3-color images created with
GLIMPSE data \citep{cyga08}.  \citet{chen09} also searched for a correlation
between EGOs and Class~I \meth\, maser emission, using the results of four
previously published \meth\, maser searches.  Green fuzzies and EGOs have
different selection criteria from one another, and from our sample of 31 green
sources, but comparisons of the \meth\, maser detection rates may provide some
insight into how these sources are related.  For simplicity, we refer to the
three types of enhanced 4.5~\um\, sources as green sources.

\citet{cham09} find that 17\% of 47 green sources in Galactic disk IRDCs are
associated with Class~I \meth\, maser emission.  \citet{cyga09} searched for
\meth\, maser emission toward a sample of $\sim$20 green sources, and report
that $\sim$65\% of their sample harbor 6.7~GHz masers, $\sim$90\% of which
also display 44~GHz maser emission.  \citet{chen09} find that $\sim$~67\% of a
sample of 61 sources are associated with Class~I \meth\, masers.  In our sample
of 31 green sources, 12 (39~$\pm$~11\%) are associated with 6.7~GHz masers, 8
(26~$\pm$~9\%) are associated with 44~GHz masers, and 5 (16~$\pm$~7\%) are
associated with both maser transitions.  Our Class~I maser detection rate
(26~$\pm$~9\%) is in rough agreement with that of \citet{cham09}.  The
similarity of the maser detection rates, along with the association of both
sets of sources with IRDCs and 24~\um\, emission, indicates that our green
sources are similar in nature to the green sources studied by \citet{cham09}.

The difference in the 6.7~GHz maser detection rate between our green sources
(39~$\pm$~11\%) and the \citet{cyga09} sample of green sources (65\%) is
fairly large.  Moreover, we find that only 5 of the 12 sources with 6.7~GHz
masers (42~$\pm$~19\%) also harbor 44~GHz maser emission, while
\citet{cyga09}\, and \citet{chen09} both find higher Class~I detection rates.
A possible reason for these discrepancies is that the \citet{cyga09} and
\citet{chen09} green sources are larger in angular extent than our green
sources.  The green sources in our sample are roughly 5-10\arcsec\, in size.
The typical green source size in the \citet{cyga09} and \citet{chen09}
samples, however, is $\sim$~10-20\arcsec, with some as large as 30\arcsec\, or
more.  If we assume that these larger green sources are located at roughly
similar distances to the sources in our sample, then their larger angular
sizes would correspond to larger physical sizes.  Because the extended
4.5~\um\, emission is associated with outflows, it stands to reason that the
\citet{cyga09} and \citet{chen09}\, green sources are associated with larger
outflows, and are therefore more evolved.  The larger, more evolved outflows
have a greater surface area of interaction with the surrounding medium,
thereby possibly explaining the increased 44~GHz detection rate toward the
\citet{cyga09} and \citet{chen09} samples.  Alternatively, the difference in
the maser detection rates could be due to the the variable nature of \meth\,
maser emission \citep{goed04}, combined with the different sensitivities of
the surveys.  Indeed, the 6$\sigma$\, sensitivity for the \citet{cyga09}\,
6.7~GHz observations is $\sim$~0.16~Jy, and the 3$\sigma$\, sensitivity of our
6.7~GHz observations is 0.15~Jy.

\subsection{Green Sources with and without 24~\um\, Emission}
Both enhanced 4.5~\um\, emission (used to identify green sources) and 24~\um\,
emission are star formation indicators, and when they are coincident with one
another, they are a powerful identifier of active star formation
\citep{beut07,cyga08,cham09}.  We find that 77\%\, (24 of 31) of the green
sources in our sample are coincident with 24~\um\, emission.  The detection
rate of 6.7~GHz \meth\, masers for green sources coincident with 24~\um\,
emission is 46$\pm$14\%, and is 29$\pm$20\% for those without 24~\um\,
emission.  The detection rate of 44~GHz \meth\, masers for green sources
coincident with 24~\um\, emission is 29$\pm$11\%.  We detect only one 44~GHz
\meth\, maser toward a green source without 24~\um\, emission (detection rate
of 14$\pm$14\%).  Again, we are limited by the small number of statistics and
the large errors, and conclusions drawn from this sample should be taken with
caution.  Nevertheless, it appears that green sources with 24~\um\, emission
are more likely to harbor both 6.7 and 44~GHz \meth\, masers than the green
sources that are not coincident with 24~\um\, emission (see
Tables~\ref{maser-summary}\, and \ref{fg-gc-summary}).  If true, these results
once again show that the combination of enhanced 4.5~\um\, emission, 24~\um\,
emission, and \meth\, maser emission are reliable protostellar locators.

Even though the green sources that are coincident with 24~\um\, emission seem
more likely to harbor \meth\, masers, we do detect masers toward green
sources that display no 24~\um\, emission.  These sources may contain 24~\um\,
emission too faint to detect using our MIPS data.  Alternatively, they may be
at a different, perhaps earlier, evolutionary state, before the dust around
the central protostar has heated sufficiently to emit brightly at 24~\um.
Finally, it is also possible that some (or all) of these green sources are not
related to star formation.  For example, planetary nebulae may also appear as
4.5~\um\, enhancement sources--recall that three green sources in our original
sample have been excluded from this analysis because their positions coincide
with those of known planetary nebulae.

\subsection{Masses of Green Sources with and without \meth\, Masers}
\begin{figure}
\centering
\includegraphics[width=0.49\textwidth]{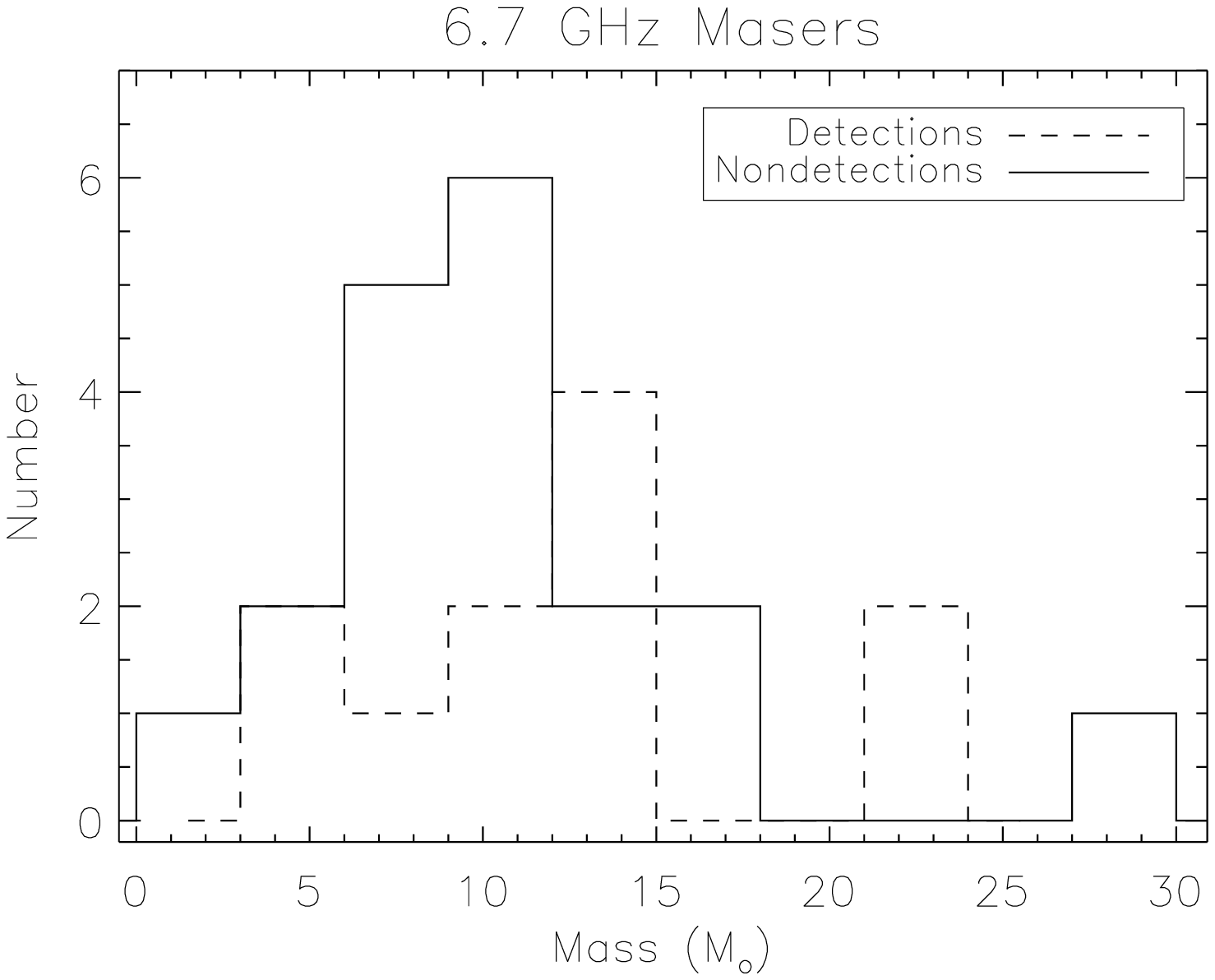}
\includegraphics[width=0.49\textwidth]{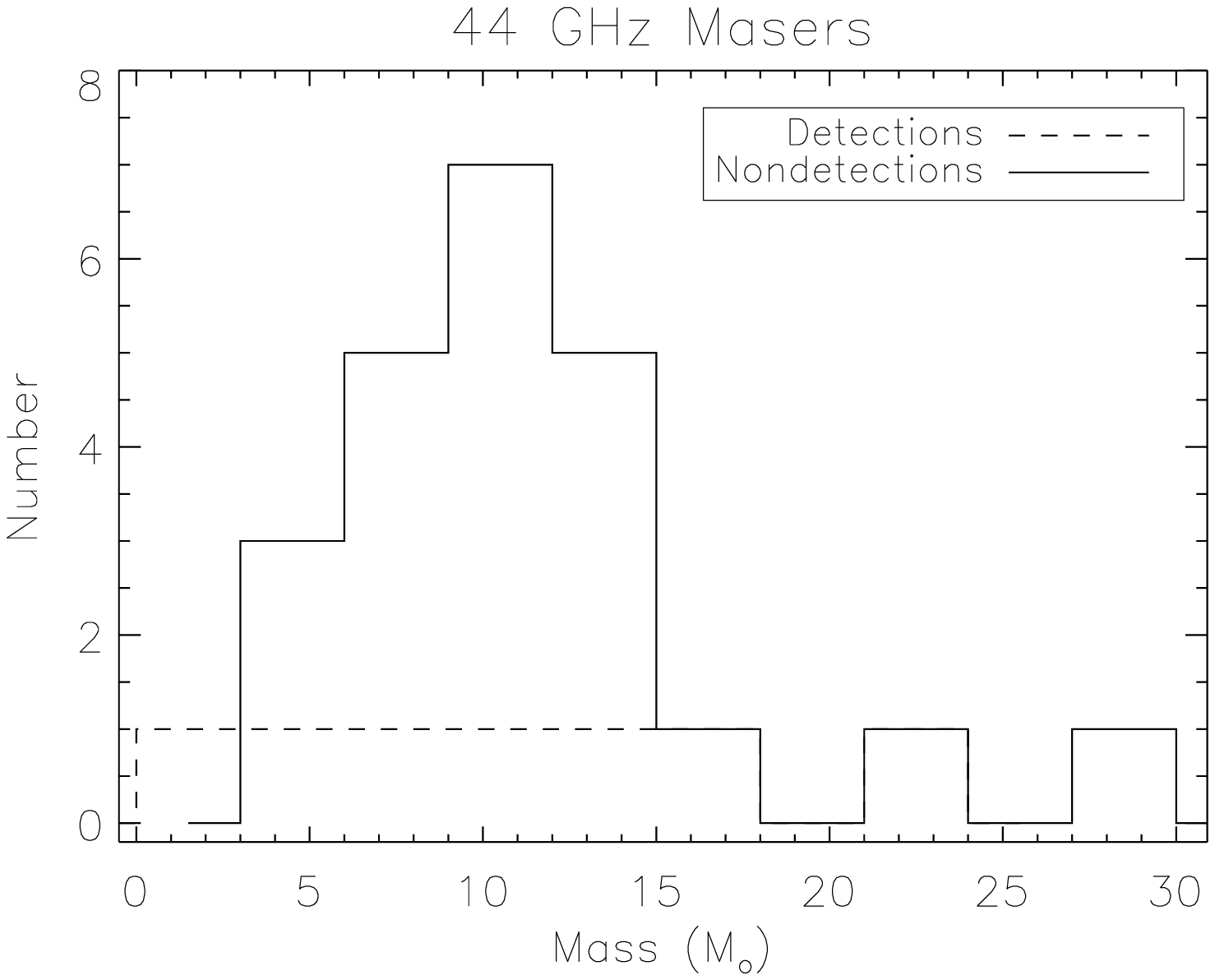}
\caption{{\it Left}:  Histograms of the masses of green sources that harbor 6.7~GHz
  \meth\, masers (dashed histogram) and that do not harbor 6.7~GHz masers (solid line
  histogram).  The median mass for green sources associated with 6.7~GHz
  masers is 12.7~\Msun, and the median mass for green sources not associated
  with 6.7~GHz masers is 9.5~\Msun.  {\it Right}:  Histograms of the masses of green sources that harbor 44~GHz
  \meth\, masers (dashed histogram) and that do not harbor 44~GHz masers (solid line
  histogram).  The median mass for green sources associated with 44~GHz
  masers is 9.9~\Msun, and the median mass for green sources not associated
  with 44~GHz masers is 10.3~\Msun.
\label{mass_histos}}
\end{figure}

We find that higher-mass sources are more likely to harbor 6.7~GHz \meth\,
masers than lower-mass sources.  The median mass of green sources associated
with 6.7~GHz \meth\, masers is 12.7~\Msun, while the median mass for those not
associated with 6.7~GHz \meth\, masers is 9.5~\Msun\, (see
Fig.~\ref{mass_histos}).  Using a Kolmogorov-Smirnov (K-S) test, we calculate
a 76\% chance that the mass distributions of green sources with and without
6.7~GHz \meth\, maser emission are drawn from a different parent distribution.
The median mass of sources with 44~GHz \meth\, masers is 9.9~\Msun, and the
median mass for those without is 10.3~\Msun\, (see Fig.~\ref{mass_histos}).
We find only a 5\% chance that the mass distributions of sources with and
without 44~GHz \meth\, masers are drawn from a different parent distribution
(according to a K-S test).  Because the mass distributions of green sources
with and without 6.7~GHz masers are likely to be different, and because the
population associated with 6.7 \meth\, masers has a larger median mass than
those without, we conclude that higher-mass sources are more likely to display
6.7~GHz maser emission than lower-mass sources.  The case is less clear for
the 44~GHz masers, the detections of which show no clear correlation with
mass.

The lowest mass that a 6.7~GHz maser is associated with is
4.9$\pm$1.8~\Msun\,(g22).  Because the masses determined by Y-Z09 are the
current masses of the sources (not the final masses), it is possible that this
source will evolve into a high-mass ($\geq$~8~\Msun) star through accretion,
consistent with the idea that 6.7~GHz \meth\, maser emission exclusively
traces high-mass star formation.  Moreover, g22 is assumed to be foreground to
the Galactic center; if it is instead at the distance of the Galactic center,
its derived mass would increase.  The lowest mass source associated with a
44~GHz maser is 2.1$\pm$1.2~\Msun\, (g12).  Because of its lower mass, the
likelihood of this source evolving into a high-mass star is smaller.  The
association of g12 with a 44~GHz maser supports with the hypothesis that this
maser transition traces low-mass star formation in addition to high-mass star
formation.

\section{Conclusions}

To study how star formation in the Galactic center's CMZ may differ from that
of the Galactic disk, we studied \meth\, maser emission toward a sample of 34
green sources.  Of the 15 Galactic center green sources, we find that 5 are
associated with radiatively excited Class~II 6.7~GHz \meth\, maser emission,
and 4 are associated with collisionally excited Class~I 44~GHz \meth\, maser
emission.  Of the 16 green sources located foreground to the
Galactic center, we find that 8 are associated with 6.7~GHz \meth\, maser
emission, and 4 are associated with 44~GHz \meth\, maser emission. Based on
these detections, we find: (1) little difference between the sources in the
Galactic center and the foreground sources, (2) that the 34 green sources in
our sample are similar to the green sources identified by \citet{cham09}
rather than the larger green sources identified by \citet{cyga08}, and (3)
that the possible relative overabundance of green sources with both 6.7 and
44~GHz masers in the Galactic center may be consistent with a recent burst of
star formation in the Galactic center.

\acknowledgments We are grateful to the anonymous referee, whose comments and
suggestions greatly improved the paper.  We also thank Vivek Dhawan at the
NRAO for his help with the 6.7 GHz data reduction.


\clearpage

\clearpage

\clearpage

\clearpage

\clearpage

\clearpage

\clearpage

\clearpage

\begin{deluxetable}{ccccccccc}
\tablecolumns{10}
\tablewidth{0pc}
\tablecaption{\label{green-summary}Summary of green sources}
\tablehead{
\colhead{Source}     & \colhead{$\ell$}       & \colhead{$b$}        & 
\colhead{Galactic}   & \colhead{24~\um}             & \colhead{Mass}  &
\colhead{6.7 GHz}      & \colhead{44 GHz} \\
\colhead{Number}     & \colhead{(\degreesym)} & \colhead{(\degreesym)} & 
\colhead{Location}   & \colhead{Emission}                             & \colhead{(\Msun)}  &
\colhead{Maser}        & \colhead{Maser}}
\startdata
g0	&	1.0407		&	$-$0.0714	&	GC	&	Y	&	16.1	$\pm$	3.5	&	N	&	N	\\
g1	&	0.9555		&	$-$0.7853	&	FG	&	Y	&	2.4	$\pm$	0.6	&	N	&	N	\\
g2	&	0.8679		&	$-$0.6963	&	FG	&	Y	&	5.5	$\pm$	1.3	&	N	&	N	\\
g3	&	0.8262		&	$-$0.2108	&	GC	&	Y	&	10.6	$\pm$	0.0	&	N	&	N	\\
g4	&	0.7800		&	$-$0.7395	&	FG	&	Y	&	6.9	$\pm$	1.9	&	N	&	N	\\
g5	&	0.7062		&	0.4075	&	FG	&	Y	&	6.0	$\pm$	3.0	&	N	&	N	\\
g6	&	0.6934		&	$-$0.0451	&	GC	&	Y	&	29.9	$\pm$	8.2	&	N	&	N	\\
g7	&	0.6796		&	$-$0.0376	&	GC	&	Y	&	9.4	$\pm$	1.2	&	N	&	N	\\
g8	&	0.6670		&	$-$0.0370	&	GC	&	Y	&		--		&	Y	&	Y	\\
g9	&	0.6667		&	$-$0.0352	&	GC	&	Y	&	21.7	$\pm$	7	&	Y	&	Y	\\
g10	&	0.6655		&	$-$0.0535	&	GC	&	N	&	9.5	$\pm$	0.7	&	N	&	N	\\
g11	&	0.5418		&	$-$0.4759	&	FG	&	N	&	5.2	$\pm$	2.5	&	N	&	N	\\
g12	&	0.5172		&	$-$0.6569	&	FG	&	Y	&	2.1	$\pm$	1.2	&	N	&	Y	\\
g13	&	0.4835		&	$-$0.7004	&	FG	&	Y	&	9.9	$\pm$	1.7	&	N	&	Y	\\
g14	&	0.4771		&	$-$0.7268	&	FG	&	Y	&	7.4	$\pm$	1.7	&	N	&	N	\\
g15	&	0.4084		&	$-$0.5042	&	FG	&	Y	&	14.3	$\pm$	4.2	&	Y	&	N	\\
g16	&	0.3763		&	0.0402	&	GC	&	Y	&	10.1	$\pm$	0.6	&	Y	&	N	\\
g17	&	0.3153		&	$-$0.2010	&	FG	&	N	&	12.7	$\pm$	2.6	&	Y	&	N	\\
g18	&	0.1667		&	$-$0.4455	&	FG	&	Y	&	14	$\pm$	4.3	&	Y	&	N	\\
g19	&	0.0915		&	$-$0.6624	&	FG	&	Y	&	5.5	$\pm$	1.3	&	Y	&	Y	\\
g20	&	0.0842		&	$-$0.6415	&	FG	&	Y	&	7.3	$\pm$	1.3	&	N	&	N	\\
g21A	&	359.9700	&	$-$0.4554	&	FG	&	Y	&	11.8	$\pm$	2.5	&	N	&	N	\\
g21B	&	359.9695	&	$-$0.4573	&	FG	&	Y	&	23.8	$\pm$	6.6	&	Y	&	N	\\
g22	&	359.9386	&	0.1709	&	FG	&	Y	&	4.9	$\pm$	1.8	&	Y	&	N	\\
g23	&	359.9316	&	$-$0.0626	&	GC	&	N	&	12.0	$\pm$	3.2	&	N	&	N	\\
g24	&	359.9072	&	$-$0.3026	&	FG	&	N	&	6.3	$\pm$	2.5	&	N	&	N	\\
g25	&	359.8412	&	$-$0.0793	&	GC	&	Y	&	10.3	$\pm$	1.2	&	N	&	N	\\
g26	&	359.6148	&	$-$0.2438	&	FG	&	Y	&	7.1	$\pm$	0.9	&	Y	&	Y	\\
g27	&	359.5995	&	$-$0.0316	&	GC	&	Y	&	17.5	$\pm$	3	&	N	&	Y	\\
g28	&	359.5701	&	0.2704	&	FG	&	N	&	4.0	$\pm$	1.3	&	N	&	N	\\
g29	&	359.4362	&	$-$0.1017	&	GC	&	N	&	14.1	$\pm$	2.4	&	Y	&	Y	\\
g30	&	359.3009	&	0.0334	&	GC	&	Y	&	8.9	$\pm$	0.7	&	N	&	N	\\
g31	&	359.1994	&	0.0419	&	GC	&	Y	&	14.4	$\pm$	4.6	&	N	&	N	\\
g32	&	358.9795	&	0.0840	&	GC	&	Y	&	10.5	$\pm$	6	&	Y	&	N	\\
\enddata
\end{deluxetable}

\begin{deluxetable}{ccccc}
\tablecolumns{5}
\tablewidth{0pc}
\tablecaption{\label{six-summary}6.7~GHz \meth\, maser detections}
\tablehead{
\colhead{Green Source} &  \colhead{Right Ascension} & \colhead{Declination}  
& \colhead{Velocity}  &  \colhead{Peak Intensity} \\
\colhead{Field}        &  \colhead{(J2000)}         & \colhead{(J2000)}      
&\colhead{(\kms)}     &  \colhead{(Jy~beam$^{-1}$)}}
\startdata
g6 & 17:47:24.7 & $-$28:21:44.2 & 68.9 & 12.2 \\
g7, g8, g9 & 17:47:19.3 & $-$28:22:16.4 & 73.8 & 1.3 \\
g7, g8, g9 & 17:47:20.1 & $-$28:23:14.0 & 60.6 & 1.9 \\
g7, g8, g9 & 17:47:20.1 & $-$28:23:48.2 & 52.4 & 1.9 \\
g7, g8, g9 & 17:47:18.7 & $-$28:22:56.0 & 71.1 & 10.3 \\
g7, g8, g9 & 17:47:20.1 & $-$28:22:42.2 & 58.5 & 2.5 \\
g10 & 17:47:22.1 & $-$28:24:43.6 & 51.3 & 1.2 \\
g10 & 17:47:18.6 & $-$28:24:25.6 & 49.7 & 19.6 \\
g10 & 17:47:21.1 & $-$28:24:19.3 & 48.6 & 5.5 \\
g15 & 17:48:33.4 & $-$28:50:51.9 & 26.0 & 0.6 \\
g17 & 17:47:9.1 & $-$28:46:16.0 & 18.4 & 26.1 \\
g17 & 17:47:9.3 & $-$28:46:16.9 & 20.0 & 2.3 \\
g19 & 17:48:25.9 & $-$29:12:6.2 & 22.2 & 13.7 \\
g22 & 17:44:48.6 & $-$28:53:59.4 & $-$0.8 & 0.7 \\
g26 & 17:45:39.1 & $-$29:23:30.8 & 22.8 & 13.8 \\
g29 & 17:44:40.6 & $-$29:28:15.7 & $-$46.4 & 46.4 \\
g29 & 17:44:40.1 & $-$29:28:13.0 & $-$56.3 & 0.8 \\
g31 & 17:43:37.4 & $-$29:36:10.3 & $-$4.1 & 1.4 \\
\enddata
\end{deluxetable}

\begin{deluxetable}{ccccc}
\tablecolumns{5}
\tablewidth{0pc}
\tablecaption{\label{forty-summary}44~GHz \meth\, maser detections}
\tablehead{
\colhead{Green Source} &  \colhead{Right Ascension} & \colhead{Declination}  & \colhead{Velocity}  &  \colhead{Peak Intensity}\\
\colhead{Field}        &  \colhead{(J2000)}         & \colhead{(J2000)}      &\colhead{(\kms)}     &   \colhead{(Jy~beam$^{-1}$)}}
\startdata
g12	&	17:49:24.431	&	$-$28:50:01.86	&	16.8	&	3.2	\\
g12	&	17:49:25.177	&	$-$28:50:04.24	&	15.5	&	1.4	\\
g13	&	17:49:31.301	&	$-$28:53:16.04	&	12.8	&	3.4	\\
g19	&	17:48:26.095	&	$-$29:12:17.76	&	16.6	&	7.2	\\
g26	&	17:45:39.408	&	$-$29:23:33.48	&	18.9	&	15.4	\\
g26	&	17:45:39.386	&	$-$29:23:33.06	&	20.2	&	9.0	\\
g27	&	17:44:47.163	&	$-$29:17:40.70	&	71.9	&	2.9	\\
g29	&	17:44:40.385	&	$-$29:28:08.64	&	$-$65.8	&	2.7	\\
\enddata
\end{deluxetable}

\begin{deluxetable}{lccc}
\tablecolumns{4}
\tablewidth{0pc}
\tablecaption{\label{maser-summary}Summary of \meth\, maser detection rates toward
  green sources}
\tablehead{
                      &  \colhead{Total} & \colhead{With 6.7 GHz}  &  \colhead{With 44 GHz}\\
                      &  \colhead{Number} & \colhead{Masers}  & \colhead{Masers}}
\startdata
All green sources \tablenotemark{a} &  31  & 12 (39~$\pm$~11\%) & 8 (26~$\pm$~9\%)  \\
Foreground sources                  &  16  & 7  (44~$\pm$~17\%) & 4 (25~$\pm$~13\%)  \\
Galactic center sources             &  15  & 5  (33~$\pm$~15\%) & 4 (27~$\pm$~13\%) \\
Sources with 24~\um\, emission      &  24  & 11 (46~$\pm$~14\%) & 7 (29~$\pm$~11\%)  \\
Sources with no 24~\um\, emission   &  7   & 2  (29~$\pm$~20\%) & 1 (14~$\pm$~14\%)   \\
\enddata
\tablenotetext{a}{Excludes the three planetary nebulae (g1, g2, and g4).}
\end{deluxetable}

\begin{deluxetable}{lccccc}
\tablecolumns{4}
\tablewidth{0pc}
\tablecaption{\label{fg-gc-summary}Properties of FG and GC sources}
\tablehead{
     & \multicolumn{2}{c}{FG Sources}  &  \multicolumn{2}{c}{GC Sources}\\
\cline{2-3} \cline{4-5}
     & \colhead{Number} & \colhead{Percent\tablenotemark{a}}  & \colhead{Number} & \colhead{Percent\tablenotemark{b}}
}
\startdata
6.7~GHz \meth\, Maser   &   7 &  44 ~$\pm$~ 17 \%  & 5  & 33 ~$\pm$~ 15 \% \\
44~GHz \meth\, Maser    &   4 &  25 ~$\pm$~ 13 \%  & 4  & 27 ~$\pm$~ 13 \% \\
24~\um\, emission       &  12 &  75 ~$\pm$~ 22 \%  & 12 & 80 ~$\pm$~ 23 \% \\
Associated with IRDC    &  16 & 100 ~$\pm$~ 25 \%  & 14 & 93 ~$\pm$~ 25 \% \\
\cline{1-5}
Median Mass (Mass Range) &   \multicolumn{2}{c}{7.3~\Msun\, (2.1 to 23.8 \Msun)}    &
\multicolumn{2}{c}{12.0~\Msun\, (8.9 to 29.9 \Msun)}   \\

\enddata
\tablenotetext{a}{Calculated using the total number of foreground sources, which is 16
  (excluding the three PNe).}
\tablenotetext{b}{Calculated using the total number of Galactic center sources, which is 15.}

\end{deluxetable}

\end{document}